\documentclass[a4paper,11pt]{article}
\pdfoutput=1
\usepackage[utf8]{inputenc}
\usepackage{jheppub}
\usepackage{graphicx}
\usepackage{caption}
\usepackage{subcaption}
\usepackage{hyperref}
\usepackage[T1]{fontenc}

\newcommand{\be}{\begin{equation}}
\newcommand{\ee}{\end{equation}}
\newcommand{\bea}{\begin{eqnarray}}
\newcommand{\eea}{\end{eqnarray}}

\newcommand{\nn}{\nonumber\\}

\def\tp{\tau_{\Pi}}

\def\CA{\mathcal{A}}
\def\CB{\mathcal{B}}

\def\CD{\mathcal{D}}

\def\CG{\mathcal{G}}

\def\CL{\mathcal{L}}

\def\CN{\mathcal{N}}
\def\CT{\mathcal{T}}
\def\CO{\mathcal{O}}

\def\CW{\mathcal{W}}
\def\CZ{\mathcal{Z}}

\def\qfr{\mathfrak{q}}
\def\wfr{\mathfrak{w}}

\def\lgb{\lambda_{\scriptscriptstyle GB}}
\def\ggb{\gamma_{\scriptscriptstyle GB}}

\title{Second-order transport, quasinormal modes and zero-viscosity limit in the Gauss-Bonnet holographic fluid}

\author[a]{Sa\v{s}o Grozdanov}
\author[b]{and Andrei O. Starinets}

\affiliation[a]{Instituut-Lorentz for Theoretical Physics, Leiden University, \\ Niels Bohrweg 2,  Leiden 2333 CA, The Netherlands }

\affiliation[b]{Rudolf Peierls Centre for Theoretical Physics, University of Oxford, \\ 1 Keble Road,  Oxford OX1 3NP, United Kingdom }

\emailAdd{grozdanov@lorentz.leidenuniv.nl}
\emailAdd{andrei.starinets@physics.ox.ac.uk}

\abstract{Gauss-Bonnet holographic fluid is a useful theoretical laboratory to study the effects of curvature-squared terms in the dual gravity action on transport coefficients, quasinormal spectra and the analytic structure of thermal correlators at strong coupling. To understand the behavior and possible pathologies of the Gauss-Bonnet fluid in $3+1$ dimensions, we compute (analytically and non-perturbatively in the Gauss-Bonnet coupling) its second-order transport coefficients, the retarded  two- and three-point correlation functions of the energy-momentum tensor in the hydrodynamic regime as well as the relevant quasinormal spectrum. The Haack-Yarom universal relation among the second-order transport coefficients is violated at second order in the Gauss-Bonnet coupling. In the zero-viscosity limit, the holographic fluid still produces entropy, while the momentum diffusion and the sound attenuation are suppressed at all orders in the hydrodynamic expansion. By adding higher-derivative electromagnetic field terms to the action, we also compute corrections to charge diffusion and identify the non-perturbative parameter regime in which the charge diffusion constant vanishes.}

\preprint{OUTP-16-21P}


\begin{document} 
\maketitle
\flushbottom

\section{Introduction}
\label{sec:intro}
Gauge-string duality has been applied successfully to explore qualitative, quantitative and conceptual issues in fluid dynamics \cite{Son:2007vk,Baier:2007ix,Rangamani:2009xk,Banerjee:2012iz,Jensen:2012jh,Kovtun:2012rj,Crossley:2015tka,Haehl:2015pja,Haehl:2016pec,Haehl:2016uah}. Although the number of quantum field theories with known dual string (gravity) descriptions is very limited, their transport and spectral function properties at strong coupling can in principle be fully determined, thus giving  valuable insights into the behavior of strongly interacting quantum many-body systems. Moreover, dual gravity methods can be used to determine coupling constant dependence of a variety of physical quantities with an ultimate goal of  
interpolating between weak and strong coupling results and describing, at least qualitatively, the intermediate coupling behavior in theories of phenomenological 
interest \cite{Romatschke:2015gic,Grozdanov:2016vgg,Heller:2016rtz,Heller:2016gbp,Grozdanov:2016zjj,Andrade:2016rln}.

For generic neutral fluids, there are two independent first-order  transport coefficients (shear viscosity $\eta$ and bulk viscosity $\zeta$), 
and fifteen second-order coefficients\footnote{The existence 
of a local entropy current with non-negative divergence implies $\eta \geq 0$, $\zeta \geq 0$ \cite{landau-6} and constrains the number of independent coefficients at second order to ten \cite{Bhattacharyya:2012nq}. Alternatively, independent "thermodynamical" \cite{Moore:2012tc} terms in the hydrodynamic expansion can be derived from the generating functional without resorting to the entropy current analysis \cite{Banerjee:2012iz,Jensen:2012jh}. A computerized algorithm determining all tensor structures appearing at a given order of the 
hydrodynamic derivative expansion has been recently proposed in ref.~\cite{Grozdanov:2015kqa}. 
Modulo constraints potentially arising from the entropy current analysis (not attempted in ref.~\cite{Grozdanov:2015kqa}), 
it identifies 68 new coefficients for non-conformal neutral fluids and 20 coefficients for conformal ones at third order 
of the derivative expansion.}  (see e.g. \cite{Romatschke:2009im}). For Weyl-invariant or "conformal" fluids, the additional symmetry constraints reduce the number of transport coefficients to one at first order (shear viscosity $\eta$) and five at second order\footnote{There are no further constraints in addition to $\eta \geq 0$ coming from the non-negativity of the divergence of the entropy current in the conformal case \cite{Bhattacharyya:2012nq}. } (usually denoted $\tau_\Pi$, $\kappa$, $\lambda_1$, $\lambda_2$, $\lambda_3$). The coefficients $\eta$, $\tau_\Pi$, $\lambda_1$, $\lambda_2$ are "dynamical", whereas $\kappa$ and $\lambda_3$ are "thermodynamical" in the classification\footnote{Essentially, the coefficient is called "dynamical" if the corresponding term in the derivative expansion vanishes in equilibrium, and "thermodynamical" otherwise.} introduced in ref.~\cite{Moore:2012tc}. In the parameter regime where the dual 
Einstein gravity description of conformal fluids  is applicable 
(e.g. at infinite 't Hooft coupling $\lambda = g^2_{YM}N_c$ and infinite $N_c$ in theories such as ${\cal N}=4$ $SU(N_c)$ supersymmetric Yang-Mills (SYM) theory in $d=3+1$ dimensions), the six transport coefficients (in $d$ space-time dimensions, $d>2$)
are given by \cite{Bhattacharyya:2008mz}
\begin{eqnarray}
\eta_{\;} &=& s/4 \pi\,, \label{viscratio} \\
\tau_\Pi &=&  \frac{d}{4\pi T} \left( 1 + \frac{1}{d} \left[ \gamma_E + \psi \left( \frac{2}{d}\right) \right] \right)\,,\label{conf-tau} \\
\kappa_{\;} &=& \frac{d}{d-2} \, \frac{\eta}{2\pi T} \,, \label{conf-kappa} \\
\lambda_1 &=&  \frac{d \eta}{8\pi T}\,, \label{conf-lam-1} \\
\lambda_2 &=& \left[ \gamma_E + \psi \left( \frac{2}{d}\right) \right]\, \frac{\eta}{2\pi T}\,,  \label{conf-lam-2} \\
\lambda_3 &=& 0  \label{conf-lam-3}\,,
\end{eqnarray}
where $s$ is the entropy density, $\psi(z)$ is the logarithmic derivative of the gamma function, 
and $\gamma_E$ is 
the Euler-Mascheroni constant. Generically, one expects corrections to these formulas in (inverse) powers of the parameters 
such as $\lambda$ and $N_c$. For ${\cal N}=4$ SYM at finite temperature, the leading $\lambda^{-3/2}$ corrections to all six coefficients are known  \cite{Buchel:2004di, Benincasa:2005qc, Buchel:2008sh, Buchel:2008ac,Buchel:2008bz,Buchel:2008kd,Saremi:2011nh,Grozdanov:2014kva} (see Appendix \ref{appendixN=4SYM}, where weak and strong coupling results are discussed). Other coupling constant corrections to the results at infinitely strong t'Hooft coupling in this theory include corrections to the entropy \cite{Gubser:1998nz,Pawelczyk:1998pb}, photon emission rate \cite{Hassanain:2012uj}, and poles of the retarded correlators of the energy-momentum tensor \cite{Stricker:2013lma,Waeber:2015oka,Grozdanov:2016vgg}. Leading 
 corrections in $1/N_c^2$, intimately related to the issue of hydrodynamic "long time tails", were discussed in refs.~\cite{Kovtun:2003vj,CaronHuot:2009iq,Kovtun:2011np}, and in refs.~\cite{Myers:2008yi,Buchel:2008ae}.

In the regime of strong coupling, theories with gravity dual description appear to exhibit robust properties of transport coefficients and relations among them. 
One of such properties is the universality of shear viscosity to entropy density ratio $\eta/s = 1/4\pi$ in the limit described by a dual  gravity with two-derivative action \cite{Kovtun:2004de}, \cite{Buchel:2004qq}, \cite{Kovtun:2003wp,Buchel:2003tz,Starinets:2008fb}. Another one seems to be the Haack-Yarom relation: following the observation in ref.~\cite{Erdmenger:2008rm}, the linear combination of the second-order transport coefficients\footnote{We use notations and conventions of \cite{Baier:2007ix}. See Appendix \ref{appendixB} and  footnote 91 on page 128 of ref.~\cite{Haehl:2015pja} for clarification of sign conventions appearing in the literature.}
\begin{equation}
\label{HY}
H \equiv 2 \eta \tp - 4\lambda_1 - \lambda_2
\end{equation}
 was proven to vanish in all conformal theories dual to two-derivative gravity\footnote{Note that all transport coefficients in $H$ are "dynamical" in terminology of ref.~\cite{Moore:2012tc}.}
   \cite{Haack:2008xx}. Eqs. ~(\ref{conf-tau}), (\ref{conf-lam-1}), (\ref{conf-lam-2}) show this explicitly. Somewhat surprisingly, the Haack-Yarom relation continues to hold to next to leading order in the strong coupling expansion, at least in ${\cal N}=4$ $SU(N_c)$ supersymmetric Yang-Mills theory in $d=3+1$ dimensions in the limit of infinite $N_c$ \cite{Grozdanov:2014kva}, in theories dual to curvature-squared gravity \cite{Grozdanov:2014kva}, in particular, in the Gauss-Bonnet holographic liquid\footnote{As advertised in ref.~\cite{Grozdanov:2014kva} and shown below (and, independently, in 
ref.~\cite{Shaverin:2015vda} using fluid-gravity duality methods), the Haack-Yarom relation does
 not hold non-pertutbatively in the Gauss-Bonnet coupling.
} (perturbatively in the Gauss-Bonnet coupling) \cite{Shaverin:2012kv}. It was shown recently that the result $H=0$ continues to hold for non-conformal liquids along 
the dual gravity RG flow \cite{Kleinert:2016nav}.\footnote{It appears that at weak coupling, 
the relation $H=0$ does not hold. We briefly review the results at weak coupling in Appendix \ref{appendixN=4SYM}.} It remains to be seen whether such robustness extends to higher-order transport coefficients and/or other properties of strongly coupled finite temperature theories and whether it is related to the presence of event horizons in dual gravity.\footnote{We would like to thank P.~Kovtun and M.~Rangamani for a discussion of these issues.}

Monotonicity and other properties of transport coefficients are of interest for studies of near-equilibrium behavior at strong coupling, in particular, thermalization, and for attempts to uncover a universality similar to the one exhibited by the ratio of shear viscosity to entropy density. Monotonicity of transport coefficients or their dimensionless combinations may seem more exotic than the monotonicity of central charges \cite{Zamolodchikov:1986gt,Komargodski:2011vj} or the free energy \cite{Klebanov:2011gs,Pufu:2016zxm}, yet it is often an observed property, at least in a given state of aggregation \cite{Kovtun:2004de,Csernai:2006zz}.

In ${\cal N}=4$ SYM 
at infinite\footnote{At large but finite $N_c$, and large $\lambda$, the result for $\eta/s$ is also corrected by the term proportional to $\lambda^{1/2}/N^2_c$ \cite{Myers:2008yi,Buchel:2008ae}.} $N_c$, the shear viscosity to entropy density ratio appears to be a monotonic function of the coupling \cite{Kovtun:2004de}, with the correction to the universal infinite coupling result being positive  \cite{Buchel:2004di,Buchel:2008sh},
\begin{equation}
\frac{\eta}{s} = \frac{1}{4\pi} \left( 1 + 15\zeta (3) \lambda^{-3/2}+ \ldots \right)\,.
\end{equation}
Subsequent calculations revealed that the corrections coming from higher derivative terms in the gravitational action can have either 
sign \cite{Kats:2007mq, Brigante:2007nu}. For the action with generic curvature squared higher derivative terms 
\begin{align}\label{R2action}
S_{R^2} = \frac{1}{2 \kappa_5^2 } \int d^5 x \sqrt{-g} \left[ R - 2 \Lambda + L^2 \left( \alpha_1 R^2 + \alpha_2 R_{\mu\nu} R^{\mu\nu} +
\alpha_3 R_{\mu\nu\rho\sigma} R^{\mu\nu\rho\sigma}  \right) \right],
\end{align}
where the cosmological constant $\Lambda = - 6 / L^2$, the 
shear viscosity - entropy density 
ratio is\footnote{All second-order transport coefficients for theories dual to the background (\ref{R2action}) have been computed in 
ref.~\cite{Grozdanov:2014kva}.} \cite{Kats:2007mq, Brigante:2007nu}
\begin{equation}
\frac{\eta}{s} = \frac{1}{4\pi} \left( 1 - 8 \alpha_3  \right) + O \left(\alpha_i^2\right).
\end{equation}
The sign of the coefficient $\alpha_3$ affects not only viscosity but also the analytic structure of  correlators in the dual thermal field theory \cite{Grozdanov:2016vgg}. 

Corrections to Einstein gravity results computed from generic higher-order derivative terms in the dual gravitational action can be trusted so long as  they remain (infinitesimally) small relative to the leading order result, as they are obtained by treating the higher-derivative terms in the equations of motion perturbatively. This limitation 
arises due to Ostrogradsky instability and other related pathologies such as ghosts associated with higher-derivative actions \cite{Ostrogradsky}, \cite{pais},  \cite{Stelle:1977ry}, \cite{Zwiebach:1985uq} (see also refs.~\cite{Woodard:2006nt}, \cite{Woodard:2015zca} for a modern 
discussion of Ostrogradsky's theorem, and ref.~\cite{ostrogradsky-bio} for an interesting historical account of Ostrogradsky's life and work). One may be tempted to lift the constraints imposed by Ostrogradsky's  theorem by considering actions in which coefficients in front of higher derivative terms conspire to give equations of motion no higher than second-order in derivatives as happens e.g. in Gauss-Bonnet gravity in dimension $D>4$ or, more generally, Lovelock gravity \cite{Lovelock:1971yv}. Gauss-Bonnet (and Lovelock) gravity 
 has been used as a  laboratory for non-perturbative studies of higher derivative curvature effects on transport coefficients of conformal fluids with holographic duals \cite{Kats:2007mq,Brigante:2007nu,Brigante:2008gz,Buchel:2009tt,Buchel:2009sk,deBoer:2009gx,Camanho:2009hu,Camanho:2010ru,Camanho:2013pda},
\cite{Grozdanov:2016vgg}. In particular, the celebrated result for the shear viscosity-entropy ratio in a (hypothetical) conformal 
fluid dual to $D=5$ Gauss-Bonnet gravity  \cite{Brigante:2007nu},
 \begin{equation}
\frac{\eta}{s} = \frac{1 - 4 \lgb}{4\pi}, 
\label{gbviscosity}
\end{equation}
has been obtained {\it non-perturbatively} in the Gauss-Bonnet coupling $\lgb$. The result would imply that there exist CFTs whose viscosity can be tuned all the way to zero in the regime described by a dual {\it classical} (albeit non-Einsteinian) gravity. It was found, however, that for $\lgb$ outside of the interval 
\begin{align}
\label{brigante-bound}
-\frac{7}{36} \leq \lgb \leq \frac{9}{100},
\end{align}
the dual field theory exhibits pathologies associated with superluminal propagation of modes at high momenta or negativity of the energy flux in a dual 
CFT \cite{Brigante:2007nu,Brigante:2008gz,Hofman:2008ar,Buchel:2009tt,Hofman:2009ug,Buchel:2009sk}. For Gauss-Bonnet gravity in $D$ dimensions ($D\geq 5$), the result \eqref{gbviscosity} generalizes to \cite{Camanho:2009vw,Buchel:2009sk}
\begin{equation}
\frac{\eta}{s} = \frac{1}{4\pi} \left[ 1 - \frac{2(D-1)}{D-3} \lgb\right]
\label{gbviscosity-D}
\end{equation}
and the inequalities corresponding to  eq.~\eqref{brigante-bound} become\footnote{Curiously, in the $D\rightarrow \infty$ limit, the range  \eqref{brigante-bound-D} is $-3/4\leq \lgb \leq 1/4$. Note that the black brane metric is well defined for $\lgb \in (-\infty,1/4]$ for any $D$. We shall only consider $D=5$ in the rest of the paper.} \cite{Camanho:2009vw,Buchel:2009sk}
\begin{align}
\label{brigante-bound-D}
-\frac{(3D-1)(D-3)}{4(D+1)^2} \leq \lgb \leq \frac{(D-3)(D-4)(D^2-3D+8)}{4(D^2-5D+10)^2}\,.
\end{align}
Given the constraints \eqref{brigante-bound-D} and {\it monotonicity} of $\eta/s$ in \eqref{gbviscosity-D}, one may conjecture a GB gravity bound on $\eta/s$ \cite{Camanho:2009vw,Buchel:2009sk},
\begin{equation}
\frac{\eta}{s} \geq \frac{1}{4\pi} \left[ 1 - \frac{(D-1)(D-4)(D^2-3D+8)}{2(D^2-5D+10)^2}\right],
\label{gbviscosity-bound-D}
\end{equation}
instead of the Einstein's gravity bound $\eta/s\geq 1/4\pi$. For $3+1$-dimensional CFTs, the GB bound 
would imply $\eta/s\geq (0.640)/4\pi$ \cite{Brigante:2008gz}. Recently, the constraints \eqref{brigante-bound} were confirmed and generalized to Gauss-Bonnet black holes with spherical (rather than planar) horizons by considering boundary causality and bulk hyperbolicity violations in Einstein-Gauss-Bonnet gravity \cite{Andrade:2016yzc}. Since these causality problems arise in the ultraviolet, one may  hope that treating Gauss-Bonnet gravity as a low energy theory with unspecified ultraviolet completion would allow one to consider its hydrodynamic (infrared) limit 
without worrying about causality violating ultraviolet modes, i.e. that  it is in principle 
possible to cure the problems in the ultraviolet without affecting the hydrodynamic (infrared) regime (one may also try to construct a theory with a low temperature phase transition breaking the link between the hydrodynamic IR and causality breaking UV modes \cite{Buchel:2010wf}). However, a reflection on the recent analysis by Camanho {\it et al.} \cite{Camanho:2014apa} of the {\it bulk} causality violation in higher derivative gravity seems to imply that, provided the relevant conclusions of  ref.~\cite{Camanho:2014apa} are correct\footnote{See refs.~\cite{Reall:2014pwa, Willison:2014era, Papallo:2015rna,Willison:2015vaa,Cheung:2016wjt,Andrade:2016yzc,Afkhami-Jeddi:2016ntf} for recent discussions.}, a reliable treatment of Gauss-Bonnet terms beyond perturbation theory for the purposes of fluid dynamics is not possible. The Einstein-Gauss-Bonnet action in $D=5$ is given by 
\begin{align}\label{GBaction}
S_{GB} = \frac{1}{2\kappa_5^2} \int d^5 x \sqrt{-g} \left[ R - 2 \Lambda + \frac{\lgb l_{\scriptscriptstyle GB}^2}{2} 
\left( R^2 - 4 R_{\mu\nu} R^{\mu\nu} + R_{\mu\nu\rho\sigma} R^{\mu\nu\rho\sigma} \right) \right],
\end{align}
where the cosmological constant $\Lambda = - 6 / L^2$, and $l_{GB}$ is the scale of the 
Gauss-Bonnet term which {\it a priori} is not necessarily related to the cosmological constant scale set by $L$.
As argued in ref.~\cite{Camanho:2014apa}, the generic bulk causality violations in Gauss-Bonnet classical gravity  can only be cured by including an infinite set of higher spin fields with masses squared
 $m_s^2 \propto 1/ \lgb l_{\scriptscriptstyle GB}^2$. Integrating out these fields to obtain a low energy effective theory would lead to an infinite series of additional higher derivative terms in the gravitation action. Schematically, the modified action would have the form
\begin{align}\label{GBaction-mod}
S_{GB, mod} = \frac{1}{2\kappa_5^2} \int d^5 x \sqrt{-g} \left[ R - 2 \Lambda + \sum\limits_{k=1}^{\infty} c_k\,   \lgb^k \, l_{\scriptscriptstyle GB}^{2k}\,  {\cal R}^{k+1} \right].
\end{align}
Considering a specific solution (e.g. a black brane whose scale is set by the cosmological constant) and 
rescaling the coordinates $x \rightarrow \bar{x}=x/L$ leads to 
\begin{align}\label{GBaction-mod-L}
S_{GB, mod} = \frac{L^3}{2\kappa_5^2} \int d^5 \bar{x} \sqrt{-\bar{g}} \left[ \bar{R} +12 + \sum\limits_{k=1}^{\infty} c_k\,  \bar{\lambda}_{\scriptscriptstyle GB}^k \, \bar{{\cal R}}^{k+1} \right],
\end{align}
where\footnote{In considering Gauss-Bonnet black hole solutions, it is convenient to set $l_{\scriptscriptstyle GB} = L$. Then 
$\bar{\lambda}_{\scriptscriptstyle GB}= \lgb$.} $\bar{\lambda}_{\scriptscriptstyle GB}= \lgb l_{\scriptscriptstyle GB}^2/L^2$. To suppress contributions (e.g. to transport coefficients) coming from the (unknown) terms with $k>1$, one has to assume $\bar{\lambda}_{\scriptscriptstyle GB} \ll 1$. This is similar to the condition $l_s/L \ll 1$ in the usual top-down holography. Thus, generically one may expect results such as (\ref{gbviscosity}) to be potentially corrected by terms $O(\lgb^2)$ and/or higher, and therefore be reliable only for $\lgb \ll 1$. It seems, therefore, that one essentially cannot escape the Ostrogradsky problem (at least not in classical gravity) 
 by engineering a specific higher-derivative Lagrangian with second-order equations of motion. An alternative view of the aspects of the analysis in ref.~\cite{Camanho:2014apa} has been advocated in refs.~\cite{Reall:2014pwa, Papallo:2015rna} (see also \cite{Willison:2014era, Willison:2015vaa} and \cite{Andrade:2016yzc}). Our approach to these problems will be purely pragmatic:\footnote{We would like to thank P.~Kovtun for his incessant criticism of using Gauss-Bonnet gravity in holography.} we shall {\it a priori} 
 ignore any existing or debated constraints on the Gauss-Bonnet coupling and explore the influence of curvature-squared terms on quasinormal spectra and transport coefficients for all range of the coupling allowing a black brane solution, i.e. for $\lgb \in (-\infty,1/4]$ (see section  \ref{Sec:TwoPointAndQNM}). In particular, we are interested in revealing any generic features the presence of higher-curvature terms in the action may have (as pointed out in ref.~\cite{Grozdanov:2016vgg}, the spectra of $R^2$ and $R^4$ backgrounds exhibit qualitatively similar novel features not present in Einstein's gravity). We use the action \eqref{GBaction} (with $l_{\scriptscriptstyle GB} = L$) to compute transport coefficients, quasinormal spectrum and thermal correlators  analytically and non-perturbatively in Gauss-Bonnet coupling, fully exploiting the advantage of having to deal with second-order equations of motion in the bulk. Different techniques will be used to compute Gauss-Bonnet transport: fluid-gravity duality, Kubo formulae applied to two- and three-point correlators, and quasinormal modes.
We find that only the three-point functions method allows to determine all the coefficients analytically: other approaches face technical difficulties we were not able to resolve. In a hypothetical dual CFT, constraints on Gauss-Bonnet coupling considered e.g. in ref.~\cite{Buchel:2009sk} manifest themselves in the superluminal propagation of high-momentum modes for $\lgb$ outside of the interval \eqref{brigante-bound-D}. In the far more stringent scenario of ref.~\cite{Camanho:2014apa}, one may expect to detect anomalous behavior in the regime of small frequencies and momenta and in 
transport coefficients. Accordingly, we shall look for pathologies in the hydrodynamic behavior of the model at finite values of 
$\lgb$ indicating the lack of ultraviolet completion and the potential need for corrections coming from the unknown terms in (\ref{GBaction-mod-L}). 

The full non-perturbative set of first- and second-order Gauss-Bonnet transport coefficients can be determined analytically and is given by\footnote{The shear viscosity $\eta$ as a function of temperature and $\ggb$ is given in Eq.~\eqref{etagb}.} 
\begin{align}
&\eta = s \ggb^2/4 \pi\,, \label{gb-visc} \\
&\tp = \frac{1}{2\pi T} \left( \frac{1}{4} \left(1+\ggb\right) \left( 5+\ggb - \frac{2}{\ggb}\right) - \frac{1}{2} \ln \left[\frac{2 \left(1+\ggb\right)}{\ggb }\right]  \right) , \label{l0}\\
&\kappa  = \frac{\eta}{\pi T} \left( \frac{\left(1+\ggb\right)\left(2\ggb^2-1\right)}{2\ggb^2} \right)\,,  \label{l4} \\
&\lambda_1 = \frac{\eta}{2\pi T } \left( \frac{ \left(1+\ggb\right)\left(3 - 4\ggb +2\ggb^3\right)}{ 2\ggb^2 } \right) , \label{l1} \\   
&\lambda_2 = - \frac{\eta}{\pi T} \left( - \frac{1}{4} \left(1+\ggb \right) \left(1+ \ggb - \frac{2}{\ggb} \right) + \frac{1}{2} \ln \left[\frac{2 \left(1+\ggb\right)}{\ggb }\right]  \right),  \label{l2} \\
&\lambda_3 = -\frac{\eta}{\pi T}   \left( \frac{\left(1+\ggb\right)\left(  3+\ggb-4\ggb^2   \right)}{\ggb^2}  \right), \label{l3} 
\end{align}
where we have defined 
\begin{align}
\label{Gamma}
\ggb \equiv \sqrt{1-4\lgb}\,.
\end{align}
An alternative way of writing the Gauss-Bonnet second-order coefficients is given by Eqs. \eqref{cosm1} -- \eqref{cosm4}. In the limit of $\lgb \rightarrow 0$ ($\ggb \rightarrow 1$), which corresponds to Einstein's gravity, one recovers the standard results for infinitely 
strongly coupled conformal fluids in $3+1$ dimensions given by eqs.~\eqref{tc1} and \eqref{tc2} in appendix \ref{appendixN=4SYM}. The result for $\eta$ was obtained in ref.~\cite{Brigante:2007nu} 
and the relaxation time $\tau_\Pi$ was first found numerically in ref.~\cite{Buchel:2009tt}. Coefficients $\tau_\Pi$ and $\kappa$ were previously computed analytically in 
ref.~\cite{Banerjee:2011tg}, and we have reported $\lambda_1$, $\lambda_2$, $\lambda_3$ 
in ref.~\cite{Grozdanov:2015asa}. To linear order in $\lgb$, the results coincide\footnote{The notations used in 
  ref.~\cite{Shaverin:2012kv} are related to the ones in this paper by
$\lambda_0=\eta \tau_\Pi$, $\delta = 4 \lgb$ and $\kappa_5^2=8\pi G_5$.} 
 with those found in ref.~\cite{Shaverin:2012kv}. 
 
Using the results (\ref{gb-visc}), (\ref{l0}), (\ref{l1}), (\ref{l2}), we find the Haack-Yarom function 
in Gauss-Bonnet gravity 
\begin{align}
\label{SecondOrderUniViolation}
H(\lgb) = - \frac{\eta}{\pi T} \frac{\left(1-\ggb\right) \left(1 - \ggb^2 \right) \left(3+2\ggb\right)}{\ggb^2}  = - \frac{40 \lgb^2 \eta}{\pi T}\, + \CO\left(\lgb^3\right).
\end{align}
Curiously, $H(\lgb)\leq 0$ for the Gauss-Bonnet holographic liquid. Whether $H(\lgb)$ is corrected beyond leading order by terms coming from (\ref{GBaction-mod-L}) remains an open question: {\it a priori}, we do not know if $H$ must vanish beyond the Einstein gravity approximation.

Computing the energy-momentum tensor correlation functions in holographic models with higher-derivative dual gravity terms, one finds a new pole on the imaginary frequency axis. This pole, first found in the quasinormal spectrum analysis of ref.~\cite{Grozdanov:2016vgg}, is  moving 
from the complex infinity closer and closer to the origin as the parameter in front of the higher-derivative term in the action 
(such as $\lgb$ in eq.~\eqref{GBaction}) increases, and can be approximated analytically in the small-frequency expansion. The poles of this type appear to be generic in higher-derivative gravity: they are present in $R^2$ and $R^4$ gravity, and their behavior is qualitatively similar \cite{Grozdanov:2016vgg}.

Another interesting feature of Gauss-Bonnet holographic liquid is the zero-viscosity limit. In ref.~\cite{Bhattacharya:2012zx}, Bhattacharya {\it et al.} 
suggested the existence of a non-trivial second-order non-dissipative hydrodynamics, i.e. a theory whose fluid dynamics derivative expansion has no contribution to entropy production while still having some of the transport coefficients non-vanishing.\footnote{The authors of \cite{Bhattacharya:2012zx} considered an effective field theory approach \cite{Dubovsky:2005xd,Dubovsky:2011sj} to non-dissipative uncharged second-order hydrodynamics. The approach relies on a classical effective action and standard variational techniques to derive the energy-momentum tensor. It is thus unable to incorporate dissipation. The inclusion of dissipation into the description of hydrodynamics, using the same effective description, was analysed in \cite{Endlich:2012vt,Grozdanov:2013dba}.} For conformal fluids, the classification of  \cite{Bhattacharya:2012zx} implies the existence of a four-parameter family of non-trivial non-dissipative fluids with $\eta =0$ and  non-vanishing coefficients $\tau_\Pi$, $\kappa$, $\lambda_1 = \kappa /2$, $\lambda_2$ and $\lambda_3$. Given the result (\ref{gbviscosity}), the hypothetical theory dual to Gauss-Bonnet gravity in the  limit of $\lgb \rightarrow 1/4$ is a natural candidate for a dissipationless fluid (ignoring for a moment any potential corrections coming from (\ref{GBaction-mod-L})). In the limit of $\lgb \rightarrow 1/4$ ($\ggb \rightarrow 0$) we find \cite{Grozdanov:2015asa}
\begin{align}
\eta\tp  =  0 ,&& \lambda_1 =   \frac{3 \pi^2 T^2}{2 \sqrt{2}\kappa_5^2} ,
&&\lambda_2 =  0 ,&&\lambda_3 = - \frac{3 \sqrt{2} \pi^2 T^2}{\kappa_5^2} ,
&& \kappa = - \frac{\pi^2 T^2}{\sqrt{2} \kappa_5^2} .
\label{zero-gamma}
\end{align}
At first glance, this result realizes the dissipationless liquid scenario outlined in ref.~\cite{Bhattacharya:2012zx}: the shear and bulk viscosities are zero while some of the second-order coefficients are not. However, the relationship $\kappa = 2 \lambda_1$, which is required for ensuring zero entropy production, does not hold among the coefficients in (\ref{zero-gamma}). We therefore conclude that the holographic Gauss-Bonnet liquid does not fall into the class of non-dissipative liquids discussed in ref.~\cite{Bhattacharya:2012zx}. This may be a hint that the corrections from (\ref{GBaction-mod-L}) must indeed be included.

The paper is organised as follows. In section~\ref{Sec:TwoPointAndQNM} we analyze the finite-temperature two-point correlation functions of energy-momentum tensor in the theory dual to Gauss-Bonnet gravity as well as the relevant quasinormal modes in the scalar, shear and sound channels of metric perturbations, including the new pole on the imaginary axis at finite coupling $\lgb$. Kubo formulas determine the coefficients $\eta$, $\kappa$ and $\tau_\Pi$. The shear channel quasinormal frequency is used to confirm the results for
 $\eta$ and $\tau_\Pi$, and to find the third-order transport coefficient $\theta_1$. We discuss the limit $\lgb \rightarrow 1/4$, where the full quasinormal spectrum can be found analytically, and the limit $\lgb \rightarrow -\infty$. In section \ref{Sec:Fluid/gravity}, we apply the fluid-gravity duality technique to compute the 
 Gauss-Bonnet transport coefficients. All coefficients except $\kappa$ can be determined in this approach. However, due to technical difficulties, all of them with the exception of $\eta$ can be found only perturbatively as series in $\lgb$. A more efficient method of three-point functions is considered  in  section~\ref{section:three-point functions}, where all the coefficients are computed analytically and non-perturbatively, and we also 
 discuss the monotonicity properties of the coefficients and the zero-viscosity limit. Finally, in section \ref{sec:ChargeDiffSection}
we discuss the influence of higher derivative terms on charge diffusion in the most general four derivative Einstein-Maxwell theory. Section  \ref{sec:conclusions} 
with conclusions is followed by several appendices:  in Appendix \ref{appendixN=4SYM}, a brief summary of second-order transport coefficients in  ${\cal N}=4$ SYM at weak and strong coupling is given. A comparison of notations and  conventions used in the literature on second-order hydrodynamics and, specifically, in the discussion of Haack-Yarom relation 
is given in Appendix \ref{appendixB}. In Appendix \ref{sec:BC-horizon} we outline the procedure of setting the boundary conditions at the 
horizon in hydrodynamic approximation. Appendices  \ref{appendixC} and \ref{sec:EMGB} contain some technical results.

\section{Energy-momentum tensor correlators and quasinormal modes of Gauss-Bonnet holographic fluid}
\label{Sec:TwoPointAndQNM}
The coefficients of the four-derivative terms in the Gauss-Bonnet action \eqref{GBaction} ensure that 
the corresponding equations of motion contain only second derivatives of the metric. The equations are given by 
\begin{align}\label{GBeom}
E_{\mu\nu} \equiv& \; R_{\mu\nu} - \frac{1}{2} g_{\mu\nu} R + g_{\mu\nu} \Lambda - \frac{\lgb L^2}{4} g_{\mu\nu} \left( R^2 - 4 R_{\mu\nu} R^{\mu\nu} + R_{\mu\nu\rho\sigma} R^{\mu\nu\rho\sigma}  \right) \nn  
&+ \lgb L^2 \left( R R_{\mu\nu} - 2 R_{\mu\alpha} R_{\nu}^{~\alpha} - 2 R_{\mu\alpha\nu\beta} R^{\alpha\beta} + R_{\mu\alpha\beta\gamma} R_\nu^{\alpha\beta\gamma}  \right)  =  0\,.
\end{align}
The equations \eqref{GBeom} admit a black brane solution\footnote{Exact solutions and thermodynamics of black branes and black holes in Gauss-Bonnet gravity were considered in \cite{Cai:2001dz} (see also refs.~\cite{Cvetic:2001bk,Nojiri:2001aj,Cho:2002hq,Neupane:2002bf,Neupane:2003vz}).}
\begin{align}\label{BB}
ds^2 = - f(r) N^2_{GB} dt^2 + \frac{1}{f(r)} dr^2 + \frac{r^2}{L^2} \left(dx^2 + dy^2 +dz^2 \right),
\end{align}
where 
\begin{align}
\label{f-black}
f(r) = \frac{r^2}{L^2} \frac{1}{2\lgb} \left[1 - \sqrt{1-4\lgb \left(1 - \frac{r^4_+}{r^4} \right) } \right].
\end{align}
The arbitrary constant $N_{\scriptscriptstyle GB}$ will be set to normalize the speed of light at the boundary  (i.e. in the dual CFT) to unity,
\begin{align}\label{NhashDef}
N_{\scriptscriptstyle GB}^2 = \frac{1}{2} \left(1+\sqrt{1-4\lgb} \right),
\end{align}
and we henceforth use this value. The solution with $r_+=0$ corresponds to the AdS vacuum metric in Poincar\'{e} coordinates 
with the AdS curvature scale squared $\tilde{L}^2=L^2/f_\infty$ \cite{Buchel:2009sk}, where 
\begin{equation} 
f_\infty = \lim_{r\rightarrow \infty} f(r) = \frac{1-\sqrt{1-4\lgb}}{2\lgb} = \frac{2}{1+\ggb}\,. 
\end{equation}
The parameter $\ggb$ is defined in Eq.~(\ref{Gamma}). We shall use $\lgb$ and $\ggb$ interchangeably, 
and set $L=1$ in the rest of the paper unless stated otherwise. The Hawking temperature, the entropy density and the energy density associated with the black 
brane background \eqref{BB}  are given, correspondingly, by 
\begin{align}
&T = N_{\scriptscriptstyle GB} \frac{r_+}{\pi L^2} = \frac{r_+}{\sqrt{2} \pi L^2 } \sqrt{ 1+\ggb} = \frac{r_+}{\pi \tilde{L}^2} \left( \frac{1+\ggb}{2}\right)^{3/2}, \label{EntropyAndEnergy} \\
&s =  \frac{2 \pi}{\kappa_5^2} \left(\frac{r_+}{L}\right)^3 = \frac{4\sqrt{2}\pi^4 L^3}{\kappa_5^2} \frac{T^3}{\left(1+\ggb\right)^{3/2} } = 
\frac{16\pi^4 \tilde{L}^3}{\kappa_5^2} \frac{T^3}{\left(1+\ggb\right)^{3} }\, , 
\label{EntropyAndEnergy3} \\
&\varepsilon =3 P = \frac{3}{4} T s\,. \label{EntropyAndEnergy2} 
\end{align}
The metric \eqref{BB} is well defined  for $\lgb \in (-\infty,1/4]$  (or $\ggb \in [0,\infty)$, with the interval of positive $\lgb$  corresponding to  the interval $\ggb \in [0,1)$). We note that $s/T^3$ is a monotonically {\it decreasing} function of $\ggb$ in the interval $\ggb \in [0,\infty)$.

The holographic dictionary relating the coupling $\lgb$ of Gauss-Bonnet gravity in $D$ dimensions to the parameters of the dual CFT has been thoroughly discussed in ref.~\cite{Buchel:2009sk} (see also the comprehensive discussion of the $D=5$ case in ref.~\cite{Buchel:2008vz}). For a class of four-dimensional CFTs (usually characterized by the central charges $c$ and $a$), there exists a parameter regime (e.g. $\lambda \gg N_c^{2/3} \gg 1$ \cite{Kats:2007mq,Buchel:2009sk}) in which the dual description is given by Einstein gravity with a negative cosmological constant plus curvature squared terms {\it treated as small perturbations}, so that e.g. the coefficient $\alpha_3$ in the action \eqref{R2action} is $\alpha_3 \sim (c-a)/c \sim 1/N_c \ll 1$, as in the discussion of the superconformal ${\cal N}=2$ $Sp(N_c)$ gauge theory with four fundamental and one antisymmetric traceless hypermultiplets by Kats and Petrov\footnote{Other examples, as well as the string theory origins of the curvature-squared terms in the effective action are discussed in ref.~\cite{Buchel:2008vz}.}
  \cite{Kats:2007mq}. For {\it finite} $\lgb$, if a dual CFT exists at all, one may relate the Gauss-Bonnet coupling to the parameters characterizing two- and three-point functions of the energy-momentum tensor in the CFT \cite{Buchel:2009sk}. In particular, the holographic calculation  \cite{Buchel:2009sk} gives the central charge $c$ 
\begin{equation} 
c = \frac{\pi^2 \tilde{L}^3}{\kappa_5^2} \ggb  .
\end{equation}
Note that the central charge is a monotonically {\it increasing} non-negative function of $\ggb$ in the interval $\ggb \in [0,\infty)$, with $c=0$ at $\ggb=0$ (i.e. at $\lgb=1/4$). Generically, we may expect $\lgb$ to be a function of both $\lambda$ and $N_c$ at large but finite values of these parameters.

 We compute the retarded two-point functions $G_{\mu\nu,\rho\sigma}^R$ of the energy-momentum tensor in a hypothetical finite-temperature $4d$ CFT  dual to the Gauss-Bonnet background \eqref{BB} following the standard holographic recipe \cite{Son:2002sd,Policastro:2002se,Policastro:2002tn,Kovtun:2005ev}. Gravitational quasinormal modes of the background corresponding to the poles of the correlators $G_{\mu\nu,\rho\sigma}^R$ \cite{Son:2002sd,Kovtun:2005ev} have been computed 
and analyzed in detail as a function of the Gauss-Bonnet parameter $\lgb$ in ref.~\cite{Grozdanov:2016vgg}. The quasinormal spectrum at  $\lgb=1/4$ is computed analytically in section \ref{sec:GBExtremeCoupling} of 
the present paper.

The full gravitational action needed to compute the correlators contains the 
Gibbons-Hawking term and the counter-term required by the holographic renormalisation,
\begin{align}\label{FullAction}
S = S_{GB} + S_{GH} + S_{c.t.},
\end{align}
where $S_{GB}$ is the Gauss-Bonnet action  \eqref{GBaction}, the modified Gibbons-Hawking term is given by
\begin{align}\label{GHterm}
S_{GH} = - \frac{1}{\kappa_5^2} \int d^4 x \sqrt{-\gamma} \left[ K + \lgb \left( J - 2 G_\gamma^{\mu\nu} K_{\mu\nu} \right)\right],
\end{align}
and the counter-term action is (see e.g. \cite{Brihaye:2008kh})
\begin{align}\label{GBct}
S_{c.t.} = \frac{1}{\kappa_5^2} \int d^4 x \sqrt{-\gamma} \left( c_1 - \frac{c_2}{2} R_\gamma   \right),
\end{align}
where
\begin{align}\label{CTCoefficients}
c_1 = - \frac{\sqrt{2} \left(2+\sqrt{1-4\lgb} \right)}{\sqrt{1+\sqrt{1-4\lgb} } }, &&c_2 = \sqrt{\frac{\lgb}{2}} \frac{\left(3 - 4\lgb -3 \sqrt{1-4\lgb} \right)}{\left(1 - \sqrt{1-4\lgb} \right)^{3/2}} .
\end{align}
Here $\gamma_{\mu\nu} = g_{\mu\nu} - n_\mu n_\nu$ is the induced metric on the boundary, $n^\mu$ is the vector normal to the boundary, i.e. $n_\mu = \delta_{\mu r} / \sqrt{ g^{rr} }$, $R_\gamma$ is the induced Ricci scalar and $G^{\mu\nu}_\gamma$ is the induced Einstein tensor on the boundary. 
The extrinsic curvature tensor is  
\begin{align}\label{KTen}
K_{\mu\nu} = - \frac{1}{2} \left( \nabla_\mu n_\nu + \nabla_\nu n_\mu \right),
\end{align}
$K$ is its trace and the tensor $J_{\mu\nu}$ is defined as
\begin{align}\label{JTen}
J_{\mu\nu} = \frac{1}{3} \left( 2 K K_{\mu\rho} K^{\rho}_{~\nu} + K_{\rho\sigma} K^{\rho\sigma} K_{\mu\nu} - 2 K_{\mu\rho} K^{\rho\sigma} K_{\sigma\nu} - K^2 K_{\mu\nu} \right).
\end{align}
Similarly, $J$ denotes the trace of $J_{\mu\nu}$.

Due to rotational invariance, we may choose the fluctuations $h_{\mu\nu}$ of the background metric to have the momentum along the $z$ axis, i.e. 
we can  set $h_{\mu\nu} = h_{\mu\nu} (r) e^{-i t \omega + i q z}$, which enables us to introduce the three independent 
gauge-invariant combinations of the metric components \cite{Kovtun:2005ev}---scalar ($Z_1$), shear ($Z_2$) and sound ($Z_3$):
\begin{align}
  &Z_1 = h^x_{~y}\,, \label{eq:GinvZ1}   \\
  &Z_2 = \frac{q }{r^2} h_{tx} + \frac{\omega}{ r^2}  h_{xz}\,, \label{eq:Ginv4g2} \\
 &Z_3 = \frac{2 q^2}{r^2 \omega^2} h_{tt} +\frac{4 q}{r^2 \omega} h_{tz}  - \left(  1 - \frac{q^2 N_{\scriptscriptstyle GB}^2 \left(4 r^3 - 2 r f(r)\right)}{2 r \omega^2 \left(r^2 - 2 \lgb f(r)\right)}   \right) \left( \frac{h_{xx}}{r^2} + \frac{h_{yy}}{r^2} \right) + \frac{2}{r^2} h_{zz}\, .  \label{eq:GinvZ3}
\end{align}
Throughout the calculation, we use the radial gauge $h_{r\mu} = 0$ and the standard dimensionless expressions for the frequency and the 
spatial momentum
\begin{align}
\wfr = \frac{\omega}{2\pi T}, &&  \qfr = \frac{q}{2\pi T}\,.
\label{wq-gothic}
\end{align}
By symmetry, the 
equations of motion obeyed by the three functions $Z_1$, $Z_2$, $Z_3$ decouple \cite{Kovtun:2005ev}. Introducing the new variable $u = r_0^2/r^2$, 
the equation of motion in each of the three channels can be written in the form of a linear second-order differential equation
\begin{align}
\partial_u^2 Z_i + A_i \partial_u Z_i + B_i Z_i = 0\,,
\label{eq:eom_GB_ginv}
\end{align}
where $i=1,2,3$ and the coefficients $A_i$ and $B_i$ are given in Appendix \ref{appendixC}.  For some applications, especially in fluid-gravity duality, it will be convenient to 
use yet another radial variable, $v$, defined by \cite{Brigante:2007nu}
\begin{align}
v = 1 - \sqrt{ 1 - \left(1-u^2\right)\left(1 - \ggb^2 \right) },
\label{var-v}
\end{align}
so that the horizon is at $v = 0$ and the boundary at $ v = 1 - \ggb$. The new coordinate is singular at zero 
Gauss-Bonnet coupling, $\lgb = 0$ ($\ggb = 1$), thus the results for $\lgb =0$, which are identical to those of  $\CN = 4$ SYM theory at infinite 't Hooft coupling and infinite $N_c$, have to be obtained independently.

On shell, the action \eqref{FullAction} reduces to the surface terms, 
\begin{align}\label{OSActDef}
S = S_{horizon} + S_{\partial M},
\end{align}
where the contribution from the horizon should be discarded \cite{Son:2002sd}, \cite{Herzog:2002pc}. In terms of the gauge-invariant variables
\eqref{eq:GinvZ1}, \eqref{eq:Ginv4g2} and \eqref{eq:GinvZ3}, the part of the action involving derivatives of the fields can be written as 
\begin{align}\label{Sonshell}
S_{\partial M} = \lim_{\epsilon\to 0} \left\{ \frac{\pi^2 T^2}{8 \kappa_5^2} \sum_{i=1}^3 \int \frac{d\omega dq}{(2\pi)^2} \,\CA_i (\epsilon, \omega, q) \CZ_i (\epsilon, -\omega, -q)  \CZ_i' (\epsilon, \omega,q) + \cdots \right\},
\end{align}
where $\CZ'$ is the derivative of $\CZ(u,\omega,q)$ with respect to the radial coordinate. The functions $\CA_i$ include the boundary 
contributions from the parts $S_{GB}$ and $S_{GH}$ of 
the action \eqref{FullAction}, but not from $S_{c.t.}$. The ellipsis in Eq.~\eqref{Sonshell} stands for the boundary terms proportional to the products $h_{\mu\nu}(\epsilon,-\omega,-q) h_{\rho\sigma}(\epsilon,\omega,q)$ arising from all the three 
parts of the action \eqref{FullAction}. In the following, we shall only need those terms in our discussion of the scalar sector\footnote{The 
full scalar channel onshell action is given by Eq.~\eqref{ScalarOnShellFull}.}. The explicit expressions for $\CA_i$ are given by
\begin{align}
\CA_1 (u,\omega,q) &=   \frac{4 \pi^2 T^2}{N_{\scriptscriptstyle GB}^5 u}\, \frac{\bar{N} \bar{f}}{  1- \bar{f} }, \\
\CA_2 (u,\omega,q) &= \frac{1}{N_{\scriptscriptstyle GB}^5 u}\, \frac{\bar{N} \bar{f} \left(1-\bar{f}\right)}{  \bar{N} \bar{f} \qfr^2 - \left(1-\bar{f}\right)^2 \wfr^2 } ,\\
\CA_3 (u,\omega,q) &=  \frac{3 \pi^2 T^2 }{ N_{\scriptscriptstyle GB}^5 u }\, \frac{\left(1-4\lgb\right)^2\bar{N} \bar{f} (1-\bar{f})^3 \wfr^4}{\left[ \bar{N} \left(\bar{f} + \bar{f}^2 + 4 \lgb - 12 \lgb \bar{f}  \right) \qfr^2 - 3 \left(1-4\lgb\right) \left(1-\bar{f}\right)^2 \wfr^2 \right]^2 },
\end{align}
where 
$$
\bar{f} = 1 - \sqrt{1- 4 \lgb (1-u^2)}\,, \;\; \qquad \;\; \bar{N} = N_{\scriptscriptstyle GB}^2\, \frac{1-4\lgb}{2\lgb}\,,
$$
and ${\cal Z}_i(u, \omega,q)$ are the solutions to Eq.~(\ref{eq:eom_GB_ginv}) obeying the incoming wave boundary condition at the horizon  and normalized to $Z_i^{(0)}(\omega,q)$ at the boundary at $u=\epsilon \rightarrow 0$  \cite{Son:2002sd}, i.e. 
\begin{equation}
\label{normalized-sol}
{\cal Z}_i (u, \omega,q) = Z_i ^{(0)}(\omega,q)\, \frac{Z_i (u,\omega,q)}{Z_i(\epsilon,\omega,q)}\,,
\end{equation}
where $Z_i (u,\omega,q)$ are the incoming wave solutions to Eq.~(\ref{eq:eom_GB_ginv}). 
\subsection{The scalar channel}
\label{Sec:Scalar}
In this section, we extend the analysis of the scalar sector of metric perturbations performed in ref.~\cite{Brigante:2007nu} to second order in the hydrodynamic expansion. To that order, the retarded two-point function of the appropriate components of  the  energy-momentum tensor obtained by considering a linear response to metric perturbation has the form  \cite{Baier:2007ix}
\begin{align}\label{GScalarHydro}
G_{xy,xy}^{R, lin.resp.} \left(\omega,q\right) = P - i\eta\omega +\eta \tp \omega^2 - \frac{\kappa}{2} 
\left( \omega^2 + q^2 \right) +\cdots.
\end{align}
Using dual gravity, we compute the retarded Green's function $G_{xy,xy}^{R}$ analytically for $\wfr\ll 1$ and $\qfr\ll 1$, 
and read off the transport coefficients $\tp$ and $\kappa$ by comparing the result with Eq.~\eqref{GScalarHydro}. 
A novel feature at finite $\ggb$ is the appearance  
of a new pole of the function $G_{xy,xy}^{R}(\omega,q)$ in the complex frequency plane \cite{Grozdanov:2016vgg}. 
The pole is moving up the imaginary axis with $\ggb$ increasing. It is entering the region $\wfr\ll 1$ at intermediate values 
of $\ggb$ and thus is visible in the analytic approximation.  

To compute the two-point function in the regime of small frequency, we need a solution  of the scalar channel differential equation 
\eqref{eq:eom_GB_ginv} for $\wfr\ll 1$ and $\qfr\ll 1$. Using the variable $v$ defined by the relation (\ref{var-v}) and imposing the in-falling boundary condition \cite{Son:2002sd} 
by isolating the leading singularity at the horizon via
\begin{align}\label{ZsolScalar}
Z_1(v) = Z_1^{(b)} \left(\frac{v}{2\lgb}\right)^{-i\wfr/2} (1+g(v))\,,
\end{align}
one can rewrite the equation \eqref{eq:eom_GB_ginv} as
\begin{align}
\label{eqg}
v \left(1-v\right) \partial^2_v g(v) + \left[1+v+i \wfr \left(v-1\right) \right] \partial_v g(v) + \CG(v)\left[ g(v) + 1\right] = 0\,,
\end{align}
where $\CG$ is a function of $\wfr$ and $\qfr$ of the form 
\begin{align}
\CG(v) = - i  \wfr + \wfr^2 \CG_\wfr (v) + \qfr^2 \CG_\qfr (v)\,
\end{align}
and
\begin{align}
&\CG_\wfr (v)= \frac{ (v-1) \left[ \left( 4 \lgb+v(v-2) \right)^{3/2} -8 \lgb^{3/2} (v-1)^2  \right]}{4 v \left(4 \lgb+v(v-2)\right)^{3/2}}, \\
&\CG_\qfr (v)=  \frac{  \left(v-1\right) \sqrt{\lgb} \left(1+\sqrt{1-4 \lgb}\right)\left(1+8 \lgb +3 v (v-2) \right)}{2 \left(4 \lgb+v (v-2)\right)^{3/2}}  .
\end{align}
The constant $Z_1^{(b)}$ in Eq. \eqref{ZsolScalar} is the normalization constant. To find a perturbative solution $g(v)$ for $\wfr \ll 1$, $\qfr \ll 1$, we 
introduce a book-keeping expansion parameter $\mu$ \cite{Kovtun:2005ev} and write 
\begin{align}
\label{expo-g}
g(v) = \sum_{n=1}^\infty \mu^n g_n(v), 
\end{align}
where the functions $g_n$ satisfy the equations
\begin{align}\label{ScalarRec}
v \left(1-v\right) \partial^2_v g_n(v) + \left(1+v\right) \partial_v g_n(v) + H_n(v) = 0.
\end{align}
The functions $H_n$ are determined recursively from $\CG$ and $g_{m}$ with $m < n$ by
\begin{align}\label{HnEquation}
H_n (v)= i \wfr \partial_v \left[ \left(1-v\right) g_{n-1} (v)\right] + \left(\wfr^2 \CG_\wfr(v) + \qfr^2 \CG_\qfr(v) \right) g_{n-2} (v),
\end{align}
where $n \geq 1$. At first order, $g_0 = 1$ and $g_{-1} = 0$ which gives $H_1 = - i \wfr$. 
A solution to Eq.~(\ref{HnEquation}) can be written in the form
\begin{align}\label{gnsol}
g_n(v) = D_n + \int^v dv' \frac{\left(1-v'\right)^2}{v'} \left(C_n - \int^{v'} dv'' \frac{H_n (v'') }{\left(1-v''\right)^3}  \right),
\end{align}
where $C_n$ and $D_n$ are the integration constants. In particular, for $n=1$ we have
\begin{align}
g_1(v) &= D_1 - \frac{1}{2} C_1 \left(4-v\right) v + \left(C_1 + \frac{i\wfr}{2} \right) \ln v  .
\end{align}
Factorization \eqref{ZsolScalar} implies that  the functions $g_n$ must be regular at the horizon (at $v=0$). In the case of $g_1$, the regularity condition leads to $C_1 = - i\wfr/2$. Furthermore, all $g_n$ with $n>1$ must vanish at the horizon (see Appendix \ref{sec:BC-horizon}). For $n=1$, this amounts to setting $D_1 = 0$. Hence, to linear order in $\wfr$ and $\qfr$ we have
\begin{align}\label{G1Scalar}
g_1(v) = \frac{i \wfr}{4} \left(4-v\right) v .
\end{align}
Repeating the procedure, we find the function $g_2(v)$:
\begin{align}\label{G2Scalar}
g_2 (v) =\,\,& \wfr^2 g^{(\wfr)}_{2} (v) + \qfr^2 g^{(\qfr)}_2 (v)  \nn
&+ \frac{\wfr^2}{4} \int^v \frac{(1-v')^2 \ln \left[\ggb ^2 -1 + v' -\sqrt{\left(\ggb ^2-1\right) \left(\ggb ^2-(1-v')^2\right)}\right]}{v'} \, dv'  .
\end{align}
The functions $g^{(\wfr)}_{2}$ and $g^{(\qfr)}_{2}$ appearing in Eq.~(\ref{G2Scalar}) are given by lengthy
but closed-form expressions. Even though we do not have a closed-form expression for the remaining integral in Eq.~(\ref{G2Scalar}), this is irrelevant for the purposes of 
computing the two-point function in the 
hydrodynamic limit, since the existing expression for $g_2$ is sufficient for fixing both the boundary conditions on $g_2$ itself and for determining the near-boundary expansion of $Z_1$. More precisely, the integral in Eq.~(\ref{G2Scalar}) comes from the outer integration in \eqref{gnsol} and does not affect the regularity at the horizon thus allowing to fix the integration constant $C_2$. The integral in \eqref{G2Scalar} can be evaluated order-by-order in the near-boundary expansion of the integrand and the constant $D_2$ can be re-absorbed into the integration constant. 

The full on-shell action \eqref{OSActDef} including the contact terms is given by  
\begin{align}\label{ScalarOnShellFull}
S =& \,- P V_4  - \lim_{\epsilon\to 0 }  \frac{\pi^4 T^4}{\kappa_5^2}  \int \frac{d\omega dq}{(2\pi)^2} \left[- \frac{2\sqrt{2}\,\ggb}{\left(1+\ggb\right)^{5/2} \epsilon}  \CZ_1 (\epsilon, -\omega, -q)  \CZ_1' (\epsilon, \omega,q)  \right. \nn
&+ \left. \left( \frac{1}{\sqrt{2} \left(1+\ggb\right)^{3/2}}  -  \frac{\ggb \left(\qfr^2 - \wfr^2\right) }{ \sqrt{2(1+\ggb)}\, \epsilon } \right)\CZ_1 (\epsilon, -\omega, -q) \CZ_1 (\epsilon, \omega,q) + \cdots \right],
\end{align}
where  we used the near-boundary regulator $u=\epsilon \rightarrow 0$. Here, the first term is minus the four-volume $V_4$ times 
the free energy density (i.e. the pressure P), where  
\begin{align}\label{Pressure}
P = \frac{\sqrt{2} \pi^4 T^4}{\left(1+\ggb\right)^{3/2} \kappa_5^2},
\end{align}
which is consistent with Eqs. \eqref{EntropyAndEnergy2} and \eqref{EntropyAndEnergy3}.  The ellipsis denotes higher-order terms in $\wfr$ and $\qfr$ and terms vanishing in the $\epsilon\to 0$ limit.

\begin{figure}[htbp]
\centering
\includegraphics[width=0.6\textwidth]{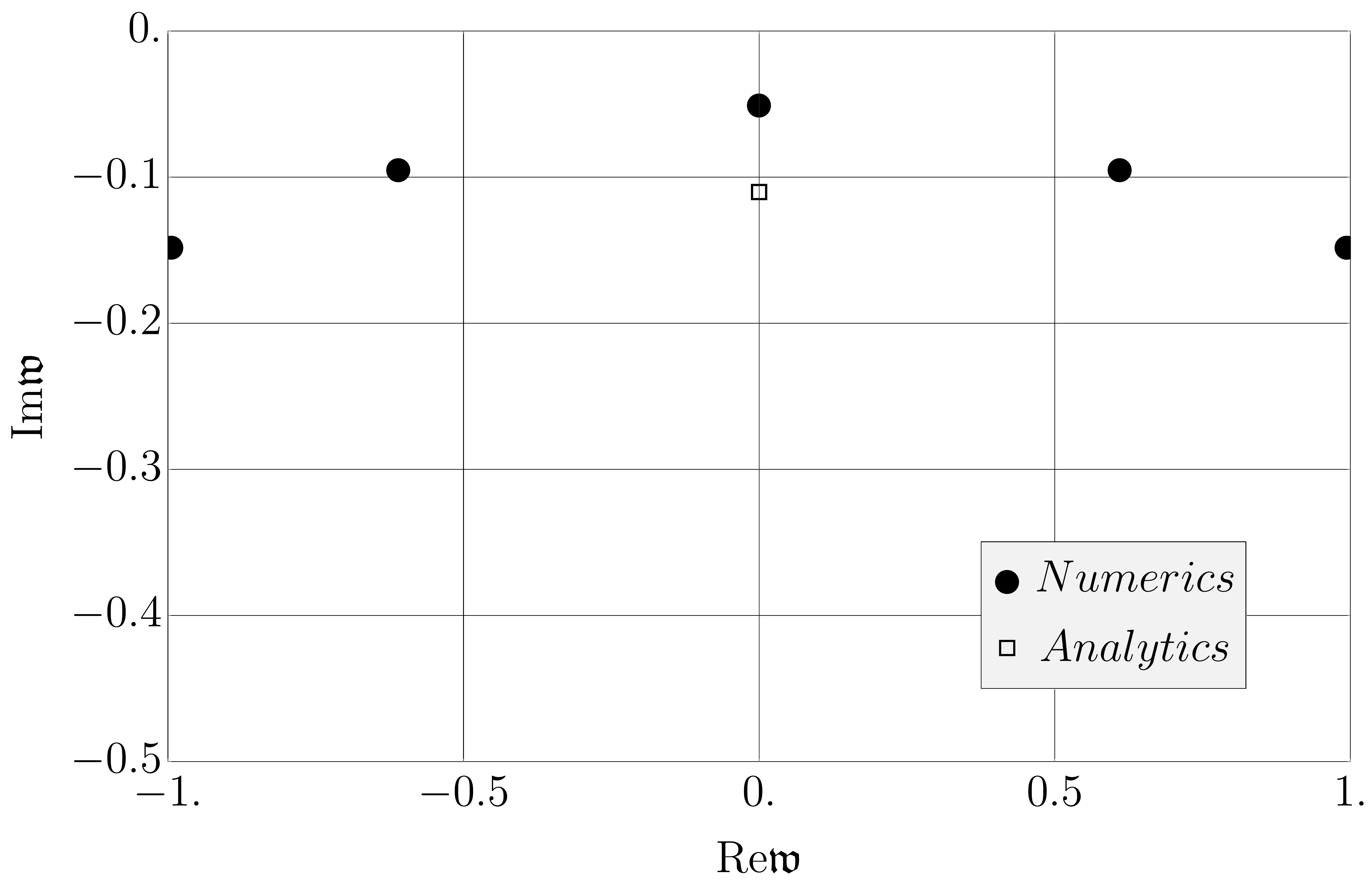}
\caption{The poles of the scalar channel Green's function $G_{xy,xy}^{R}(\wfr,\qfr)$ in the vicinity of origin in the complex frequency plane at $\qfr = 0.1$ and $\lgb \approx - 7.3125$ (corresponding to $\ggb\approx 5.5$). The poles found numerically are shown by black circles. The white square shows
 the analytic approximation \eqref{GapScalar} to the location of the pole on the imaginary axis.}
\label{Fig:scalar-chanel-green}
\end{figure}

The retarded two-point function $G_{xy,xy}^{R}(\omega,q)$ can then be computed by evaluating the boundary action \eqref{ScalarOnShellFull}. Using the solution \eqref{ZsolScalar} to first order in $\wfr$ and $\qfr$ (i.e. including only the function $g_1$ in the expansion \eqref{expo-g}) we find
\begin{align}
\label{GxyxyHol}
G_{xy,xy}^{R}(\omega,q) =&\,  \frac{\sqrt{2} \pi^4 T^4}{\left(1+\ggb \right)^{5/2} \kappa_5^2} \biggr[ \ggb +1 -4 i \ggb  \wfr  \nn
&  +\frac{8 (\ggb -1) (\ggb +2) \ggb  \wfr }{\wfr  \left[ \ggb  (\ggb +2)-3+2 \ln 2 -2  \ln (\ggb +1) \right]+4 i} \biggr].
\end{align}
The Green's function has a pole on the imaginary axis at 
\begin{align}\label{GapScalar}
\wfr \equiv \wfr_{\mathfrak{g}} = - \frac{4 i}{\ggb \left(\ggb + 2 \right) - 3 + 2 \ln \left(\frac{2}{ \ggb+1}\right )} \approx - \frac{4i}{\ggb^2}\,.
\end{align}
The approximation in Eq.~\eqref{GapScalar} assumes $\ggb \gg 1$. The pole is absent from the spectrum at $\lgb =0$ ($\ggb =1$) or, rather, it is located at complex infinity. At non-vanishing $\lgb$ of either sign, the pole moves up the imaginary axis with $|\lgb|$ increasing. For positive $\lgb$, it reaches the quasinormal frequency value at $\lgb = 1/4$ in that limit, determined analytically in section \ref{sec:GBExtremeCoupling}. For negative $\lgb$, the pole moves up to the origin. Its location is correctly captured by the small 
frequency perturbative expansion of the solution $g(v)$ only for sufficiently large $\ggb$ (see Fig.~\ref{Fig:scalar-chanel-green} and  
ref.~\cite{Grozdanov:2016vgg} for details). 

A small frequency expansion of Eq.~\eqref{GxyxyHol} is
\begin{align}
\label{GxyxyHolsmall-w}
G_{xy,xy}^{R}(\omega,q) =&\,  \frac{\sqrt{2} \pi^4 T^4}{\left(1+\ggb \right)^{5/2} \kappa_5^2} \biggr[ \ggb +1 -2 i  \wfr  \ggb^2 (\ggb +1) \biggr]
+ O(\wfr^2)\,.
\end{align}
A comparison with Eq.~\eqref{GScalarHydro} gives the familiar expression for pressure \eqref{Pressure} and the shear viscosity \cite{Brigante:2007nu}
\begin{align}\label{etagb}
\eta = \frac{\sqrt{2} \pi^3 T^3 L^3}{\kappa_5^2} \frac{\ggb^2}{\left(1+\ggb\right)^{3/2}} = 
\frac{4 \pi^3 T^3 \tilde{L}^3}{\kappa_5^2} \frac{\ggb^2}{\left(1+\ggb\right)^{3}} \,,
\end{align}
where we have reinstated $L$ (or $\tilde{L}$) momentarily. To compute the second-order coefficients $\tp$ and $\kappa $, we need to include the function $g_2$ in the expansion \eqref{expo-g} and the solution
\eqref{ZsolScalar}. The resulting expressions for $g_2$ and the corresponding Green's function are very cumbersome and are not shown here explicitly. 
The small frequency expansion of the Green's function, however, matches the hydrodynamic result \eqref{GScalarHydro} perfectly. Combining the equations \eqref{ZsolScalar}, \eqref{G1Scalar} and \eqref{G2Scalar} and comparing with \eqref{GScalarHydro}, we can read off the coefficients $\tau_{\Pi}$ and $\kappa$ given by Eqs.~(\ref{l0}) and (\ref{l4}), respectively. They coincide with the expressions found earlier in  ref.~\cite{Banerjee:2010zd} by using a different method.

The full quasinormal spectrum of metric fluctuations in the scalar channel as a function of $\ggb$ has been analyzed in detail in ref.~\cite{Grozdanov:2016vgg}. The spectrum qualitatively differs from the one at  $\lgb =0$ in a number of ways, depending on the sign of $\lgb$. For $\lgb >0$, there is an inflow of new quasinormal frequencies (poles of $G_{xy,xy}^{R}(\omega,q)$ in the complex frequency plane), rising up from complex infinity  
along the imaginary axis. At the same time, the poles of the two symmetric branches recede from the finite complex plane as $\lgb$ is increased from $0$ to $1/4$, and disappear altogether in the limit $\lgb\rightarrow 1/4$. The spectrum in this limit coincides with the one obtained analytically at $\lgb =1/4$ in section 
\ref{sec:GBExtremeCoupling} of the present paper. For $\lgb <0$, on the contrary, the poles in the symmetric branches become more dense with the magnitude of $\lgb$ increasing, and the two branches gradually lift up towards the real axis. They appear to form branch 
cuts $(-\infty,-q] \cup [q,\infty)$ in the limit $\ggb\rightarrow \infty$. For small $\qfr$ and very large $\ggb$, this would imply accumulation of poles of the Green's function in the region $|\wfr|\ll1$. We have not investigated this limit in detail. Also, as noted above, there is at least one new pole (seen in  Fig.~\ref{Fig:scalar-chanel-green}) rising up the imaginary axis.  The residue and the position of the pole $\wfr_{\mathfrak{g}}$ contribute to the shear viscosity and to the position of the corresponding transport peak of the spectral function. A qualitatively similar phenomenon has been observed in the case of ${\cal N}=4$ SYM at large but finite 't Hooft coupling \cite{Grozdanov:2016vgg}.

\subsection{The shear channel}
The energy-momentum tensor two-point functions $G_{zx,zx}$, $G_{tx,tx}$, $G_{tx,zx}$ in the shear channel can be expressed through the single scalar function $G_2$ as explained in ref.~\cite{Kovtun:2005ev}. For example,\footnote{\label{ftn} Our notations $Z_1$, $Z_2$, $Z_3$ correspond to $Z_3$, $Z_1$, $Z_2$ of
 ref.~\cite{Kovtun:2005ev}, and the same holds for $G_{1,2,3}$.}
\begin{equation}
\label{corr-norm}
G_{xz,xz}(\omega,q) = \frac{\omega^2}{2(\omega^2-q^2)}\, G_2 (\omega,q) + \cdots ,
\end{equation}
where the ellipsis represents the contact terms. In holography, the function $G_2$ is determined by the solution ${\cal Z}_2(u, \omega,q)$  (\ref{normalized-sol}) of the equation  \eqref{eq:eom_GB_ginv} obeying the appropriate boundary conditions, and by the relevant part of the on-shell boundary action \eqref{Sonshell}.

The retarded correlators in the shear channel are characterized by the presence of the hydrodynamic diffusive mode whose dispersion relation is given by
\begin{align}
\omega = - i \frac{\eta}{\varepsilon + P}\, q^2 - i  \left[ \frac{\eta^2\tau_\Pi}{(\varepsilon + P)^2}-\frac{\theta_1}{2(\varepsilon + P)}\right] q^4 + \cdots ,
\label{diff-GB}
\end{align}
where $\theta_1$ is the transport coefficient of the third-order hydrodynamics introduced in ref.~ \cite{Grozdanov:2015kqa}. Higher terms in the momentum expansion of the shear mode depend on the (unclassified) fourth- and higher-order transport coefficients. Since the Gauss-Bonnet fluid is Weyl-invariant ("conformal"), we have 
$\varepsilon = 3 P$ and thus $\eta/(\varepsilon + P)  =  \left(1 - 4\lgb\right)/4\pi T =\ggb^2/4\pi T$. 
In holography, the quasinormal mode \eqref{diff-GB} can be found analytically by solving the equation  \eqref{eq:eom_GB_ginv} 
perturbatively for $\wfr\ll 1$, $\qfr\ll 1$:
\begin{align}
\label{disp-rel-shear-mode}
\wfr = &- i \frac{\ggb^2}{2}\, \qfr^2 - i  \,     \frac{\ggb^4}{16} \left[ (1+\ggb)^2 + 2 \ln \left( \frac{\ggb}{2(1+\ggb)} \right) \right]\,  \qfr^4+ \cdots .
\end{align}
The coefficient in front of the term quadratic in momentum coincides with the one predicted by hydrodynamics of the holographic Gauss-Bonnet fluid with known shear viscosity. Since the coefficient $\tau_\Pi$ is also known (e.g. from Eq.~\eqref{GScalarHydro}), the quartic term in \eqref{disp-rel-shear-mode} allows one to read off the coefficient $\theta_1$:
\begin{align}
\theta_1 = \frac{\eta}{8\pi^2 T^2} \ggb \left(2 \ggb^2 + \ggb - 1 \right).
\end{align}
In the dissipationless limit $\ggb\rightarrow 0$ we have $\theta_1 \sim \ggb^3 \rightarrow 0$. In fact, it can be seen numerically \cite{Grozdanov:2016vgg}  that the full shear mode 
\eqref{diff-GB} approaches zero in the limit  $\ggb\rightarrow 0$. At $\ggb=0$ ($\lgb=1/4$), this mode disappears from the spectrum altogether due to the vanishing residue which is consistent with our analytic results for the spectrum at $\lgb=1/4$ in section  \ref{sec:GBExtremeCoupling}. 

The full quasinormal spectrum 
was investigated numerically and partially analytically in ref.~\cite{Grozdanov:2016vgg}. Its behavior as a function of $\lgb$ is qualitatively similar to the one in the scalar channel, with the exception of one curious phenomenon: at fixed $\qfr$, the new pole rising up the imaginary axis with (negative) $\lgb$ increasing in magnitude, collides with the hydrodynamic pole \eqref{diff-GB} at some $\lgb = \lgb^c (\qfr )$, and the two poles move off the imaginary axis. 
This is interpreted as breakdown of the hydrodynamic regime at a given $\qfr = \qfr_c (\lgb )$. Curiously, the range of applicability of the hydrodynamic regime 
(i.e. the range $\qfr \in [0,\qfr_c])$ increases with the field theory "coupling" (understood as the inverse of $|\lgb|$) increasing \cite{Grozdanov:2016vgg}.  

The retarded correlation functions of the energy-momentum tensor in the shear channel can be computed from the boundary action \eqref{Sonshell}. For the function $G_2$ in Eq.~\eqref{corr-norm} we find\footnote{As in refs.~\cite{Policastro:2002se,Kovtun:2005ev}, we ignore
possible contact terms coming from $S_{c.t.}$. See remarks in Appendix A of ref.~\cite{Kovtun:2005ev}.}
\begin{equation}
G_2 (\omega,q) = 4\left( \omega^2 - q^2\right) \frac{\pi^2 T^2}{8 \kappa_5^2} \lim_{\epsilon \rightarrow 0}  \CA_2  (\epsilon, \omega, q)\frac{Z_2 '(\epsilon,\omega,q)}{Z_2(\epsilon,\omega,q)}\,.
\end{equation}
In the hydrodynamic approximation, to first non-trivial order in $\wfr$, $\qfr$, with both $\wfr \sim\mu\ll 1$ and $\qfr\sim \mu\ll 1$ scaling the same way, the shear channel solution to Eq.~\eqref{eq:eom_GB_ginv} 
obeying the incoming wave boundary condition is 
\begin{align}\label{Z2Solution}
&Z_2 (u) = Z^{(b)}_2 \left(1 - u^2 \right)^{-i \wfr / 2}  \Biggr( 1 + \frac{i \qfr^2}{2 \wfr}  \frac{\ggb^2}{ 1 - \ggb }  
\left( 1-\sqrt{\ggb ^2-\ggb ^2 u^2+u^2} \right)  + 
 \frac{ i \wfr}{4} \Biggl[3 - \ggb ^2  \nn &+\left(\ggb ^2-1\right) u^2-2 \sqrt{\ggb ^2-\ggb ^2 u^2+u^2}+2 \ln
    \frac{ 1 + \sqrt{\ggb ^2-\ggb ^2 u^2+u^2}}{2}    \Biggr]\Biggr),
\end{align} 
where $Z^{(b)}_2$ is the normalization constant. We note that in order to obtain the hydrodynamic dispersion relation \eqref{disp-rel-shear-mode} that includes  information about the second and the third order transport coefficients, we need to find $Z_2$ to one order higher, but using 
the scaling $\omega \sim \mu^2$ and $\qfr \sim \mu$ is sufficient to extract  the diffusive pole.

For the correlation function $G_2$ in the regime $\wfr\ll 1$, $\qfr \ll 1$  we thus find the following expression
\begin{equation}
\label{G2corr}
G_2 =  \frac{2 \sqrt{2} \pi^3 T^3 \ggb^2}{(1+\ggb)^{3/2} \kappa_5^2} \left( \frac{ \omega^2 - q^2 }{i \omega  - i \omega^2 / \omega_{\mathfrak{g}} -  \ggb^2 q^2 / 4\pi T }\right),
\end{equation}
where $\omega_{\mathfrak{g}} = 2\pi T \wfr_{\mathfrak{g}}$ (see Eq.~\eqref{GapScalar}). 
At vanishing Gauss-Bonnet coupling $\lgb=0$ ($\ggb = 1$) one has  $|\wfr_{\mathfrak{g}}| \rightarrow \infty$ and 
we formally recover\footnote{Upon the identification $N_c^2 = 4\pi^2/\kappa_5^2$.}  the standard result for ${\cal N}=4$ SYM at infinitely strong 't Hooft 
coupling and infinite $N_c$ \cite{Policastro:2002se,Kovtun:2005ev} but it should be noted that the formula \eqref{G2corr} is accurate only 
for $|\wfr_{\mathfrak{g}}| \ll 1$, i.e. for sufficiently large $\ggb$. The correlator \eqref{G2corr} has two poles with the following dispersion relations, expanded to $q^2$:
\begin{align}
\omega_1&= - i \frac{ \ggb^2}{4\pi T} q^2 ,\\
\omega_2 & = \omega_{\mathfrak{g}} + i \frac{\ggb^2}{4\pi T} q^2 . 
\end{align}
The first is the usual diffusive pole, corresponding to  quadratic part of the dispersion relation \eqref{disp-rel-shear-mode}, while the second pole is a new non-hydrodynamic pole coming from complex infinity at non-zero $\lgb$. This pole moves up the imaginary axis with $\ggb$ increasing and is responsible for the breakdown of hydrodynamics in the large $\ggb$ limit for any fixed non-zero value of $q$ (see  ref.~\cite{Grozdanov:2016vgg} for details). 

The above expression for the Green's function and the dispersion relations are only valid in a (double expansion) regime in which not only $\wfr \sim \qfr \ll 1$ but also $\ggb \gg 1$. The latter condition is required for the gapped mode on the imaginary axis to satisfy $|\wfr | \ll 1$. Note also that the form of the dissipative corrections implies that $\ggb \qfr \ll 1$. Obviously, these restrictions are only necessary if we are interested in analytic expressions.
 
 The location of the momentum density diffusion pole confirms the result \eqref{gbviscosity} for the shear viscosity of
  Gauss-Bonnet holographic fluid. We note that in the limit $\lgb \rightarrow 1/4$ ($\ggb \rightarrow 0$) the residue of the diffusion pole vanishes. 
  The full Green's function can be determined numerically. The corresponding spectral function in the shear channel for various values of $\ggb$ has been computed numerically in ref.~\cite{Grozdanov:2016vgg}.

\subsection{The sound channel}

The correlation functions in the sound channel can be expressed through the single scalar function\footnote{See footnote \ref{ftn}.}
 $G_3$ \cite{Kovtun:2005ev}. For example, for the energy density two-point function in the conformal case we have
\begin{equation}
\label{corr-norm-3}
G_{tt,tt}(\omega,q) = - 4 \frac{ \delta^2 S_{\partial M}}{  \delta H_{tt}^{(0)} (\omega ,q )  \delta H_{tt}^{(0)} (-\omega ,- q )  }  =   \frac{2 q^4}{3(\omega^2-q^2)^2}\, G_3 (\omega,q) + \cdots,
\end{equation}
and similar expressions are available for other components of the energy-momentum tensor in the sound channel 
\cite{Policastro:2002tn,Kovtun:2005ev}. To compute $G_3$ in holography, one needs the solution ${\cal Z}_3(u, \omega,q)$  (\ref{normalized-sol}) of the equation  \eqref{eq:eom_GB_ginv} and the relevant part of the on-shell boundary action \eqref{Sonshell}. As in Eq.~\eqref{Sonshell}, the ellipsis  represents
 the contribution from the contact terms. The function $H_{tt}^{(0)}$ denotes the boundary value of the fluctuation $H_{tt} = h_{tt} / r^2 = h_{tt} u (1+\ggb) / 2 \pi^2 T^2$.

The hydrodynamic modes in the sound channel are the pair of sound waves whose dispersion relation is predicted by 
relativistic hydrodynamics up to a quartic term in spatial momentum:
\begin{equation}
\label{disp-sound-hydro}
\omega = \pm c_s\, q - i \Gamma\, q^2 \mp \frac{\Gamma}{2c_s} \left( \Gamma - 2 c_s^2 \tau_\Pi\right)\, q^3
- i \left[ \frac{8 \eta^2 \tau_\Pi}{9 (\varepsilon +P)^2} - \frac{\theta_1 + \theta_2}{3(\varepsilon +P)}\right]\, q^4 +\cdots\,,
\end{equation}
where $c_s = 1/\sqrt{3}$ is the speed of sound, $\Gamma = 2\eta/3(\varepsilon +P)$, $\varepsilon +P = s T$ in the absence of chemical potential,  and $\tau_\Pi$, $\theta_1$, $\theta_2$ are transport coefficients of the second- and third-order (conformal) hydrodynamics in four space-time dimensions.

Solving the equation \eqref{eq:eom_GB_ginv}  
for $Z_3$ perturbatively for $\wfr \ll 1$, $\qfr \ll 1$, imposing the incoming wave boundary 
condition at the horizon and the Dirichlet condition at the boundary, we find
the hydrodynamic quasinormal mode\footnote{Here it is tacitly assumed that $\ggb$ is small enough. For moderate and large $\ggb$, 
in addition to the mode (\ref{sound-qnm}),  there exists another mode moving up the imaginary axis with $\ggb$ increasing. This mode enters the hydrodynamic domain $\wfr \ll 1$, $\qfr \ll 1$ for $\ggb \sim 2 - 4$ and can be seen analytically, as discussed in ref.~\cite{Grozdanov:2016vgg}.}
\begin{align}
\wfr_{1,2} =& \pm \frac{1}{\sqrt{3}} \qfr - \frac{1}{3} i \ggb^2\qfr^2  \nn
& \mp  \frac{1}{12 \sqrt{3}} \ggb  \left(2 + \ggb ^3-6 \ggb ^2-3 \ggb +2 \ggb  \ln \left[ \frac{2 (1+\ggb )}{\ggb }\right] \right)  \qfr^3 + \ldots \, . \label{sound-qnm}
\end{align}
Comparing the expansion \eqref{sound-qnm} to the prediction \eqref{disp-sound-hydro} of conformal hydrodynamics one finds the same expressions for the 
shear viscosity - entropy density ratio and the second-order transport coefficient $\tp$ as the ones reported in  Eqs.~\eqref{gb-visc} and \eqref{l0}. This agreement is gratifying but more analytic work is needed to extend the expansion  \eqref{sound-qnm} to quartic order and determine the 
coefficient $\theta_2$ of the third-order hydrodynamics. Other features of the quasinormal spectrum are qualitatively similar to the scalar case and are discussed in full detail in ref.~\cite{Grozdanov:2016vgg}.
   
The coefficients in front of the quadratic, qubic and possibly\footnote{Possibly, because the expression for $\theta_2$ remains unknown.} quartic terms in the dispersion relation  \eqref{disp-sound-hydro} vanish in the limit $\ggb\rightarrow 0$. This limit is hard to study numerically but it is conceivable that the higher terms vanish as well leaving the linear propagating mode $\wfr = \pm \qfr /\sqrt{3}$. Such a mode, however, is absent in the exact spectrum at $\ggb=0$ (see section \ref{sec:GBExtremeCoupling}). 
 
To first order in the hydrodynamic expansion, the gauge-invariant mode is given by
\begin{align}
Z_3 (u) =&\,\, Z_3^{(b)} \left(1-u^2\right)^{-i \wfr / 2 } \Biggr(   \frac{\ggb ^2-\sqrt{\ggb ^2-\ggb ^2 u^2+u^2}}{(\ggb -1) \ggb ^2 \sqrt{\ggb ^2-\ggb ^2
   u^2+u^2}}-\frac{3 \wfr ^2}{\ggb ^2 \qfr^2}       \nn
&+ \frac{i  \wfr \left( \Xi_\wfr \wfr^2 + \Xi_\qfr \qfr^2 \right) }{4  \qfr^2 \ggb ^2 \left(1-\ggb ^2\right) \sqrt{\ggb ^2-\left(\ggb ^2-1\right) u^2}}  \Biggr) ,
\end{align} 
where
\begin{align}
\Xi_\wfr =& -3 \left(\ggb ^2-1\right) U \left(\ggb ^2-\left(\ggb ^2-1\right) u^2+2 U-2 \ln (U+1)-3+ 2 \ln 2 \right), \\
\Xi_\qfr =& \,\, (\ggb +1) \left(\ggb ^2 \left(9 \ggb ^2-5+ 2\ln 2\right)+\left(\ggb ^2-1\right) u^2 \left(-9 \ggb
   ^2+U+2\right)\right) \nn
&+ (\ggb +1) \left(-U \left(7 \ggb ^2-3+ 2\ln 2 \right)+2 \left(U-\ggb ^2\right) \ln (U+1)\right),    
\end{align} 
and we have used $U^2 = u^2 + \ggb^2 - u^2 \ggb^2$. The correlation function $G_3$ can then be computed from
\begin{equation}\label{G3Full}
G_3 (\omega,q) = -\frac{ 48 \left(\omega^2 - q^2 \right)^2 }{ \omega^4}  \frac{\pi^2 T^2}{8 \kappa_5^2} \lim_{\epsilon \rightarrow 0}  \CA_3  (\epsilon, \omega, q)\frac{Z_3 '(\epsilon,\omega,q)}{Z_3(\epsilon,\omega,q)},
\end{equation}
giving
\begin{align}\label{G3corr}
G_3 (\omega,q) =  \frac{8 \sqrt{2} \pi^4 T^4}{(1+\ggb)^{3/2} \kappa_5^2} \left( \frac{q^2 - \omega q^2 / \omega_{\mathfrak{g}} - i \ggb^2 \omega \left(3 \omega^2 - 5 q^2 \right) / 4\pi T  }{ \left(3 \omega^2 - q^2\right) \left( 1 - \omega/ \omega_{\mathfrak{g}} \right) +  i \ggb^2 \omega q^2 / \pi T }\right).
\end{align}
As required by rotational invariance, $G_1(\omega, 0) = G_2(\omega, 0)=G_3 (\omega, 0)$ \cite{Kovtun:2005ev}. The contact term in the on-shell action \eqref{Sonshell} relevant for the computation of  $G_{tt,tt} (\omega, q)$ is
\begin{align}
S_{\partial M} = \cdots +  \frac{\pi^2 T^2}{8 \kappa_5^2} \int \frac{d\omega dq}{(2\pi)^2}  \,\frac{ \sqrt{2} \pi^2 T^2  }{ 3 (1+\ggb)^{3/2} }  \frac{ 29 q^4 - 30 \omega^2 q^2 + 9 \omega^4  }{  \left( \omega^2 - q^2 \right)^2 } H^{(0)}_{tt} ( -\omega, -q) H^{(0)}_{tt} (\omega,q)  .
\end{align}
The full retarded energy density two-point function is then
\begin{align}\label{Gtttt}
G_{tt,tt} (\omega, q ) =  \frac{3 \sqrt{2} \pi^4 T^4}{(1+\ggb)^{3/2} \kappa_5^2} \left( \frac{     \left(5 q^2 - 3 \omega^2 \right) \left( 1 - \omega/ \omega_{\mathfrak{g}} \right) -  i \ggb^2 \omega q^2 / \pi T      }{ \left(3 \omega^2 - q^2\right) \left( 1 - \omega/ \omega_{\mathfrak{g}} \right) +  i \ggb^2 \omega q^2 / \pi T }\right).
\end{align}
The thermodynamic (equilibrium) contribution has been omitted from this expression. To this order in 
the hydrodynamic expansion, the spectrum contains three modes,
\begin{align}
\omega_{1,2} &= \pm \frac{1}{\sqrt{3}} q - i \frac{\ggb^2}{6\pi T} q^2, \\
\omega_3 &= \omega_{\mathfrak{g}} +  i \frac{\ggb^2}{3\pi T} q^2.
\end{align}
The first two are the attenuated sound modes \eqref{sound-qnm} and the third mode is the gapped mode similar to those in the scalar and shear channels. As in the shear channel, these results require the following scalings to be respected: $\wfr \sim \qfr \ll 1$, $\ggb \gg 1$ and hence, $\ggb \qfr \ll 1$.

Second-order corrections to the two hydrodynamic sound modes were given by Eq.~\eqref{sound-qnm}. To study the spectrum beyond second-order hydrodynamics and investigate higher-frequency spectrum, we must again resort to numerics. We note that for better control over the numerics, it is useful to follow \cite{Kovtun:2006pf} and write 
\begin{align}\label{Z3Frob}
Z_3(u) = \CA \left[ 1 + a_1 u + \cdots\right] + \left( \CA \, h \ln u + \CB \right) u^2 \left[ 1 + b_1 u + \cdots \right],
\end{align}
which is a standard Fr\"{o}benius expansion result. The retarded Green's function is then proportional to $\CB / \CA$. Because of the logarithmic term in $Z_3$, it is beneficial to the precision of our numerics to seek the poles of $\CB / \CA$ (or zeros of $\CA/\CB$) as opposed to the zeros of $\CA$. Furthermore, the full Green's function includes information about the values of the residues at the poles. By writing 
\begin{align}
\CB = \frac{1}{2} \lim_{u\to 0} \left( Z_3''(u) - 2 \CA\, h \ln u \right) - \frac{3}{2} \CA \, h,
\end{align}
we obtain the following expression convenient for the computation of quasinormal modes:
\begin{align}
\frac{\CB}{\CA} = \lim_{u\to 0 } \left[ \frac{Z_3''(u)}{2 Z_3(u)}  - h \ln u - \frac{3}{2} h \right].
\end{align}
The coefficient $h$ can be found analytically, $h = - \left( 1 + \ggb\right)^4 \left(\wfr^2 - \qfr^2\right)^2 / 32 $. For a detailed 
discussion of the quasinormal spectrum, see ref.~\cite{Grozdanov:2016vgg}. A comprehensive analysis of the large spatial 
momentum asymptotics similar 
to the one accomplished for the strongly coupled ${\cal N}=4$ SYM
in refs.~\cite{Festuccia:2008zx,Fuini:2016qsc} would be of interest but has not been attempted neither in ref.~\cite{Grozdanov:2016vgg} 
nor in the present paper.

\subsection{Exact quasinormal spectrum at $\lgb = 1/4$}
\label{sec:GBExtremeCoupling}

At $\lgb = 1/4$, the equations of motion  \eqref{eq:eom_GB_ginv}  for all channels simplify drastically. They reduce to the following system
\begin{align}
&\text{Scalar channel:}& &Z_1'' - \frac{2-u}{u(1-u)} Z_1' + \frac{\wfr^2-3(1-u)\qfr^2}{4 u (1-u)^2} Z_1 = 0, \label{crit-z1} \\
&\text{Shear channel:}& &Z_2'' - \frac{2-u}{u(1-u)} Z_2' + \frac{\wfr^2}{4 u (1-u)^2} Z_2 = 0  , \label{crit-z2} \\
&\text{Sound channel:}& &Z_3'' - \frac{2-u}{u(1-u)} Z_3' + \frac{\wfr^2+(1-u)\qfr^2}{4 u (1-u)^2} Z_3 = 0  \label{crit-z3} .
\end{align}
Solutions to these equations can be written in terms of the hypergeometric function. The indicial exponents of Eqs.~(\ref{crit-z1}) - (\ref{crit-z3})
at the horizon at $u=1$ are equal to $\pm i\wfr/2$, as expected. Curiously, the exponents at the boundary singular point $u=0$ are $\alpha_{1,2} = \{0,3\}$ 
and {\it not} $\alpha_{1,2} = \{0,2\}$, which are their values for any  $\lgb <1/4$ (and in fact for all five-dimensional bulk fluctuations dual to operators of conformal dimension $\Delta=4$ of a $3+1$-dimensional boundary theory). The standard holographic dictionary then implies that at $\lgb=1/4$ the 
dual theory operators scale as the energy-momentum tensor in {\it six} rather than four dimensions. Technically, the reason for this "dimensional transmutation" is related to the fact that the "standard" terms in the wave equations  \eqref{eq:eom_GB_ginv} 
most singular in the limit $u\rightarrow 0$ are multiplied by the coefficients proportional to $(1-4\lgb)$ and thus vanish at  $\lgb=1/4$. 
At this value of the Gauss-Bonnet coupling, the theory becomes ``topological gravity'' \cite{Chamseddine:1989nu} with a number of 
curious properties.\footnote{We thank the referee for bringing refs.~\cite{Chamseddine:1989nu} and  \cite{Crisostomo:2000bb} to our attention.} In particular, thermodynamic properties of the black  brane solution at $\lgb=1/4$ are 
different from the ones at $\lgb<1/4$ \cite{Crisostomo:2000bb}. The underlying physical reasons and significance of this limit 
are not entirely clear to us, although they might be related to the issues discussed in ref.~\cite{Banados:2005rz} and refs.~\cite{Verlinde:1995mz,Witten:2007ct,Witten:2009at}. 

We note that the Gauss-Bonnet black brane metric is regular at  $\lgb=1/4$:
\begin{align}\label{BBB}
ds^2 = - \frac{r^2}{L^2} \left(1-  \frac{r^2_+}{r^2} \right)\, dt^2 + \frac{L^2}{2 r^2 \left(1-  \frac{r^2_+}{r^2} \right)} dr^2 + \frac{r^2}{L^2} \left(dx^2 + dy^2 +dz^2 \right).
\end{align}
Rescaling  the coordinates $t,x,y,z$ and the parameter $L$, it can be brought into the form 
\begin{align}\label{BBBB}
ds^2 = - \frac{r^2}{L^2} \left(1-  \frac{r^2_+}{r^2} \right)\, dt^2 + \frac{L^2}{r^2 \left(1-  \frac{r^2_+}{r^2} \right)} dr^2 + \frac{r^2}{L^2} \left(dx^2 + dy^2 +dz^2 \right)\,.
\end{align}
For fluctuations depending on $r,t,z$ only, the metric \eqref{BBBB} is nothing but the BTZ metric with $T_L = T_R$ (see e.g. Eq.~(4.1) of \cite{Son:2002sd} with $\rho_- =0$) which explains the emergence of the hypergeometric equations in the system of equations (\ref{crit-z1}) - (\ref{crit-z3}). The zero temperature limit of the metric 
\eqref{BBBB}  is the standard $AdS_5$ solution in Poincar\'{e} patch coordinates. Note, however, that the action at  $\lgb=1/4$ is obviously not the standard Einstein-Hilbert action, 
and thus the fluctuation equations are {\it not} the "usual" fluctuation equations around $AdS_5$ but rather are given by the zero-temperature limit of 
Eqs.~(\ref{crit-z1}) - (\ref{crit-z3}).

 The solutions to Eqs.~(\ref{crit-z1}) - (\ref{crit-z3}) obeying the incoming wave boundary conditions are given by
\begin{align} 
& \text{Scalar:}& &Z_1 = \left(1-u\right)^{-\frac{i\wfr}{2}} {}_2 F_1 \left[\Omega - \frac{\sqrt{4-3\qfr^2}}{2}  , \Omega + \frac{\sqrt{4-3\qfr^2}}{2}  ,1-i\wfr, 1-u  \right], \\
& \text{Shear:}& &Z_2  = \left(1-u\right)^{-\frac{i\wfr}{2}} {}_2 F_1 \, \Big[\Omega - 1  , \Omega + 1, 1-i\wfr, 1-u  \Big],\\
& \text{Sound:}& &Z_3 = \left(1-u\right)^{-\frac{i\wfr}{2}} {}_2 F_1 \left[\Omega - \frac{\sqrt{4+\qfr^2}}{2}  , \Omega + \frac{\sqrt{4+\qfr^2} }{2}  ,1-i\wfr, 1-u  \right],  
\end{align}
where $\Omega \equiv -1-\frac{i\wfr}{2}$. Given the three solutions, the quasinormal spectrum is determined analytically by imposing the Dirichlet condition $Z_i (0)=0$ at the boundary. We find  
\begin{align}
&\text{Scalar:}& &\wfr = -i \left(4+2n_1 - \sqrt{4-3\qfr^2}\right) , & &\wfr = -i \left(4+2n_2 + \sqrt{4-3\qfr^2}\right) , \label{QNMScalar} \\
&\text{Shear:}& &\wfr = -2 i \left(1+n_1\right), & &\wfr = -2 i \left(3+n_2\right),\label{QNMShear} \\
&\text{Sound:}& &\wfr = -i \left(4+2n_1 - \sqrt{4+\qfr^2}\right) , & &\wfr = -i \left(4+2n_2 + \sqrt{4+\qfr^2}\right)  ,\label{QNMSound}
\end{align}
where $n_1$ and $n_2$ are independent non-negative integers. The numerical study of the Gauss-Bonnet quasinormal spectrum in ref.~\cite{Grozdanov:2016vgg} shows that in the limit  $\lgb \rightarrow 1/4$ the quasinormal frequencies approach the ones found above. The spectrum in the shear channel is $\qfr$-independent. In the scalar and sound channels, for sufficintly large $\qfr$ the modes cross into the upper half plane of frequency thus signaling an instability.
This is perhaps not surprising given the causality problems in the boundary theory observed for sufficiently large spatial momentum in 
ref.~\cite{Brigante:2007nu} and other publications. 

Finally, let us address the questions of what happens to the hydrodynamic poles in the limit of $\lgb \to 1/4$ ($\ggb \to 0$). By examining the limit of the sound correlator $G_{tt,tt}(\omega, q)$ given by Eq. \eqref{Gtttt} computed for any generic value of $\ggb$ (or the limit of $G_3$ given by Eq.~\eqref{G3corr}), we find a non-vanishing Green's function with an unattenuated sound mode, $\omega = \pm q / \sqrt{3}$. On the other hand, the sound spectrum computed analytically at $\ggb = 0$ (cf. Eq.~\eqref{QNMSound}) contains no such mode. This situation can be contrasted with the shear channel: there, the correlator $G_2$ (cf. \eqref{G2corr}) vanishes in the same limit and there is no remaining diffusive mode in the spectrum. Consistently, the exact quasinormal spectrum at $\ggb = 0$ (cf. \eqref{QNMShear}) contains no mode at $\wfr = 0$, either.

We do not have a full understanding of this phenomenon but can offer the following comments. Examine more closely the limit $\ggb \to 0$ of the sound correlator $G_3$ \eqref{G3Full}. First, we notice that its $Z_3$-independent prefactor gives different expressions depending on which of the two limits, $\ggb \to 0$ or $\epsilon \to 0$, is taken first. Namely,
\begin{align}
 \lim_{\ggb \to 0 }\left[ \lim_{\epsilon \to 0} \frac{ 48 \left(\omega^2 - q^2 \right)^2 }{ \omega^4}  \frac{\pi^2 T^2}{8 \kappa_5^2}   \CA_3  (\epsilon, \omega, q) \right] &=  \frac{8 \sqrt{2} \pi^4 T^4}{\kappa_5^2} \frac{\ggb}{\epsilon} + \cdots, \label{G3PrefactorLim1}  \\
 \lim_{\epsilon \to 0 }\left[ \lim_{\ggb \to 0} \frac{ 48 \left(\omega^2 - q^2 \right)^2 }{ \omega^4}  \frac{\pi^2 T^2}{8 \kappa_5^2}   \CA_3  (\epsilon, \omega, q) \right] &=  \frac{72 \sqrt{2} \pi^4 T^4}{\kappa_5^2} \frac{  \left(\omega^2 - q^2 \right)^2   }{ \left( 3\omega^2 - q^2\right)^2  } \frac{ \ggb^2 }{ \epsilon^2  } +\cdots, \label{G3PrefactorLim2}
\end{align}
where the ellipses denote terms subleading in the expansions of $\epsilon$ and $\ggb$ around zero. Now, in the expansion around $\ggb = 0$, the Fr\"{o}benius series \eqref{Z3Frob} becomes 
\begin{align}\label{Z3Frobgamma0}
Z_3 (u) = \CA + \cdots + \CB u^2 + \left(  \frac{\CA \left(3\omega^2 - q^2 \right) }{ 144 \pi^2 T^2 \ggb^2 } + \cdots - \frac{\CB \left(\omega^2 - q^2\right)}{48 \pi^2 T^2}  \right) u^3 + \cdots .
\end{align}
By first taking $\epsilon$ and then $\ggb$ to zero (the order of limits we took to find $G_3$ in Eq.~\eqref{G3corr}), one again recovers the leading order hydrodynamic expression
\begin{align}
G_3 (\omega,q) = - \frac{16 \sqrt{2} \pi^4 T^4 \ggb}{\kappa_5^2} \frac{ \CB }{\CA} = \frac{8 \sqrt{2} \pi^4 T^4}{\kappa_5^2} \frac{q^2}{3\omega^2 - q^2} + \cdots.
\end{align}
With the opposite order of limits, the prefactor \eqref{G3PrefactorLim2} and the solution \eqref{Z3Frobgamma0} yields
\begin{align}
G_3 \left(\omega, q\right) = -  \frac{3 \sqrt{2} \pi^2 T^2}{2 \kappa_5^2} \frac{  \left(\omega^2 - q^2 \right)^2   }{ \left( 3\omega^2 - q^2\right)  } \left( 1- \frac{ \left(\omega^2 - q^2\right) }{ \left(\omega^2 - q^2  /3 \right) } \lim_{\ggb \to 0 } \ggb^2    \frac{\CB}{\CA}   \right),
\end{align} 
where $\CA$ and $\CB$ depend on $\ggb$. What this expression reveals is that it is possible for the unattenuated sound mode to be a pole of the Green's function, having entered into the expression from the prefactor, not the ratio of $\CB / \CA$. Thus, such a pole would not appear as a part of the quasinormal spectrum.

\subsection{The limit  $\lgb\to -\infty$}
It is tempting to investigate the limit $\lgb\to -\infty$ analytically to confirm the observations based on numerical simulations. However, taking this limit is problematic for two reasons. First, on a technical level, the equations of motion for fluctuations contain products of the type $\lgb (r-r_+)$ which remain finite 
for $r$ sufficiently close to the horizon $r_+$, even at large $|\lgb|$. This can possibly be dealt with by a variable redefinition but the second problem is more serious. The Kretschmann  curvature invariant evaluated on the black brane solution \eqref{BB} is  
\begin{align}
\label{kretschmann}
R_{\mu\nu\rho\sigma} R^{\mu\nu\rho\sigma} \propto \frac{1}{ r^4 \left( r^4 \left(1 - 4\lgb\right) + 4 r_+^4 \lgb   \right)^3 }. 
\end{align}
For $\lgb \in [0,1/4]$, the curvature singularity in Eq.~\eqref{kretschmann} is at $r = 0$. However, for $\lgb < 0$ the curvature singularity is located at
\begin{align}
r = \frac{r_+ }{ \left( 1 - \frac{1}{4\lgb}   \right)^{1/4}} .
\end{align}
Thus, as $\lgb$ is tuned from $0$ to $-\infty$, the curvature singularity moves continuously from $r = 0$ to the horizon $r = r_+$ and becomes a naked singularity\footnote{The appearance of naked singularities in the solutions of Lovelock gravity has been investigated in ref.~\cite{Camanho:2011rj}.} in the limit  $\lgb\to -\infty$. Because the classical background geometry is singular at the horizon, considering classical metric fluctuations in the limit  $\lgb\to -\infty$ would be meaningless. In some sense, the need for an ultraviolet completion of gravity in this limit is in accord 
with the observations made  in ref.~\cite{Grozdanov:2016vgg} and in the present paper that the regime of large negative $\lgb$ qualitatively corresponds to the regime of weak coupling in the  field theory which generically requires the full dual stringy rather than dual gravity description.

As a curious observation, we note the following. In the large (negative) $\lgb$ expansion, the Ricci scalar evaluated on the solution \eqref{BB} to leading order becomes
\begin{align}
\lim_{\lgb \to -\infty} R = \frac{2 \left(15 r^4 r_+^4 - 10 r^8 - 3 r_+^8 \right)}{L^2 r^2 \left(r^4 - r_+^4 \right)^{3/2}}  \sqrt{ - \frac{1}{\lgb} }  , 
\end{align}
and the leading order contribution to the Kretschmann scalar is
\begin{align}
\lim_{\lgb \to -\infty} R_{\mu\nu\rho\sigma} R^{\mu\nu\rho\sigma} = \frac{4 \left( 10 r^{16}-30 r^{12} r_+^4+33 r^8 r_+^8-12 r^4 r_+^{12}+3 r_+^{16}   \right)}{L^4 r^4 \left(r^4 - r_+^4\right)^3 }   \left(  \frac{1}{-\lgb} \right).
\end{align}
In fact, all three curvature scalars that appear in the Gauss-Bonnet term, $R_{\mu\nu\rho\sigma} R^{\mu\nu\rho\sigma}$, $R_{\mu\nu} R^{\mu\nu}$ and $R^2$, are singular at $r = r_+$ and scale as $1 / \lgb$, while their combination that appears in the action remains finite and independent of $r$: 
\begin{align}
\lim_{\lgb \to -\infty} \left( R^2 - 4 R_{\mu\nu} R^{\mu\nu} + R_{\mu\nu\rho\sigma} R^{\mu\nu\rho\sigma} \right)  = - \frac{120}{\lgb L^4}.
\end{align}
As as result of these scalings, the Einstein-Gauss-Bonnet action \eqref{GBaction} to leading order in $\lgb$ 
reduces to  the Gauss-Bonnet term and the cosmological constant $\Lambda = - 6 / L^2$:
\begin{align}
\lim_{\lgb \to -\infty} S_{GB} = \frac{\lgb L^2}{4\kappa_5^2} \int d^5 x \sqrt{-g} \left[   
 R^2 - 4 R_{\mu\nu} R^{\mu\nu} + R_{\mu\nu\rho\sigma} R^{\mu\nu\rho\sigma}  - \frac{4 \Lambda}{\lgb L^2} \right].
\end{align}
This theory has a black brane solution that coincides with the $\lgb \to -\infty$ limit of the solution \eqref{BB},
\begin{align}
ds^2 = \sqrt{-\lgb} \left[ - \frac{\tilde r^2}{L^2} \sqrt{1 - \frac{\tilde r_+^4}{\tilde r^4} } \, dt^2 + \frac{L^2}{\tilde r^2 \sqrt{1 - \frac{\tilde r_+^4}{\tilde r^4} }} d \tilde r^2+\frac{ \tilde r^2}{L^2} \left(dx^2 + dy^2 +dz^2 \right) \right],
\end{align}
where we have introduced a rescaled radial coordinate $r = \left( -\lgb \right)^{1/4} \tilde r$. 

\section{Gauss-Bonnet transport coefficients from fluid-gravity correspondence}
\label{Sec:Fluid/gravity}

From the analysis of quasinormal spectra and retarded two-point functions in section \ref{Sec:TwoPointAndQNM}, we were able to determine non-perturbative expressions for the Gauss-Bonnet transport coefficients $\eta$, $\tau_\Pi$, $\kappa$ (and also $\theta_1$ of the third-order hydrodynamics). To find the remaining transport coefficients, one can use either the fluid-gravity correspondence or the Kubo formulae applied to three-point functions. 
In this section, we shall use the fluid-gravity methods  \cite{Bhattacharyya:2008jc,Rangamani:2009xk}. Previously, fluid-gravity approach has been used to determine the shear viscosity  \cite{Dutta:2008gf}  and second-order hydrodynamic coefficients \cite{Shaverin:2012kv} of Gauss-Bonnet holographic liquid perturbatively in $\lgb$.

Fluid-gravity correspondence uses the fact that the bulk metric perturbations $h_{\mu\nu}$ source the energy-momentum tensor $T^{\mu\nu}$ in the generating functional of the boundary quantum field theory \cite{Gubser:1998bc,Witten:1998qj}. Gravitational bulk action should thus be able to capture all of the energy-momentum properties of the dual theory. 
The procedure for computing the holographic energy-momentum tensor, inspired by the old prescription of Brown and York \cite{Brown:1992br}, 
was proposed in ref.~\cite{Balasubramanian:1999re}. One expects then that in the appropriate variables a gradient expansion of the bulk 
metric should capture the hydrodynamic gradient expansion of the dual field theory's energy-momentum tensor. 

Following ref.~\cite{Bhattacharyya:2008jc}, we write the Gauss-Bonnet black brane background solution \eqref{BB} in the Eddington-Finkelstein coordinates,
\begin{align}\label{FGbrane1}
ds^2 = - r^2 f(br) dv^2 +2 N_{\scriptscriptstyle GB} dv dr + r^2 d x^i dx^i \,,
\end{align}
where $N_{\scriptscriptstyle GB}$ is given by Eq.~\eqref{NhashDef}. We  set $L=1$ for convenience and defined $b\equiv 1/r_+$ to be consistent with the notations used in 
ref.~\cite{Bhattacharyya:2008jc}. The function $f(br)$ is 
\begin{align}
f(br) = \frac{N_{\scriptscriptstyle GB}^2}{2 \lgb} \left[ 1 - \sqrt{1-4\lgb \left(1-\frac{1}{b^4 r^4} \right) }\right].
\end{align}
The energy-momentum tensor is given by the expression 
\begin{align}
\label{emtensorfg}
T_{\mu\nu} =&\, \frac{r^2}{\kappa_5^2} \left[ K_{\mu\nu} - K \gamma_{\mu\nu} + \lgb \left(3 J_{\mu\nu} - J \gamma_{\mu\nu} \right) +  c_1 \gamma_{\mu\nu} +  c_2 G^{(\gamma)}_{\mu\nu} \right],
\end{align}
where all the ingredients are defined just below Eq.~\eqref{CTCoefficients}.

The next step is to boost the brane solution \eqref{FGbrane1} along a space-time dependent velocity four-vector $u^a(x)$, where 
\begin{align}
u^a = \frac{1}{\sqrt{1-\beta^2}} \left(1, \beta^i \right), 
\end{align} 
with $i=1,2,3$ corresponding to the spatial boundary coordinates. Note that $x^a = (v,x,y,z)$ in Eddington - Finkelstein coordinates. The boosted black brane metric, which we denote by $g^{(0)}_{\mu\nu}$, becomes
\begin{align}\label{g0}
ds_{(0)}^2 = &- 2 N_{\scriptscriptstyle GB} u_a \left(x^c\right) dx^a dr - r^2 f\left(b\left(x^c\right) r\right) u_a \left(x^c\right) u_b  \left(x^c\right) dx^a dx^b \nonumber \\
&+ r^2 \Delta_{ab} \left(x^c\right) dx^a dx^b.
\end{align}
Generically, the metric \eqref{g0} is no longer a solution of the Einstein-Gauss-Bonnet equations of motion \eqref{GBeom}. In fluid-gravity correspondence, assuming a slow-varying dependence of the coefficients on the coordinates $x^a$ and making a gradient expansion, one imposes the equations of motion 
\eqref{GBeom} as the condition each term in the expansion must satisfy. We make a gradient expansions in the derivatives of the fields $\beta^i \left(x^a \right)$ and $b\left(x^a \right)$ to second order, in agreement with the boundary theory's standard second-order hydrodynamic gradient expansion in velocity and temperature fields (see e.g. Appendix \ref{appendixB}). To second order, the metric will have the form
\begin{align}
g_{\mu\nu} = g_{\mu\nu}^{(0)} + \epsilon g^{(1)}_{\mu\nu} + \epsilon^2 g^{(2)}_{\mu\nu},
\end{align}
where $g_{\mu\nu}^{(0)}$ and $g_{\mu\nu}^{(1)}$ are expanded up to terms involving two derivatives of $b$ and $\beta^i$ inclusive. We shall use
$\epsilon$ as a book-keeping parameter in the derivative expansion.

The procedure of solving  equations \eqref{GBeom} order by order is greatly simplified, if one notices that it is sufficient to solve 
the equations of motion locally around some point $x^a = X^a$. The global metric can be obtained from 
these data alone \cite{Bhattacharyya:2008jc}. The local expansions of the fields $b$ and $\beta^i$ are given by
\begin{align}
&b = b_{(0)} |_{X^a} + \epsilon x^a \partial_a b_{(0)} |_{X^a} + \epsilon b_{(1)} |_{X^a} + \frac{\epsilon^2}{2} x^a x^b\partial_a \partial_b b_{(0)} |_{X^a} + \epsilon^2 x^a \partial_a b_{(1)} |_{X^a} ,   \label{ExpB} \\ 
&\beta^i  =  \beta^i_{(0)} |_{X^a} + \epsilon x^a \partial_a \beta^i_{(0)} |_{X^a} + \frac{\epsilon^2}{2} x^a x^b\partial_a \partial_b \beta^i_{(0)} |_{X^a} \label{ExpBeta}. 
\end{align} 
We choose to work in a local frame at the origin, $X^a = 0$, where
\begin{align}
\label{fixedb0}
b_0 = 1 &&\text{and} && \beta^i = 0.
\end{align}
Furthermore, it is consistent to choose a gauge with $\beta^i_{(1)} = 0$ at $x^a = X^a$ \cite{Bhattacharyya:2008jc}.

\subsection{First-order solution}

The most general expression for  the first-order metric $g_{\mu\nu}^{(1)}$ can be conveniently written in a scalar-vector-tensor form
\begin{align}\label{FGMetricFirstOrder}
ds^2_{(1)} =\; & \frac{k_1(r)}{r^2} dv^2 - 3 N_{\scriptscriptstyle GB} h_1(r) dv dr  + \frac{2}{r^2} \left(\sum_{i=1}^3 j_{1}^i (r) dx^i \right)  dv       \nn 
& + r^2 h_2(r) \left(dx^2+dy^2+dz^2\right) + r^2  \CA_{ab} dx^a dx^b,
\end{align}
where $x^i = (x,y,z)$, $k_1$ and $h_1$ are scalars, $j_1^i$ is a three-vector and $\CA_{ab}$ is a tensor. As discussed above, we proceed by using the expanded forms of $b$ and $\beta^i$ given in \eqref{ExpB} and \eqref{ExpBeta} to write the order-$\epsilon$ metric as $g_{\mu\nu} = g^{(0)}_{\mu\nu} + \epsilon g^{(1)}_{\mu\nu}$. Then the equations of motion  \eqref{GBeom} generate the following set 
of constraints and dynamical equations:
\begin{align}
\text{Scalar}: \; & \nn
& \text{Constraint 1:} ~ & r^2 f_0(r) E_{vr} + N_{\scriptscriptstyle GB} E_{vv} = 0, \label{Con1} \\
& \text{Constraint 2:} ~ & r^2 f_0(r) E_{rr} + N_{\scriptscriptstyle GB} E_{vr} = 0, \label{Con2} \\
& \text{Dynamical equation 1:} ~ & E_{rr} = 0, \label{Dyn1} \\
\text{Vector}: \; & \nn
& \text{Constraint 3:} ~ & r^2 f_0(r) E_{ri} + N_{\scriptscriptstyle GB} E_{vi} = 0, \label{Con3}  \\
& \text{Dynamical equation 2:} ~ & E_{ri} = 0 , \label{Dyn2} \\
\text{Tensor}: \; & \nn
& \text{Dynamical equation 3:} ~ & E_{ij} = 0. \label{Dyn3}
\end{align}

First, we solve the Dynamical equation 1 in \eqref{Dyn1} for $h_1(r)$. We then use Constraint 2 in \eqref{Con2} which relates $k_1'(r)$ to $h_1(r)$ to solve for $k_1(r)$. Constraints 1 and 3 in \eqref{Con1} and \eqref{Con3} give 
\begin{align}
\partial_v b_0  = \frac{1}{3} \partial_i \beta^i & &\text{and}& & \partial_i b_0 = \partial_v \beta^i.
\end{align}
Finally, we can solve the remaining Dynamical equations 2 and 3 in \eqref{Dyn2} and \eqref{Dyn3} to find $j_1(r)$ and the tensor $\CA_{ab}$ which contains information about  shear viscosity. 

The global first-order metric, $g_{\mu\nu} = g_{\mu\nu}^{(0)} + \epsilon g_{\mu\nu}^{(1)}$, can be written as a sum  \cite{Bhattacharyya:2008jc},
\begin{align}\label{FGMetricFirstOrderSol}
ds^2 = \sum_{n=1}^6 \CA_n ,
\end{align}
where  the six line elements $\CA_n$ are defined as  
\begin{align}
&\CA_1 = - 2 N_{\scriptscriptstyle GB} u_a dx^a dr, && \CA_2 = - r^2 f_0 (b r) u_a u_b dx^a dx^b , \\
&\CA_3 = r^2 \Delta_{ab} dx^a dx^b ,&&\CA_4 = 2 r^2 b F_0(br) \sigma_{ab} dx^a dx^b, \label{A34} \\
&\CA_5 = \frac{2}{3} N_{\scriptscriptstyle GB} r u_a u_b \partial_c u^c dx^a dx^b, && \CA_6 = - N_{\scriptscriptstyle GB} r u^c \partial_c \left(u_a u_b \right) dx^a dx^b.
\end{align}
The last step is to find the function  $F_0(r)$ entering Eq.~\eqref{A34}. This function is part of the tensor $\CA_{ab}$ satisfying Eq.~\eqref{Dyn3}. 
Explicitly, the second-order differential equation for $F_0$ is
\begin{align}\label{F0Eq}
\frac{\partial}{\partial r} \left[ \left( r^5 - \frac{r^7}{\sqrt{1 - \left(1 - r^4 \right) \ggb^2  }} \right) \frac{\partial F_0}{\partial r}   \right] =  \frac{ \left(1-\ggb^2\right) \left( 5  - \left(5-3 r^4 \right)\ggb^2  \right)  r^4 }{2 \sqrt{2} \sqrt{1+\ggb}  \left(  1 - \left(1-r^4\right)\ggb^2   \right)^{3/2} } .
\end{align}
A  pleasant feature of fluid-gravity duality is that the kernel (the part involving the derivatives) of dynamical equations remains the same for all unknown functions at all orders in the gradient expansion. This was manifest in Eq.~\eqref{ScalarRec} and we expect the same from the equations such as 
Eq.~\eqref{F0Eq}. A solution to Eq.~\eqref{F0Eq} regular at the horizon and vanishing at the boundary is given by
\begin{align}
F_0 (r) =&~ \frac{1}{8\sqrt{2}} \Biggr\{   \frac{ (1+i) \left(1-\ggb ^2\right)^{1/4} \left[(1-i) \; \text{arctanh}(\ggb )+\pi - (1- i) \ggb       \right]}{(1-\ggb )^{1/4} (1+\ggb)^{3/4}}   \nn
& + \frac{\ggb ^{3/2} \Gamma \left(\frac{1}{4}\right)^2 \, _2F_1\left[\frac{1}{4},1;\frac{1}{2};\frac{1}{1-\ggb^2}\right]}{\sqrt{\pi }(1-\ggb )^{1/4} (1+\ggb)^{3/4} }   +  \frac{1 -\ggb ^2-i \pi  r^4+2 r^2 \sqrt{1 - \left(1-r^4\right) \ggb ^2 } }{\sqrt{\ggb } r^4}   \nn
&+  \frac{1}{ \sqrt{1+\ggb} } \ln \left[ \frac{(1+r)^2 \left(1+r^2 \right) \left(r^2-\sqrt{1 - \left(1-r^4\right) \ggb ^2 }\right)}{r^4
   \left(r^2 + \sqrt{1 -  \left(1-r^4\right) \ggb ^2  }\right)}\right]       \nn
& -  \frac{2}{\sqrt{1+\ggb}}    \; \text{arctan}(r) + \frac{4r \sqrt{1-\ggb ^2}}{\sqrt{1+\ggb }}  F_1 \left[ \frac{1}{4}, -\frac{1}{2}, 1; \frac{5}{4} ; -\frac{\ggb^2 r^4}{1-\ggb^2 }, r^4 \right]         \Biggr\},
\end{align}
where $F_1(a,b,b';c;w,z)$ is the Appell hypergeometric function of two variables and where $_2F_1 (a,b;c;z)$ is 
the Gauss hypergeometric function. The expansion of the Appell function at $r\rightarrow \infty$ (explicitly written here for $0 < \ggb < 1$) can be found by using the theorems in ref.~\cite{ferreira2004asymptotic}:   
\begin{align}\label{AppellExp}
&F_1 \left[ \frac{1}{4}, -\frac{1}{2}, 1; \frac{5}{4} ; -\frac{\ggb^2 r^4}{1-\ggb^2 }, r^4 \right] = - \frac{\Gamma \left(\frac{1}{4}\right) \Gamma \left(\frac{5}{4}\right) \,
   _2F_1\left[\frac{1}{4},1;\frac{1}{2};\frac{1}{1-\ggb ^2}\right]}{\sqrt{\pi }} \left(\frac{\ggb^2}{1-\ggb ^2 }\right)^{3/4} \frac{1}{r} \nn
&+\left(\frac{\ggb^2}{1-\ggb ^2 }\right)^{1/2}  \frac{1}{r^2} + \frac{27 \left(8-\ggb ^2\right) \Gamma \left(-\frac{3}{4}\right)^3    }{         2048 \sqrt{\pi } \ggb ^{5/2} \left(1-\ggb
   ^2\right)^{7/4} \Gamma \left(\frac{1}{4}\right)}   \nn
&\times \left\{  \left(1-\ggb ^2\right) \left(\,
   _2F_1\left[-\frac{3}{4},1;\frac{1}{2};\frac{1}{1-\ggb ^2}\right]+2\right)  +  3 \ggb ^2 \,
   _2F_1\left[\frac{1}{4},1;\frac{1}{2};\frac{1}{1-\ggb ^2}\right]\right\}  \frac{1}{r^5}  + \cdots  .
\end{align}
The result  \eqref{AppellExp} allows us to find the expansion of $F_0(r)$ near the boundary,
\begin{align}
F_0 (r) = \frac{\sqrt{1+\ggb }}{2\sqrt{2} r}-\frac{\ggb  \sqrt{1+\ggb}}{8 \sqrt{2} r^4} + \CO\left(r^{-5}\right),
\end{align}
valid to order $\CO(r^{-4})$ which is sufficient for the purposes of computing the boundary energy-momentum tensor. Substituting $F_0 (r)$ 
into the first-order metric $g^{\mu\nu}_{(1)}$ and computing the energy-momentum tensor 
\eqref{emtensorfg} 
with the full first-order solution
we recover the {\it non-perturbative} result for the shear viscosity $\eta$ presented in \eqref{etagb}.

\subsection{Second-order solution}

The second-order correction  $g^{(2)}_{\mu\nu}$ is computed in a similar way: first, we perturb $g^{(0)}_{\mu\nu} + \epsilon g^{(1)}_{\mu\nu}$ 
to second order in derivative expansion and then find $g^{(2)}_{\mu\nu}$ requiring that the Einstein-Gauss-Bonnet equations of motion \eqref{GBeom} are satisfied.

To find the second-order transport coefficients non-perturbatively, we would need to solve differential equations with the differential operator given by the left-hand side
of Eq.~\eqref{F0Eq} and the right-hand sides involving integrals over the Appell function \eqref{AppellExp}. This program faces a certain technical challenge, and we 
were not able to find closed-form expressions for the transport coefficients in this way. It is possible, however, to obtain terms of 
the perturbative expansion of transport coefficients in $\ggb$ and thus check the fully non-perturbative results \eqref{l0},  \eqref{l1},  \eqref{l2} 
found by
 using the method of three-point functions (see ref.~\cite{Grozdanov:2015asa} and 
 section \ref{section:three-point functions}), as well as perturbative results by Shaverin \cite{Shaverin:2012kv}. 

A convenient {\it ansatz} for the  line element of the second-order metric $g_{\mu\nu}^{(2)}$ is suggested by the tensor structure of  second-order hydrodynamics 
(see e.g. Appendix \ref{appendixB}):
\begin{align}\label{FGMetricSecondOrder}
ds^2_{(2)} =\; & \frac{k_2(r)}{r^2} dv^2 - 3 N_{\scriptscriptstyle GB} h_2(r) dv dr  + \frac{2}{r^2} \left(\sum_{i=1}^3 j_{2}^i (r) dx^i \right)  dv       \nn 
& + r^2 h_2(r) \left(dx^2+dy^2+dz^2\right) + r^2 \sum_{n=0}^3 P_n(r) \CB_n ,
\end{align}
where $x^i = (x,y,z)$, $k_2$ and $h_2$ are scalars, $j_2^i$ is a three-vector. We have also defined
\begin{align}
&\CB_0 =\left( {}_{\langle}D\sigma_{ab\rangle} + \frac{1}{3} \sigma_{ab} \left(\nabla\cdot u\right)  \right) dx^a dx^b , \label{2ndOrderTensor1} \\
&\CB_1 =  \sigma_{\langle a}^{~~c} \sigma_{b\rangle c} \; dx^a dx^b , \label{2ndOrderTensor2}\\
&\CB_2 = \sigma_{\langle a}^{~~c} \Omega_{b\rangle c} \; dx^a dx^b ,\label{2ndOrderTensor3}  \\
&\CB_3 = \Omega_{\langle a}^{~~c} \Omega_{b\rangle c} \; dx^a dx^b.\label{2ndOrderTensor4}
\end{align}
At this point, we can focus only on the four functions $P_n$,  $n=\{0,1,2,3\}$, which will give us the four second-order coefficients, 
$\lambda_0 \equiv \eta\tp$, $\lambda_1$, $\lambda_2$ and $\lambda_3$, respectively. 
Since the boundary theory is defined in flat space-time, this procedure will not allow us to find the coefficient $\kappa$. Furthermore, 
we know that in  Landau frame there are no other transport coefficients coming from either the scalar or the vector sector. Still, we need to use the constraint equation $r^2 f_0(r) E_{rr} + N_{\scriptscriptstyle GB} E_{vr} = 0$ and the dynamical equation $E_{rr} = 0$ to eliminate $h_2$, $k_2$ and their derivatives from the dynamical equations for $P_n$.  

The remaining differential equations for $P_n$ can be solved perturbatively  to an arbitrarily high order in $\lgb$. Here we outline what we believe is 
the most efficient 
way to extract information from the functions $P_n$ necessary to recover the four transport coefficients  $\lambda_{0,1,2,3}$. 
First, the functions $P_n$ are expanded in series near the boundary as
\begin{align}\label{PnExp}
P_n(r)  = \sum_{i=1}^{\infty} \frac{p^{(i)}_n}{r^i}.
\end{align} 
Then the metric \eqref{FGMetricSecondOrder} with  $P_n$ expanded  as in Eq.~\eqref{PnExp} is substituted into the full second-order metric, and the energy-momentum tensor \eqref{emtensorfg} is computed. The main observation is that in the limit  $r\to\infty$, finite contributions to $T_{\mu\nu}$ only depend on the coefficients of $P_n$ proportional to $r^{-4}$, i.e. $T_{\mu\nu}$ depends on $p_1^{(4)}$, $p_2^{(4)}$, $p_3^{(4)}$ and $p_4^{(4)}$. 

In order to find the four coefficients, we use the fact that all four differential equations for $P_n (r)$ can be written in the form of Eq.~\eqref{F0Eq}, i.e. as
\begin{align}
\partial_r \left[ Q(r) \partial_r P_n(r)  \right] - R_n(r) = 0 ,
\end{align}
where $Q$ and $R$ expanded to the desired order in $\lgb$, and the function $Q$ is the same in all four cases. 
The differential equations can be formally solved, as in Eq. \eqref{gnsol}, by writing
\begin{align}\label{Pnsol}
P_n(r) = D_n + \int^r dr' \frac{1}{Q(r')} \left(C_n - \int^{r'} dr'' R_n(r'') \right).
\end{align}
Fortunately, in Eq.~\eqref{Pnsol} it is sufficient to take the inner integral over $r''$ whose integrand depends on $F_0(r)$ expanded to the desired order of $\lgb$. 
Integration constants $C_n$ are fixed by requiring regularity at the horizon. The coefficients $D_n$ may remain undetermined since we only need the specific terms in the 
 $r\to\infty$ expansion. Thus, using the expansion \eqref{PnExp} in the differential equations \eqref{Pnsol} we find all $p^{(4)}_n$. For example, 
 from the equation obeyed by $P_0$ we obtain
\begin{align}
&\frac{p^{(1)}_0}{r^2} + \frac{2 p^{(2)}_0}{r^3} + \frac{3 p^{(3)}_0 }{r^4} + \frac{1}{r^5} \left[ 4 p^{(4)}_0 + \left(-1+\frac{\ln 2}{2}\right) \right.  \nn
&\left. + \left(\frac{19}{4}-\ln 2 \right) \lgb+  \left(\frac{1}{8}-\ln 2\right) \lgb^2+ \ldots \right] + \CO\left( r^{-6} \right)  = 0
\end{align}
which allows us to find $p^{(4)}_0$ to the desired order in $\lgb$ by setting to zero the coefficient in front of $r^{-5}$. 

\subsection{Transport coefficients}

Once the coefficients $p^{(4)}_n$ are known, the full second-order metric can be  used to determine the expansion of the energy-momentum tensor \eqref{emtensorfg} near $r\to\infty$ and read off the transport coefficients $\eta\tp$, $\lambda_1$, $\lambda_2$ and $\lambda_3$ 
from the coefficients of tensors \eqref{2ndOrderTensor1} - \eqref{2ndOrderTensor4}. The results are in exact agreement with the corresponding terms of the $\lgb$-expansions of the four non-perturbative second-order transport coefficients given by Eqs.~\eqref{l0}, \eqref{l1}, \eqref{l2} and \eqref{l3}, as well as with those computed in ref.~\cite{Shaverin:2012kv} to linear order\footnote{In matching those expressions, one should recall that the horizon scale $r_+$ in the fluid/gravity calculation is promoted to a field $b(r)$, with $b_0$ fixed by Eq.~\eqref{fixedb0}.}.

The conclusion of this section is that fluid-gravity duality applied to Gauss-Bonnet holographic fluid allows to determine all the 
transport coefficients of second-order hydrodynamics, except $\kappa$, but only the shear viscosity $\eta$ is determined non-perturbatively in $\lgb$: the coefficients $\tau_\Pi$ 
and $\lambda_{1,2,3}$ are found only as series in $\lgb$, due to technical problems related to evaluating integrals of Appell function. Finally, we note that within the fluid-gravity approach one is able to check the Haack-Yarom relation order by order in $\lgb$ and find that it is violated at quadratic order as shown in Eq.~\eqref{SecondOrderUniViolation}.

\section{Gauss-Bonnet transport from three-point functions}
\label{section:three-point functions}

The full non-perturbative expressions for the Gauss-Bonnet transport coefficients can be found by computing the three-point 
functions\footnote{In holography, the first equilibrium real-time three-point and four-point functions in strongly 
coupled ${\cal N}=4$ SYM at finite temperature were computed in refs.~\cite{Barnes:2010jp,Barnes:2010ev,Arnold:2011hp}.}  
of the energy-momentum tensor in the hydrodynamic approximation and using the Kubo-type formulae derived in refs.~\cite{Moore:2010bu,Saremi:2011nh,Arnold:2011ja}. The 
retarded three-point functions are defined following the recipes of the Schwinger-Keldysh closed time-path formalism \cite{Schwinger:1960qe,Keldysh:1964ud}. Part of 
the material in this section has some overlap with refs.~\cite{Grozdanov:2014kva,Grozdanov:2015asa} and is included here for convenience and continuity.

\subsection{An overview of the method}

In the Schwinger-Keldysh formalism, given a  Lagrangian $\CL \left[\phi, h\right]$, where $\phi$ collectively denotes  matter fields and $h$ is a 
metric perturbation around a fixed background $g$, the degrees of freedom are doubled: $\phi \to \phi^\pm$, $g \to g^\pm$, $h \to h^\pm$, where 
the index $\pm$ labels the fields defined either on a ``$+$"-time contour running from $t_0$ towards the final time $t_f > t_0$ or the ``$-$"-time contour,  where the time runs
 from  $t_f$ backwards to $t_0$. When the theory is considered at finite temperature $T = 1/\beta$, the two real-time contours can be joined by a third, imaginary 
time, contour running between $ t_f$ and $t_f - i \beta$. Fields defined  on this imaginary time contour will be denoted by $\varphi$. 
The generating functional of the energy-momentum tensor correlation functions is given by
\begin{align}
W\left[h^+, h^-\right] =& \ln \int \CD\phi^+\CD\phi^-\CD\varphi \exp \left\{i\int d^4x^+ \sqrt{-g^+} \CL\left[\phi^+(x^+), h^+\right] \right.  \nn
& \left.  - \int_0^\beta d^4 y \CL_E \left[\varphi(y)\right] - i \int d^4 x^- \sqrt{-g^-}  \CL\left[\phi^-(x^-), h^-\right] \right\}.
\end{align}
For all fields, it will be convenient to use
 Keldysh basis $\phi_R = \frac{1}{2} \left( \phi^+ + \phi^- \right)$ and $\phi_A = \phi^+ - \phi^-$. Upon computing the variation, 
 classical expectation values  obey $\phi^+ = \phi^-$. Thus, all fields with an index $A$ will vanish and one can define $T^{ab} \equiv T^{ab}_R$: 
\begin{align}
\left\langle T^{ab}_R (x) \right\rangle = -\frac{2i}{\sqrt{-g}} \frac{\partial W }{ \partial h_{A\;ab}(x) } \biggr|_{h = 0}.
\end{align}
The expectation value of $T_R$ at $x=0$ can  be expanded as 
\begin{align}
\left\langle T^{ab}_R (0) \right\rangle = & \,\,G^{ab}_R (0) - \frac{1}{2} \int d^4 x \, G^{ab,cd}_{RA} (0,x) h_{cd}(x)   \nn
& + \frac{1}{8} \int d^4xd^4y\, G^{ab,cd,ef}_{RAA}(0,x,y) h_{cd}(x)h_{ef}(y) + \ldots,
\end{align}
where $G_{RAA...}$ denote the fully retarded Green's functions \cite{Wang:1998wg} obtained by\footnote{See e.g. ref.~\cite{Moore:2010bu}.}
\begin{align}
\label{green-t}
G^{ab,cd,\ldots}_{RA\ldots} (0, x, \ldots) = \frac{(-i)^{n-1} (-2 i)^n \partial^n W   }{\partial h_{A\;ab} (0) \partial h_{R\;cd} (x) \ldots } \biggr|_{h=0}  = (-i)^{n-1} \left\langle T_R^{ab} (0) T_A^{cd}(x) \ldots  \right\rangle,
\end{align}
where the ellipses indicate further insertions of $\partial h_R$ in the expression with the derivatives as well as the $T_A^{ab}$ insertions into the n-point function on the right-hand side of Eq.~\eqref{green-t}.

We follow refs.~\cite{Moore:2010bu,Saremi:2011nh} and use Kubo formulae for pressure and transport coefficients of a conformal fluid derived
 by exciting fluctuations of the relevant metric components. Choosing the spatial momentum along the $z$ direction, one turns on 
 $h_{xy}$, $h_{xz}$ and $h_{yz}$ perturbations to obtain 
\begin{align}
&\eta = i \lim_{p,q \to 0 } \frac{\partial}{\partial q^0} G^{xy,xz,yz}_{RAA} (p,q), \label{3ptKuboEta}\\
&2 \eta \tp - \kappa = \lim_{p,q\to 0} \frac{\partial^2}{\partial \left(p^0\right)^2} G^{xy,xz,yz}_{RAA}(p,q), \label{3ptKuboTpKappa} \\
&\lambda_1 = \eta \tp - \lim_{p,q\to 0} \frac{\partial^2}{\partial p^0 \partial q^0 } G^{xy,xz,yz}_{RAA}(p,q).
\end{align}
By turning on $h_{xy}$, $h_{tx}$ and $h_{ty}$ components, we find
\begin{align}
&\lambda_3 = 4 \lim_{p,q\to 0} \frac{\partial^2}{\partial p^z \partial q^z} G^{xy,tx,ty}_{RAA}(p,q) , \\
&\kappa = \lim_{p,q\to 0} \frac{\partial^2}{\partial \left(p^z\right)^2} G^{xy,tx,ty}_{RAA}(p,q), 
\end{align}
and, finally, by considering  $h_{xy}$, $h_{ty}$ and $h_{xz}$ perturbations, we obtain
\begin{align}\label{L2KuboThreePt}
\lambda_2 = 2\eta\tp - 4 \lim_{p,q\to 0} \frac{\partial^2}{\partial p^0 \partial q^z} G^{xy,ty,xz}_{RAA} (p,q).
\end{align}
A consistency check on our calculations is provided by the following two Kubo formulae which both give the expression for pressure:
\begin{align}\label{KuboP}
P = \lim_{p^0\to 0} \lim_{q^0\to 0} G^{xy,xz,yz}_{RAA} (p,q) = -  \lim_{p^z\to 0} \lim_{q^z\to 0} G^{xy,tx,ty}_{RAA} (p,q). 
\end{align}
Note that our definitions of transport coefficients are the same as in ref.~\cite{Baier:2007ix} (see Appendix \ref{appendixB}
 for a digest of notations and conventions used in the literature).

\subsection{The three-point functions in the hydrodynamic limit}
\label{sec:3-pt-f}
The three-point functions are computed by solving the Einstein-Gauss-Bonnet equations of motion \eqref{GBeom} to second order in relevant perturbations,
\begin{align}
g_{\mu\nu} \rightarrow g_{\mu\nu} + \epsilon r^2 h^{(1)}_{\mu\nu} + \epsilon^2 r^2 h^{(2)}_{\mu\nu},
\end{align}
where the book-keeping parameter $\epsilon$ is used to indicate the order of perturbation. The Dirichlet condition $h^{(2)}_{\mu\nu} = 0$ is imposed at the boundary 
\cite{Saremi:2011nh}. Once the bulk solutions are found, one should take the triple variation of 
the on-shell action with respect to the boundary values  $h^{(b)}_{\mu\nu} = h^{(1)}_{\mu\nu} \left(r\to\infty\right)$ to find the correlators. A 
simplifying feature of this procedure is that since equations of motion 
are solved to order $\epsilon^2$, only the boundary term contributes to the three-point function, and hence no bulk-to-bulk propagators appear in the calculation. 

To compute the three-point functions used in the Kubo formulae above, we need to 
 turn on the following sets of metric perturbations:
\begin{align}
&1) & &h_{xy} = h_{xy} (r) e^{-i\left(p^0 + q^0 \right) t }, & & h_{xz} = h_{xz} (r) e^{-i p^0  t }, & & h_{yz} = h_{yz} (r) e^{- i q^0  t }, \label{Pol1}\\
&2) & &h_{xy} = h_{xy} (r) e^{ i\left(p^z + q^z \right) z }, & & h_{tx} = h_{tx} (r) e^{ i p^z  z }, & & h_{ty} = h_{ty} (r) e^{i q^z  z }, \label{Pol2}\\
&3) & &h_{xy} = h_{xy} (r) e^{- i p^0 t + i q^z z }, & & h_{xz} = h_{xz} (r) e^{-i p^0  t }, & & h_{ty} = h_{ty} (r) e^{i q^z  z }. \label{Pol3}
\end{align}
Here, we outline the steps leading to obtaining the three-point function $G^{xy,xz,yz}_{RAA}$. First, we find the 
bulk solutions for $h_{xy}^{(1)}$, $h_{xz}^{(1)}$ and $h_{yz}^{(1)}$ imposing the standard incoming wave boundary condition at the horizon 
and the condition  $h^{(1)}_{\mu\nu}  = h^{(b)}_{\mu\nu}$ at the boundary. As in section \ref{Sec:Scalar}, it will be convenient to work with the 
radial variable $v$ defined by Eq.~\eqref{var-v}. 

Since the metric fluctuations in the set \eqref{Pol1} are independent of the spatial momentum, all three of them 
obey the same\footnote{Using the explicit expressions for the coefficients  $A_i$ and $B_i$ given in Appendix \ref{appendixC}, one 
can check that at vanishing spatial momentum they are the same in all channels.}
differential equation \eqref{eq:eom_GB_ginv}, and thus 
$h_{xy}^{(1)}$, $h_{xz}^{(1)}$ and $h_{yz}^{(1)}$ will have the same functional dependence on $v$. Moreover, we can use the 
solution to the equation already obtained in section \ref{Sec:Scalar}, with $\qfr$ set to zero and the 
 relevant frequencies, $p^0+q^0$, $p^0$ and $q^0$, inserted instead of $\wfr$, respectively. Thus, 
for $h_{xy}^{(1)}$ we find the expression
\begin{align}
\label{argoff}
&h_{xy}^{(1)} (v) =  h^{(b)}_{xy} \, \left(\frac{v}{2\lgb} \right)^{-\frac{i (p^0+q^0)}{4\pi T}} \Biggr[ 1 + \frac{i (p^0+q^0)}{8\pi T} \left( 4- v \right) v  + \frac{(p^0+q^0)^2}{4 \pi^2 T^2} g^{(\wfr)}_{2} (v)
 \nn 
&+ \frac{(p^0+q^0)^2}{16 \pi^2 T^2} \int^v \frac{(1-v')^2 \ln \left[\ggb ^2 -1 + v' -\sqrt{\left(\ggb ^2-1\right) \left(\ggb ^2-(1-v')^2\right)}\right]}{v'} \, dv' \Biggr],
\end{align}
and similar formulas for $h_{xz}^{(1)}$ and $h_{yz}^{(1)}$. We can deal with the remaining integral in Eq.~\eqref{argoff} in the same way as in section \ref{Sec:Scalar}, by integrating order-by-order in the near-boundary expansion. 

Next, we need to look for the second-order solution $h^{(2)}_{xy}$, which includes the first-order metric back-reaction. The differential equation again has the form of Eq. \eqref{ScalarRec} and can be solved using the same methods. The relevant part of the solution takes the following form:
\begin{align}
h^{(2)}_{xy} = h^{(b)}_{xz} h^{(b)}_{yz} \, \left(\frac{v}{2\lgb} \right)^{-i (p^0+q^0)/(4\pi T)} \, \frac{p^0 q^0}{4\pi^2 T^2} h(v),
\end{align}
with a complicated and unilluminating expression for $h(v)$ not shown here explicitly.

With the second-order solution in hand, we substitute the resulting formula for 
 $g_{\mu\nu} + \epsilon r^2 h^{(1)}_{\mu\nu} + \epsilon^2 r^2 h^{(2)}_{\mu\nu}$ into the expression for the holographic energy-momentum 
 tensor \eqref{emtensorfg} to compute $T^{xy}$. Finally, taking derivatives with respect to $h^{(b)}_{xz}$ and $h^{(b)}_{yz}$, we
  obtain $G^{xy,xz,yz}_{RAA}$:
\begin{align}
&G^{xy,xz,yz}_{RAA}(p,q) = \frac{\sqrt{2} \pi^4 T^4}{\left(1+\ggb\right)^{3/2} \kappa_5^2} - i \left(p^0 + q^0\right)  \frac{\sqrt{2} \pi^3 T^3}{\kappa_5^2} \frac{\ggb^2}{\left(1+\ggb\right)^{3/2}}  \nn
&+ \frac{(p^0)^2 + (q^0)^2}{2} \left[ \frac{\pi ^2 T^2}{ 2 \sqrt{2}\kappa_5^2 } \frac{ (\ggb +1) \left(\ggb  \left(\ggb ^2+\ggb -2\right)+2\right)+2 \ggb ^2 \ln \left[\frac{\ggb }{2 (1 + \ggb )}\right]     }{(1 + \ggb)^{3/2} } \right]  \nn
& + p^0 q^0 \left [    \frac{\pi^2T^2}{4\sqrt{2} \kappa_5^2}  \frac{ \left(-3 \ggb ^2+2 \ggb +11\right) \ggb ^2    -6  +2
   \ggb ^2 \ln \left[\frac{\ggb }{2 (1+\ggb )}\right]   }{(1+\ggb )^{3/2}} \right]  .
\end{align}

The other three-point functions,  $G^{xy,tx,ty}_{RAA}$ and $G^{xy,ty,xz}_{RAA}$, are computed using the same procedure, with the differential equations always taking the form of \eqref{eq:eom_GB_ginv}. The only difference is that we cannot impose the in-falling boundary conditions on perturbations $h_{tx}$ or $h_{ty}$
in Eq.~\eqref{Pol2}, and similarly on $h_{ty}$ in Eq.~\eqref{Pol3}, because they only fluctuate in the $z$-direction and not time. Regularity then demands 
setting $h_{tx} = h_{ty} = 0$ at the horizon. Consequently, $h_{xy}$ in Eq.~\eqref{Pol2} also needs to vanish at the horizon.\footnote{The full expressions for the other two three-point functions are very cumbersome and will not be written here explicitly. For an example of a technically 
simpler but conceptually identical  calculation in $\CN=4$ SYM theory, see refs. \cite{Saremi:2011nh,Grozdanov:2014kva}.}

\subsection{Second-order transport coefficients and the zero-viscosity limit}
\label{sec:SubSecSecondOrderTransport}
Having computed in the hydrodynamic approximation the three-point functions $G^{xy,xz,yz}_{RAA}$, $G^{xy,tx,ty}_{RAA}$ and $G^{xy,ty,xz}_{RAA}$, we can use the Kubo formulae to compute pressure \eqref{KuboP}, shear viscosity \eqref{3ptKuboEta} and all second-order transport coefficients \eqref{3ptKuboTpKappa} -- \eqref{L2KuboThreePt}. The result for pressure coincides with the one in Eq.~\eqref{Pressure}, and the shear viscosity is 
confirmed to be given by Eq.~\eqref{etagb}. For the second-order transport coefficients we find $(L=1)$:
\begin{align}
&\eta\tp = \frac{\pi^2 T^2}{4\sqrt{2}\kappa_5^2}  \frac{\ggb}{\left(1+\ggb\right)^{\frac{3}{2}}}  \Biggl[ \left(1+\ggb\right) \left(5\ggb+\ggb^2 -2\right)- 2 \ggb  \ln \left(\frac{2 \left(1+\ggb\right)}{\ggb }\right)\Biggr], \label{cosm1}\\
&\kappa  = \frac{\pi^2 T^2}{\sqrt{2} \kappa_5^2}\left( \frac{2\ggb^2 - 1}{\sqrt{1+\ggb}} \right)\,, \label{cosm2} \\ 
&\lambda_1 = \frac{\pi^2 T^2}{2\sqrt{2}\kappa_5^2} \left( \frac{3 - 4\ggb +2\ggb^3}{\sqrt{1+\ggb}} \right) , \\   
&\lambda_2 = - \frac{\pi^2 T^2}{2\sqrt{2}\kappa_5^2}  \frac{\ggb}{\left(1+\ggb\right)^{\frac{3}{2}}}  \left( \left(1+\ggb \right) \left(2-\ggb - \ggb^2 \right) + 2 \ggb  \ln \left[\frac{2 \left(1+\ggb\right)}{\ggb }\right]\right), \label{cosm3}\\
&\lambda_3 = - \frac{\sqrt{2} \pi^2 T^2}{\kappa_5^2} \left( \frac{3+\ggb-4\ggb^2}{\sqrt{1+\ggb}}  \right). \label{cosm4}
\end{align}
Alternatively, the coefficients $\lambda_1$, $\lambda_2$, $\lambda_3$ can be expressed in terms of the shear viscosity, as in Eqs.~\eqref{l0} -- \eqref{l4}. In the absence of the Gauss-Bonnet term in the action, i.e. for  $\lgb = 0$ ($\ggb = 1$), the results reduce to those obtained for ${\cal N}=4$ SYM \cite{Baier:2007ix,Bhattacharyya:2008jc}:
\begin{align}
\eta\tp  =  \frac{\eta\left(2-\ln 2\right)}{2\pi T},& &\lambda_1 =   \frac{\eta}{2\pi T},& &\lambda_2 =  - \frac{\eta \ln 2}{\pi T},&&\lambda_3 = 0 ,&& \kappa = \frac{\eta}{\pi T}.
\end{align}
In the limit of zero viscosity, i.e. for $\lgb=1/4$ ($\ggb=0$), we find 
\begin{align}
\eta\tp  =  0 ,&& \lambda_1 =   \frac{3 \pi^2 T^2}{2 \sqrt{2}\kappa_5^2} ,&&\lambda_2 =  0 ,&&\lambda_3 = - \frac{3 \sqrt{2} \pi^2 T^2}{\kappa_5^2} ,&& \kappa = - \frac{\pi^2 T^2}{\sqrt{2} \kappa_5^2} .
\end{align}
Thus, three of the five second-order transport coefficients do not vanish in the limit of zero viscosity. However, the criteria for the liquid to be dissipationless (i.e. producing no entropy) analyzed in ref.~\cite{Bhattacharya:2012zx},
\begin{align}
\eta  =  0 ,&& \kappa = 2 \lambda_1\,, && 2\eta\tp - 4 \lambda_1 - \lambda_2  =  0\,,
\end{align}
are not satisfied in this limit \cite{Grozdanov:2015asa}.

\begin{figure}[htbp]
\centering
\includegraphics[width=1\textwidth]{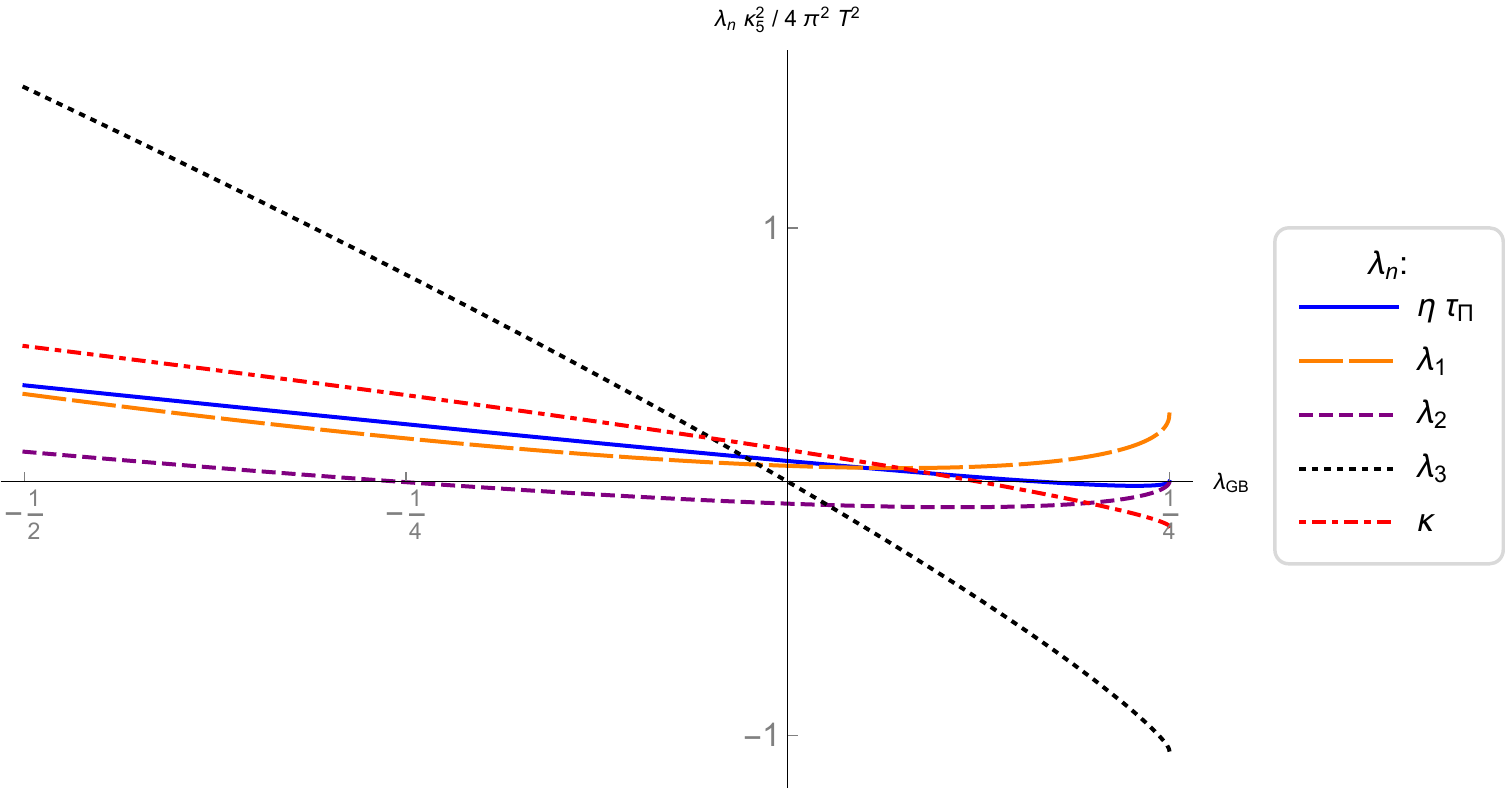}
\caption{Second-order coefficients $\lambda_n = \{\eta\tp,\lambda_1,\lambda_2,\lambda_3,\kappa\}$ of Gauss-Bonnet holographic liquid, 
in units of $4 \pi^2 T^2 / \kappa_5^2$, as functions of $\lgb$.}
\label{Fig:CoeffsAll}
\end{figure}

The five second-order coefficients $\lambda_n = \{\eta\tp,\lambda_1,\lambda_2,\lambda_3,\kappa\}$ (represented by the dimensionless ratios, $\lambda_n \kappa_5^2 / 4\pi^2 T^2$) 
are shown as functions of $\lgb$ in Fig.~\ref{Fig:CoeffsAll}. 
While $\lambda_1$ is positive-definite for all $\lgb$, other coefficients can have either sign. 

\begin{figure}[htbp]
\centering
\includegraphics[width=1\textwidth]{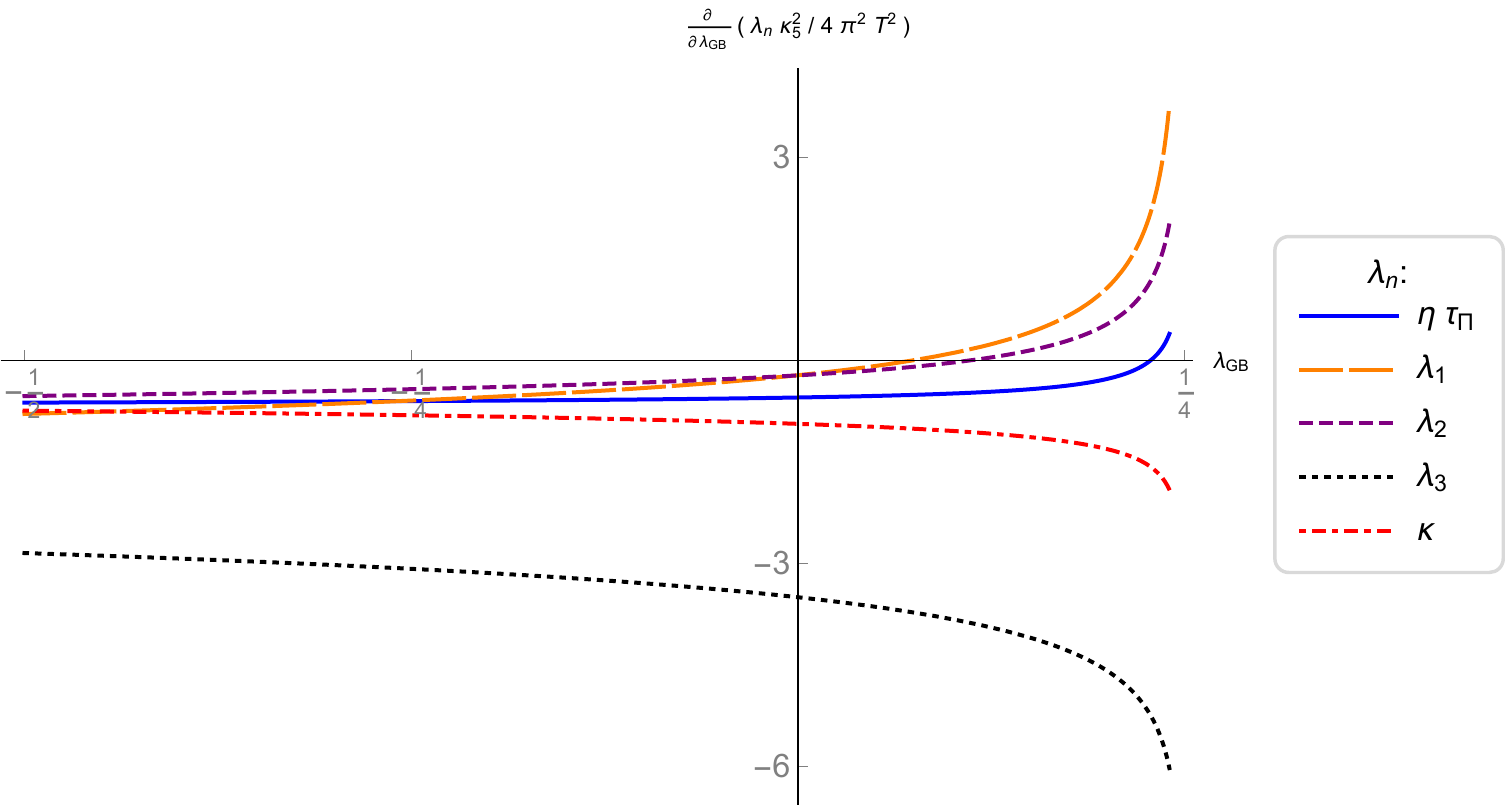}
\caption{Derivatives of the second-order coefficients $\lambda_n = \{\eta\tp,\lambda_1,\lambda_2,\lambda_3,\kappa\}$ 
with respect to $\lgb$, in units of $4 \pi^2 T^2 / \kappa_5^2$, as functions of $\lgb$.}
\label{Fig:CoeffsDerAll}
\end{figure}

The derivatives of the coefficients with respect to $\lgb$ are shown in Fig.~\ref{Fig:CoeffsDerAll}. The coefficients $\lambda_3$ and $\kappa$ are monotonically decreasing functions of $\lgb$ as can be seen from
\begin{align}
&\frac{ \kappa^2_5}{4\pi^2T^2} \frac{\partial \lambda_3}{\partial\lgb} = - \frac{1+15\ggb+12\ggb^2}{2\sqrt{2}\ggb(1+\ggb)^{3/2}} < 0 ,  \label{l3mon} \\
&\frac{ \kappa^2_5}{4\pi^2T^2} \frac{\partial \kappa}{\partial\lgb} = - \frac{1+8\ggb+6\ggb^2}{4\sqrt{2}\ggb(1+\ggb)^{3/2}} < 0, \label{kappamon}
\end{align} 
whereas the coefficients $\eta\tp$, $\lambda_1$ and $\lambda_2$ are not. 

\section{Charge diffusion from higher-derivative Einstein-Maxwell-Gauss-Bonnet action}\label{sec:ChargeDiffSection}

Can first-order transport coefficients other than shear viscosity be tuned to zero with a suitable choice of higher derivative bulk terms, 
and can this be done simultaneously with tuning to zero the viscosity? In this section, we compute non-perturbative corrections to the well known result for the $U(1)$ charge diffusion constant at infinite coupling \cite{Policastro:2002se} in a hypothetical boundary theory dual to
Einstein-Maxwell-Gauss-Bonnet gravity with the  charge neutral black brane background \eqref{BB}. 
\subsection{The four-derivative action}
\label{sec:FourDerivativeMaxwell}
We are interested in the four-derivative Einstein-Maxwell-Gauss-Bonnet action whose equations of motion  involve at most second derivatives. Such theories 
were previously considered in refs.~\cite{Anninos:2008sj,Kats:2006xp}, and in the context of an effective target-space heterotic string theory 
action in \cite{Gross:1986mw}.\footnote{See \cite{Liu:2008kt} for a discussion of field redefinitions in higher derivative Einstein-Maxwell theories.} The higher-derivative Maxwell terms may appear as a result of compactification, e.g. of a higher-dimensional 
Gauss-Bonnet action. Here we construct the necessary action directly.

We begin by considering the Einstein-Maxwell-Gauss-Bonnet theory with the most general four-derivative Maxwell field Lagrangian, 
\begin{align}
\label{emgb-action}
S = \frac{1}{2\kappa^2_5} \int d^5 x \sqrt{-g} \left[ R - 2 \Lambda + \CL_{GB} \right] + \int d^5 x \sqrt{-g} \CL_A ,
\end{align}
where $\Lambda = -6/L^2$, the Gauss-Bonnet Lagrangian $\CL_{GB}$ is given by Eq.~\eqref{GBaction}, and
\begin{align}
&\CL_A =- \frac{1}{4} F_{\mu\nu} F^{\mu\nu} + \alpha_4 R F_{\mu\nu} F^{\mu\nu} + \alpha_5 R^{\mu\nu} F_{\mu\rho} F_\nu^{~\rho} + \alpha_6 R^{\mu\nu\rho\sigma} F_{\mu\nu} F_{\rho\sigma} + \alpha_7 \left(F_{\mu\nu} F^{\mu\nu} \right)^2 \nn
&  + \alpha_8 \nabla_\mu F_{\rho\sigma} \nabla^{\mu} F^{\rho\sigma} + \alpha_9 \nabla_\mu F_{\rho\sigma} \nabla^{\rho} F^{\mu\sigma} +  \alpha_{10}  \nabla_\mu F^{\mu\nu} \nabla^{\rho} F_{\rho\nu} + \alpha_{11} F^{\mu\nu} F_{\nu\rho} F^{\rho\sigma} F_{\sigma\mu} .
\end{align}
The coupled equations of motion for  $g_{\mu\nu}$ and $A_\mu$ following from the action \eqref{emgb-action} are written in Appendix \ref{sec:EMGB}. 
To make third- and fourth-order derivatives of the fields vanish in the equations of motion \eqref{EINeom}, we must impose the following constraints on the coefficients $\alpha_n$
\begin{align}
&\alpha_4 = \alpha_6, && 8 \alpha_4 + \alpha_5 - 4 \alpha_6 = 0, \\
&4 \alpha_4 + \alpha_5 - 2 \alpha_8 - \alpha_9 = 0, &&2 \alpha_8 + \alpha_9 + \alpha_{10} = 0  \label{ConMax}.  
\end{align}
The second constraint in \eqref{ConMax} also ensures that all higher-order derivatives vanish from the Maxwell's equations \eqref{MAXeom}. 
The constraints can be solved by setting
\begin{align}
\alpha_6 = \alpha_4 ,&&\alpha_5 = -4 \alpha_4 , &&\alpha_9 = -2 \alpha_8 ,&&\alpha_{10} = 0.
\end{align} 
Coefficients $\alpha_7$ and $\alpha_{11}$ are left undetermined by this procedure. The vector field Lagrangian becomes
\begin{align}
\CL_A =& - \frac{1}{4} F_{\mu\nu} F^{\mu\nu} + \beta_1 L^2 \left( R F_{\mu\nu} F^{\mu\nu} - 4 R^{\mu\nu} F_{\mu\rho} F_\nu^{~\rho} + R^{\mu\nu\rho\sigma} F_{\mu\nu} F_{\rho\sigma} \right)  \nn
&  + \beta_4 L^2 \nabla_\mu F_{\rho\sigma}  \left( \nabla^{\mu} F^{\rho\sigma} -  2 \nabla^{\rho} F^{\mu\sigma} \right)  + \beta_2 L^2 \left(F_{\mu\nu} F^{\mu\nu} \right)^2 + \beta_3 L^2 F^{\mu\nu} F_{\nu\rho} F^{\rho\sigma} F_{\sigma\mu},
\end{align}
where we have defined the dimensionless couplings $\beta_1 \equiv \alpha_4/L^2$, $\beta_2 \equiv \alpha_7/L^2$, $\beta_3 \equiv \alpha_{11}/L^2$ and $\beta_4 \equiv \alpha_8/L^2$. To simplify the Lagrangian further, we notice that the term proportional to $\beta_4$ can be rewritten as
\begin{align}
\nabla_\mu F_{\rho\sigma}  \left( \nabla^{\mu} F^{\rho\sigma} -  2 \nabla^{\rho} F^{\mu\sigma} \right) = - 2 \nabla^\mu \nabla^\rho A^\sigma \left( R^\lambda_{~\mu\rho\sigma} + R^\lambda_{~\rho\sigma\mu} +R^\lambda_{~\sigma\mu\rho} \right) A_\lambda = 0,
\end{align}
hence the entire expression vanishes due to the cyclic property of the Riemann tensor. Thus the  Lagrangian $\CL_A$ leading to second-order equations of motion  is given by
\begin{align}\label{HDerMaxLag}
\CL_A =& - \frac{1}{4} F_{\mu\nu} F^{\mu\nu} + \beta_1 L^2 \left( R F_{\mu\nu} F^{\mu\nu} - 4 R^{\mu\nu} F_{\mu\rho} F_\nu^{~\rho} + R^{\mu\nu\rho\sigma} F_{\mu\nu} F_{\rho\sigma} \right)  \nn
&  + \beta_2 L^2 \left(F_{\mu\nu} F^{\mu\nu} \right)^2 + \beta_3 L^2 F^{\mu\nu} F_{\nu\rho} F^{\rho\sigma} F_{\sigma\mu}.
\end{align}
Therefore, there are altogether four parameters, $\lgb$, $\beta_1$, $\beta_2$ and $\beta_3$, entering the second-order equations of motion of the 
theory. One may wonder if a black hole (brane) solution with non-perturbative values of these parameters exists. The black brane 
metric \eqref{BB} is automatically a solution of the theory when $A_\mu = 0$. It is also possible  to find perturbative corrections in $\beta_1$, $\beta_2$ and $\beta_3$ to the five-dimensional AdS-Reissner-Nordstr\"{o}m metric. However, we were not able to find a generalization of the 
solution  \eqref{BB} with non-trivial $A_\mu$ and fully non-perturbative non-vanishing $\beta_1$, $\beta_2$ and $\beta_3$.\footnote{An asymptotically AdS black hole solution to the theory considered in this section with $\beta_1 = 0$ was found in an integral form and studied in \cite{Anninos:2008sj}. Unfortunately, for $\beta\neq 0$ the equations are significantly more complicated. In particular, in the relevant metric ansatz, $ds^2 = - e^{2\lambda} dt^2 + e^{2\nu} dt^2 + \ldots$, the relation $\lambda = - \nu$ is no longer true.} 

\subsection{The $U(1)$ charge diffusion constant}

To compute the charge diffusion constant in a hypothetical neutral liquid dual to the bulk action constructed in the previous section we follow the procedure outlined in \cite{Kovtun:2005ev}. We begin by perturbing the trivial $A_\mu = 0$ background vector field as $A_\mu \to A_\mu + \epsilon a_\mu$ and writing the electromagnetic field strength corresponding to the linearized perturbation as $F =  \epsilon \text{d} a$. Given the trivial background $A_\mu = 0$, the metric fluctuations decouple from $a_\mu$ and can be set to zero.\footnote{Charge diffusion in a three dimensional boundary theory, including the  $\beta_1$ term, was computed in a neutral Einstein-Hilbert black brane background in \cite{Myers:2010pk}.}

In the equations of motion, the terms proportional to $\alpha_7$ and $\alpha_{11}$ (i.e. $\beta_2$ and $\beta_3$) only contribute to quadratic or 
higher orders in the expansion in $\epsilon$. Hence, they will not contribute to the charge diffusion constant. The linearized equations obeyed by $a_\mu$ read
\begin{align}
\nabla_\nu F^{\mu\nu} & =  4 \beta_1 L^2 \nabla_\nu \left(R F^{\mu\nu} + R^{\mu\nu\rho\sigma} F_{\rho\sigma} - R^{\mu\rho} F_{\rho}^{~\nu} + R^{\nu\rho}F_\rho^{~\mu}   \right)  .
\end{align}
Vector field fluctuations can be decomposed into transverse and longitudinal modes, with charge diffusion coming from the 
low-energy hydrodynamic excitations in the longitudinal sector. By choosing the spatial
 momentum along the $z$-direction, the relevant gauge-invariant variable in the longitudinal sector is
\begin{align}
Z_4 = \qfr a_0 + \wfr a_4.
\end{align} 
We use the variable $u = r_+^2 / r^2$, with the boundary  at $u=0$ and horizon at $u=1$. Then we impose the incoming wave
 boundary condition required for the calculation of retarded correlators \cite{Son:2002sd} by writing
\begin{align}
Z_4 = \left(1 - u^2\right)^{-i\wfr/2}  \CZ_4(u)\,,
\end{align}
where the function $\CZ_4(u)$ regular at the horizon can be found perturbatively in $\mu\ll 1$, with $\qfr$ and $\wfr$ scaling as $\wfr \to \mu^2 \wfr$ and $\qfr \to \mu \qfr$. We find it useful to introduce a new variable ${\it w}$, so that $u = \sqrt{{\it w}^2 - \ggb^2}/ \sqrt{1-\ggb^2}$. The boundary is now at ${\it w}=\ggb$ and horizon at ${\it w}=1$. At order $\CO(\mu^0)$, the function $\CZ_4$ can be written as $\CZ_4 = C_1 + C_2\, z({\it w})$, where $z({\it w})$ is a solution of the equation
\begin{align}
\frac{d^2 z}{d{\it w}^2} -\frac{48 \beta_1 \left({\it w}^3-\ggb ^2\right)-\ggb ^2 \left(1-\ggb^2\right)}{{\it w} \left({\it w}^2-\ggb ^2\right) \left(1-\ggb ^2+48 \beta_1 (1-{\it w}) \right)} \frac{dz}{d{\it w}} = 0.
\end{align}
We solve for $z({\it w})$ and impose the boundary conditions $z(\ggb) = 1$ and $z(1) = 0$. The constant $C_2$ can then be expressed as a function of $C_1$, $\wfr$, $\qfr$ and other parameters of the theory by substituting $z({\it w})$ into the original differential equation, expanding to order $\CO(\mu^2)$ and imposing regularity at the horizon.

The hydrodynamic quasinormal mode can be found by solving the equation $Z_4 (\wfr,\qfr) = 0$ at the boundary for $\wfr$. The dispersion relation has the form
\begin{align}
\wfr = -i \CD \qfr^2 + \CO(\qfr^4)\,,
\end{align}
where $\CD$ is the charge diffusion constant of the dual theory. For the Gauss-Bonnet coupling in the interval $\lgb \in [0,1/4]$ ($1 \geq  \ggb \geq 0$) we find\footnote{It is also possible to write an explicit formula for  $\CD$ in the interval  $\lgb < 0$ ($\ggb >1$) but here we are mostly interested in the dissipatinless limit $\lgb \rightarrow 1/4$.}
\begin{align}\label{FullDiffusion}
\CD =&~ \frac{(1+\ggb)  (1+2 \beta) \left( \beta + \sqrt{\beta ^2-\ggb^2} \right)    }{6 (\beta -1) \left[\beta 
   \left(\beta + \sqrt{\beta ^2-\ggb^2} \right)-\ggb^2\right] }    
  \Biggl\{ \sqrt{ \left(1-\ggb^2\right) \left(\beta ^2-\ggb^2\right) } \,\ln \left[\frac{\ggb }{1+\sqrt{1-\ggb^2}}\right] \nn
&-\left(\beta
   -\ggb^2\right) \ln \left[\frac{\ggb }{\beta + \sqrt{\beta ^2-\ggb^2} }\right]  \Biggr\} \,,
\end{align}
where $\beta \equiv 1 + 48 \beta_1$ and $\ggb \equiv \sqrt{1-4\lgb} $. 

We can now consider various limits. For the two-derivative Maxwell field in Gauss-Bonnet background, i.e. for $\beta_1 = 0$ ($\beta=1$), we find the expression
\begin{align}\label{MaxwellDiffusion}
\CD = \frac{1}{2} \left( 1 + \sqrt{1 - 4  \lgb} \right).
\end{align}
For $\CN=4$ SYM theory, where $\lgb = 0$ and $\beta_1 = 0$, Eq.~\eqref{MaxwellDiffusion} reproduces the well-known result \cite{Policastro:2002se}, 
\begin{align}
\CD = 1.
\end{align}
In the zero viscosity limit $\lgb = 1/4$ ($\ggb=0$), Eq.~\eqref{MaxwellDiffusion} gives
\begin{align}
\CD = 1/2.
\end{align}
In the presence of higher-derivative vector-field terms in the Lagrangian \eqref{HDerMaxLag}, we find the diffusion constant in the two important 
limits of $\lgb$ to be
\begin{align}
&\lgb = 0: && \CD = \left( \frac{1+ 32  \beta_1}{4\sqrt{6} \sqrt{\beta_1 \left( 1+ 24 \beta_1 \right)  } } \right)  \ln  \left[ 1+ 48 \beta_1 + \sqrt{ \left( 1 + 48\beta_1 \right)^2 - 1  }  \right],  \\
&\lgb = 1/4: && \CD= \left( \frac{1+32 \beta_1}{96 \beta_1 } \right)  \ln \left( 1 + 48 \beta_1 \right) .
\label{diss-lim}
\end{align}
From Eq.~\eqref{diss-lim} one can see that for $\lgb = 1/4$, the diffusion constant $\CD$ remains a real function of the 
parameter $\beta_1$ as long as $\beta_1 > - 1/48$ and, moreover, this function is strictly positive for all $\beta_1$ in that interval. In this sense, we cannot have vanishing shear viscosity and diffusion constant simultaneously. The diffusion constant can vanish for other values of $\lgb$: 
for example, $\CD = 0$ for  $\beta_1 = - 1/32$. However, such a solution for $\CD$ is not smoothly connected to the theory which has a vanishing shear viscosity. More precisely, for any $\beta_1 = -1 /32 + \epsilon$, where $\epsilon \ll 1$, $\CD$ is complex near $\ggb = 0$.

\section{Conclusions}
\label{sec:conclusions}
Together with refs.~\cite{Grozdanov:2015asa,Grozdanov:2014kva,Grozdanov:2016vgg}, the present paper is an attempt at a 
comprehensive investigation of the second-order transport properties, energy-momentum tensor correlation functions and quasinormal spectrum in 
the Gauss-Bonnet holographic fluid in $D=3+1$ 
dimensions non-perturbatively in Gauss-Bonnet coupling $\lgb$. The existence of a strongly coupled CFT dual to classical non-Einsteinian
gravity such as Gauss-Bonnet gravity at finite $\lgb$ would be an interesting alternative to the standard scenario of gauge-gravity duality. 
However, the work of Camanho {\it et al.} \cite{Camanho:2014apa} appears to cast a serious doubt on such a possibility, reducing the status of the curvature-squared terms to that of a perturbative correction. At the same time, we have not found any obvious pathology in hydrodynamic properties of the 
hypothetical dual field theory at finite $\lgb$. 

The curvature-squared terms are interesting even as corrections to the Einstein's gravity description of a dual field theory, the second-order nature of the Gauss-Bonnet equations of motion making it easier to search for the new features such as the extra poles of the correlators not seen at $\lgb =0$. 
The analysis of gravitational quasinormal spectrum in ref.~\cite{Grozdanov:2016vgg} and in 
the present paper shows that  the analytic structure of  dual thermal correlators is qualitatively different depending on the sign of $\lgb$ (understood as inverse coupling), with the $\lgb < 0$  case showing "normal" (e.g. qualitatively similar to  ${\cal N}=4$ SYM at finite 't Hooft coupling and infinite $N_c$ 
and having a potential to connect to the kinetic regime) 
features, and the  $\lgb > 0$  case demonstrating various anomalies (whose precise meaning remains to be understood, possibly invoking various monotonicity arguments).\footnote{As shown in ref.~\cite{Grozdanov:2016vgg}, for $\lgb > 0$ (i.e. for $\eta / s < 1/4\pi$), 
the two symmetric branches of non-hydrodynamic quasinormal modes gradually move out to complex infinity with $\lgb$ increasing in the interval $[0,1/4]$. This implies that the relaxation times associated with the modes, $\tau_R = 1 / \left| \text{Im}\left( \omega \right) \right| $, tend to zero, thereby violating any conjectured lower bound on $\tau_R$ \cite{sachdev-book-2}. Furthermore, new quasinormal modes appear along the imaginary axis, approaching the analytically known results from Section \ref{sec:GBExtremeCoupling} in the limit $\lgb \rightarrow 1/4$, which can cause an instability in the system at a finite spatial momentum above certain critical value. Hydrodynamic poles move towards the real axis with vanishing dissipative parts in the limit.} On the other hand, constraints on $\lgb$ may come from different considerations such as the recent argument for $\lgb > 0$ in ref.~\cite{Cheung:2016wjt} based on unitarity. Fortunately, corrections coming from 
$R^2$ and $R^4$ higher derivative terms seem to be very similar \cite{Grozdanov:2016vgg} in uncovering 
a qualitative picture of transport and other properties at large but finite coupling.

\acknowledgments{We would like to thank Andres Anabalon, Tomas Andrade, Alex Buchel, Jos\'{e} Edelstein, Pavel Kovtun, Shiraz Minwalla, 
Rob Myers, Andy O'Bannon, Giuseppe Policastro, Massimo Porrati, Rakibur Rahman, Mukund Rangamani, Harvey Reall, Larry Yaffe, 
Amos Yarom and Alexander Zhiboedov for discussions. The work of A.O.S. was 
supported by the European Research Council under the
European Union's Seventh Framework Programme (ERC Grant agreement 307955). A.O.S. would like to thank the Institute for Nuclear Theory at the 
University of Washington (Seattle, USA) and the Mainz Institute for Theoretical Physics (Mainz, FRG) for their kind hospitality 
and partial support during the completion of this work.}


\appendix

\section{Second-order transport coefficients of ${\cal N} =4$ SYM at weak and strong coupling}
\label{appendixN=4SYM}
For the finite-temperature ${\cal N}=4$ $SU(N_c)$ supersymmetric Yang-Mills (SYM) theory in $d=3+1$ dimensions in the limit of infinite $N_c$ and infinite 't Hooft coupling $\lambda = g^2_{YM}N_c$, first- and second-order transport coefficients were computed, correspondingly, in \cite{Policastro:2001yc} and \cite{Baier:2007ix,Bhattacharyya:2008jc} using methods of gauge-gravity and fluid-gravity dualities:
\begin{align}
\eta &= \frac{\pi}{8} N^2_c T^3\,, \label{tc1} \\
\tau_{\Pi}  &= \frac{\left( 2 - \ln{2}\right)}{2\pi T}\,, \qquad \kappa = \frac{\eta}{\pi T}\,, \qquad
\lambda_1 = \frac{\eta}{2\pi T}\,, \qquad \lambda_2  =- \frac{\eta \ln{2}}{\pi T}\,, \qquad \lambda_3 = 0\,. \label{tc2}
 \end{align}
Coupling constant corrections to the coefficients (\ref{tc1}), (\ref{tc2}) can be determined from the higher-derivative terms in the low-energy effective action of type IIB string theory
\begin{align}
S = \frac{1}{2\kappa_5^2} \int d^5 x \sqrt{-g} \left(R  + \frac{12}{L^2} + \gamma \CW \right),
\label{hd-action}
\end{align}
where $\gamma = \alpha'^3 \zeta(3) / 8$, $L$ is the AdS curvature scale, and the ratio $\alpha'/L^2$ is related to the value 
of the 't Hooft coupling $\lambda$ in ${\cal N}=4$ SYM 
via $\alpha' / L^2 = \lambda^{-1/2}$. The effective five-dimensional gravitational 
constant is connected to $N_c$ by $\kappa_5 = 2\pi /N_c$. The term $\CW$ is given in terms of the Weyl tensor $C_{\mu\nu\rho\sigma}$ by
\begin{align}
\CW = C^{\alpha\beta\gamma\delta}C_{\mu\beta\gamma\nu} C_{\alpha}^{~\rho\sigma\mu} C^{\nu}_{~\rho\sigma\delta} + \frac{1}{2} C^{\alpha\delta\beta\gamma} C_{\mu\nu\beta\gamma} C_{\alpha}^{~\rho\sigma\mu} C^\nu_{~\rho\sigma\delta}.
\end{align}
Corrections to all first and second-order transport coefficients are known \cite{Buchel:2004di, Benincasa:2005qc, Buchel:2008sh, Buchel:2008ac,Buchel:2008bz,Buchel:2008kd,Saremi:2011nh,Grozdanov:2014kva}:
\begin{align}
\eta &= \frac{\pi}{8} N^2_c T^3\, \left( 1 + \frac{135 \zeta (3)}{8}\, \lambda^{-3/2} + \ldots \,  \right)\,, \label{tcsym1} \\
\tau_{\Pi}  &= \frac{ \left( 2 - \ln{2}\right)}{2\pi T}   + \frac{375 \zeta(3)}{32 \pi T} \, \lambda^{-3/2} +\ldots \,, \label{tcsym2} \\
\kappa &= \frac{N_c^2 T^2}{8}  \left( 1 - \frac{5 \zeta (3)}{4}\,  \lambda^{-3/2} + \ldots \,  \right)\,, \label{tcsym3} \\  
\lambda_1 &=  \frac{N_c^2 T^2}{16}  \left( 1 + \frac{175\zeta (3)}{4} \, \lambda^{-3/2} + \ldots \,  \right)\,,  \label{tcsym4}  \\
\lambda_2 &=- \frac{N_c^2 T^2}{16} \left( 2\ln{2}   + \frac{5 \left(97+54 \ln{2}\right) \zeta(3)}{ 8 } \lambda^{-3/2}  + \ldots \,\right), \label{tcsym5} \\
\lambda_3 &=  \frac{ N_c^2 T^2}{16} \, 25 \zeta (3) \,  \lambda^{-3/2} +\ldots \, . \label{tcsym6}
\end{align}
Leading order results for the third order coefficients $\theta_1$ and $\theta_2$ entering the hydrodynamic dispersion relations are known as well \cite{Grozdanov:2015kqa}:
\begin{align}
 \theta_1 &= \frac{N_c^2 T}{32 \pi} + O(\gamma)\,, \\
\theta_2 &= \frac{N_c^2 T}{384 \pi} \left( 22 - \frac{\pi^2}{12} - 18 \ln{2} +\ln^2 2 \right)+ O(\gamma)\,.
\end{align}
Additional explicit results for the linear combinations of ${\cal N}=4$ SYM third order coefficients can be found in ref.~\cite{Grozdanov:2015kqa}. Other coupling constant corrections to the results at infinitely strong t'Hooft coupling in finite temperature ${\cal N}=4$ SYM include corrections to the entropy \cite{Gubser:1998nz,Pawelczyk:1998pb}, photon emission rate \cite{Hassanain:2012uj}, and poles of the retarded correlator of the energy-momentum tensor \cite{Stricker:2013lma}. 

In ${\cal N}=4$ SYM at weak coupling, the shear viscosity has been computed  in ref.~\cite{Huot:2006ys}. The second-order transport coefficients in various theories at weak coupling (QCD with either 0 or 3 flavours, QED, $\lambda \phi^4$) were determined by York and Moore \cite{York:2008rr}. In conformal kinetic theory (at weak coupling) one finds $2 \eta \tau_{\Pi} + \lambda_2 =0$ \cite{Baier:2007ix,York:2008rr,Betz:2008me}. Curiously, in the theories considered in  \cite{York:2008rr} 
the Haack-Yarom relation (\ref{HY}) at weak coupling can be expressed as
\begin{equation}
H = \frac{4 \eta^2}{\epsilon + P} \left (C_1 - C_2\right) ,
\end{equation}
where $\epsilon$ and $P$ are energy density and  pressure, correspondingly, and $C_1$ and $C_2$ are theory-specific constants (e.g. $C_1 \approx 6.10517$,  $C_2 \approx 6.13264$ for $\lambda \phi^4$ theory). It appears that at weak coupling one finds $H\neq 0$  also in other (nearly conformal) 
examples \cite{GMoore}. It would be interesting to compute $H(\lambda)$ directly in ${\cal N}=4$ SYM at weak coupling. Another interesting finding of ref.~\cite{York:2008rr} is that at weak coupling the coefficients $\kappa$ and $\lambda_3$ vanish to order $\propto T^2/\lambda^4$ (but may be non-zero at $\propto T^2/\lambda^2$). We note that $\lambda_3 =0$ in the limit $\lambda \rightarrow \infty$ but has a non-trivial coupling dependence as can be seen from (\ref{tcsym6}).

\section{Notations and conventions in formulas of relativistic hydrodynamics}
\label{appendixB}
For the convenience of the reader, here we compare notations and sign conventions used in the present paper with those used in refs.~\cite{Baier:2007ix}, 
\cite{York:2008rr}, \cite{Grozdanov:2014kva},  \cite{Bhattacharyya:2008jc}, \cite{Bhattacharya:2012zx}, \cite{Haack:2008xx},  \cite{Shaverin:2012kv}.

\subsection*{Notations and conventions used in the present paper and in refs.~\cite{Baier:2007ix}, 
\cite{York:2008rr}, \cite{Grozdanov:2014kva}.}
The energy-momentum tensor of a neutral conformal relativistic fluid considered in the Landau frame is written as
\begin{align}
T^{ab} = \varepsilon u^a u^b + P \Delta^{ab} + \Pi^{ab},
\end{align}  
where $\Delta^{ab} \equiv g^{ab} + u^a u^b$, pressure $P$ and energy density $\varepsilon$ are related by the conformal fluid equation of state in four dimensions, $P = \varepsilon/3$, 
and 
\begin{align}
\Pi^{ab} =&~ -\eta \sigma^{ab} + \eta \tp \left[ {}^{\langle}D\sigma^{ab\rangle} + \frac{1}{d-1} \sigma^{ab} \left(\nabla\cdot u\right) \right] + \kappa \left[ R^{\langle ab \rangle} - (d-2)u_c R^{c \langle ab \rangle d} u_d  \right]  \nn
&+\lambda_1 \sigma^{\langle a}_{~~c} \sigma^{b\rangle c}  +\lambda_2 \sigma^{\langle a}_{~~c} \Omega^{b\rangle c}  +\lambda_3 \Omega^{\langle a}_{~~c} \Omega^{b\rangle c},
\end{align}
where $D \equiv u^a \nabla_a$. We use the following definitions (in our case, $d=4$)
\begin{align}
A^{\langle ab \rangle} \equiv \frac{1}{2} \Delta^{ac} \Delta^{bd} \left(A_{cd} + A_{dc}\right) - \frac{1}{d-1} \Delta^{ab} \Delta^{cd} A_{cd} \equiv {}^{\langle} A^{ab\rangle}, 
\end{align}
where by construction the resulting  tensors are transverse, $u_a A^{\langle ab \rangle} = 0$, traceless, $g_{ab} A^{\langle ab \rangle} = 0$, and symmetric. The tensor $\sigma^{ab}$ is a symmetric, transverse and traceless tensor involving first derivatives of the velocity field
\begin{align}
\sigma^{ab} = 2 {}^\langle \nabla^{a}u^{b\rangle}.
\end{align}
The vorticity $\Omega^{\mu\nu}$ is defined as an anti-symmetric, transverse and traceless one-derivative tensor
\begin{align}
\Omega^{ab} = \frac{1}{2} \Delta^{ac} \Delta^{bd} \left(\nabla_c u_d - \nabla_d u_c \right).
\end{align}
The Haack-Yarom relation in our notations reads
\begin{align}
\label{HY-conv}
2 \eta \tau_\Pi - 4 \lambda_1 -\lambda_2 = 0\,,
\end{align}
whereas the conformal kinetic theory result \cite{York:2008rr} is 
\begin{align}
2 \eta \tau_\Pi  + \lambda_2 = 0\,.
\end{align}
\subsection*{Notations and conventions used in refs.~\cite{Bhattacharyya:2008jc}, \cite{Bhattacharya:2012zx}}
We label the objects used in refs.~\cite{Bhattacharyya:2008jc, Bhattacharya:2012zx} with the letter "R", e.g.
\begin{align}
\sigma^{\mu\nu}_R = P^{\mu\alpha}P^{\nu\beta} \partial_{(\alpha}u_{\beta )} -\frac{1}{3} \partial_\alpha u^\alpha\,,
\end{align}
where $P^{\mu\nu} = \eta^{\mu\nu} + u^\mu u^\nu$,
 $a^{( \alpha }b^{\beta )} = (a^\alpha b^\beta + a^\beta b^\alpha)/2$, 
 $a^{[ \alpha }b^{\beta ]} = (a^\alpha b^\beta - a^\beta b^\alpha)/2$.
The vorticity is defined as
\begin{align}
\omega^{\mu\nu}_R = - \frac{1}{2} P^{\mu \alpha} P^{\nu \beta} \left(\partial_\alpha u_\beta - \partial_\beta u_\alpha \right).
\end{align}
It is clear that $\sigma_{\mu\nu}^R = \frac{1}{2} \sigma_{\mu\nu}$ and $\omega_{\mu\nu}^R = - \Omega_{\mu\nu}$.
The energy-momentum tensor is written as
\begin{align}
T^{\mu\nu} = \varepsilon u^\mu u^\nu + p P^{\mu\nu} + \Pi^{\mu\nu},
\end{align}  
where 
\begin{align}
\Pi^{\mu\nu} =&~ -\eta \sigma^{\mu\nu}_R + \dots + \lambda_2^R  
\sigma^{\langle \mu}_{~~\lambda, R} \omega^{\nu\rangle \lambda}_R  + \cdots \,.
\end{align}
Therefore, $\lambda_2^R = - 2 \lambda_2$. Similar relations hold for other coefficients. In summary,
\begin{align}
\eta &= \eta^R\,, \\
 \eta \tau_\Pi &= \frac{1}{2} \tau_R\,, \\
 \kappa &= \kappa_1^R = \frac{1}{2} \kappa_R\,, \\
 \lambda_1 &= \frac{1}{4} \lambda_1^R\,, \\
\lambda_2 &= - \frac{1}{2} \lambda_2^R\,, \\
 \lambda_3 &= - \lambda_3^R\,.
\end{align}
The Haack-Yarom relation reads
\begin{align}
2  \tau^R - 2 \lambda_1^R + \lambda_2^R = 0\,
\end{align}
or, equivalently, for liquids with $\lambda_1^R = \kappa^R$ (i.e. $\kappa = 2 \lambda_1$ in our notations)
\begin{align}
2  \tau^R - 2 \kappa^R + \lambda_2^R = 0\,.
\end{align}
The conformal kinetic theory result \cite{York:2008rr} in these notations reads 
\begin{align}
2  \tau^R  - \lambda_2^R = 0\,.
\end{align}
\subsection*{Notations and conventions used in refs.~\cite{Haack:2008xx}, \cite{Shaverin:2012kv}}
In ref.~\cite{Haack:2008xx}, the tensor $\sigma^{\mu\nu}_{HY}$ is 
defined as 
\begin{align}
\sigma^{\mu\nu}_{HY} = 2 {}^\langle \nabla^{\mu}u^{\nu\rangle}
\end{align}
and the vorticity $\omega^{\mu\nu}_{HY}$ is 
\begin{align}
\omega^{\mu\nu}_{HY} = \frac{1}{2} P^{\mu\lambda} P^{\nu\sigma} \left(\nabla_\lambda u_\sigma - \nabla_\sigma u_\lambda \right),
\end{align}
which coincides with the definitions in \cite{Baier:2007ix}, 
\cite{York:2008rr}, \cite{Grozdanov:2014kva}. The term 
in the expression for the energy-momentum tensor multiplying $\lambda_2^{HY}$,
\begin{align}
T^{\mu\nu} =&~  \lambda_2^{HY}\,  \sigma^{\langle \mu}_{~~\lambda , \mbox{\tiny{HY}}} \omega^{\lambda \nu\rangle}_{HY}  + \cdots \,,
\end{align}
is different in the order of indices from the one used in \cite{Baier:2007ix}, 
\cite{York:2008rr}, \cite{Grozdanov:2014kva}, where $T^{ab} = \lambda_2\,  \sigma^{\langle a}_{~~c} \Omega^{b\rangle c}  + \dots$, and, since vorticity is antisymmetric, we could have concluded that $\lambda_2 = -  \lambda_2^{HY}$ (?). Then the original Haack-Yarom relation as stated in  ref.~\cite{Haack:2008xx},
\begin{align}
2 \eta^{HY} \tau_\Pi^{HY} - 4 \lambda_1^{HY} -\lambda_2^{HY} = 0\,,
\label{hywrong}
\end{align}
would translate to our notations as (all other coefficients coincide with ours) 
 \begin{align}
2 \eta \tau_\Pi - 4 \lambda_1 + \lambda_2 = 0\, \;\; \mbox{\bf (incorrect)}
\end{align}
which does not agree with Eq.~\eqref{HY-conv} and is difficult to reconcile e.g. with the explicit results for ${\cal N}=4$ SYM given by \eqref{tc1} - \eqref{tc2}. We believe that there is a typo in  ref.~\cite{Haack:2008xx}, either in the arrangement of indices (it should be the same as in \cite{Baier:2007ix}, 
\cite{York:2008rr}, \cite{Grozdanov:2014kva}) or, alternatively, in the definition of vorticity (it should have an extra minus sign in front), or perhaps in the sign in front of $\lambda_2$ in the equation (\ref{hywrong}). The same observation has been recently made in ref.~\cite{Haehl:2015pja}. Correcting this typo, we have $\lambda_2 = \lambda_2^{HY}$ and then notations in  \cite{Haack:2008xx} would give the same signs of transport coefficients as the ones in refs.~\cite{Baier:2007ix}, 
\cite{York:2008rr}, \cite{Grozdanov:2014kva}.

We note that in the paper by Shaverin and Yarom \cite{Shaverin:2012kv}, the notations for  $\sigma^{\mu\nu}_{SY}$,  vorticity $\omega^{\mu\nu}_{SY}$ and their coupling $ \lambda_2^{SY}\,  \sigma^{\langle \mu}_{~~\alpha , \mbox{\tiny{SY}}} \omega^{\alpha \nu\rangle}_{SY} $ are the same as in  ref.~\cite{Haack:2008xx}. The relations between our transport coefficients (i.e. the ones in \cite{Baier:2007ix}, 
\cite{York:2008rr}, \cite{Grozdanov:2014kva}) and the ones used in  \cite{Shaverin:2012kv} are
\begin{align}
\eta^{SY} = \eta\,, \\
 \lambda_0^{SY} = \eta \tau_\Pi\,, \\
  \lambda_1^{SY} = \lambda_1\,, \\
\lambda_2^{SY} = - \lambda_2\,, \\
 \lambda_3^{SY} = \lambda_3\,.
\end{align}
The Haack-Yarom relation as written in  \cite{Shaverin:2012kv} reads
\begin{align}
- 2 \lambda_0^{SY}  + 4 \lambda_1^{SY} -\lambda_2^{SY} = 0\,,
\label{sy}
\end{align}
which translates in our notations into Eq.~\eqref{HY-conv}, as expected.

\section{Boundary conditions at the horizon in the hydrodynamic regime}
\label{sec:BC-horizon}
In this Appendix, we clarify the procedure of imposing the incoming wave boundary condition at the horizon on a 
(gauge-invariant) fluctuation
$Z$ given by a perturbative series in the hydrodynamic regime  ($\wfr \ll 1$ and $\qfr \ll 1$). Consider such a solution $Z_1$ 
near the horizon $u=1$:
\begin{equation}
Z_1 = (1-u)^{-i \wfr/2} F(u,\wfr)\,,
\end{equation}
where $\qfr$ is ignored for simplicity. Here, the function $F$ (regular at $u=1$ by Fr\"{o}benius construction) is found perturbatively as a series in $\wfr\ll 1$,
\begin{equation}
 F(u,\wfr) = F_0 (u) + \wfr F_1 (u) + \wfr^2 F_2 (u) + \cdots\,,
\end{equation}
where $F_i (u)$ satisfy the equation of motion obeyed by $Z$ to a given order in $\wfr$, 
with $F_i (1) = S_i$ for $i\geq 0$, and $S_i$ are constants independent of $u$ and $\wfr$. Now consider another solution, $Z_2$, near $u=1$,
\begin{equation}
Z_2 = (1-u)^{-i \wfr/2} C(\wfr ) G(u,\wfr)\,,
\end{equation}
where $C(\wfr)$ is a function of $\wfr$ only, and $G$ is found perturbatively by solving the differential equation obeyed by $Z$ by a series in $\wfr\ll 1$,
\begin{eqnarray}
&\,& G(u,\wfr) = G_0 (u) + \wfr G_1 (u) + \wfr^2 G_2 (u) + \cdots\,, \label{expa1} \\
&\,& G_0(1) = 1, G_i (1) = 0 \; \mbox{for} \;  i > 0\,. \label{expa2}
\end{eqnarray}
Now, the solution $Z_1$ with its boundary condition at the horizon can always be written as $Z_2$ with the appropriate choice of the 
function $C(\wfr)$. Indeed, 
expanding $C(\wfr)$ in Taylor series at $\wfr=0$, $C(\wfr)=C(0) + C'(0) \wfr+...$, we get
\begin{eqnarray}
Z_2 &= (1-u)^{-i \wfr/2} \Biggl\{  G_0 (u) C(0) + \wfr \left[ G_0 (u) C'(0) + G_1(u) C(0) \right]  \nonumber \\
&+ \wfr^2 \left[ \frac{1}{2} \, G_0(u) C''(0)\,  + G_1(u) C'(0)  + C(0) G_2(u) \right] + O(\wfr^3) \Biggr\}\,.
\end{eqnarray}
Comparing $Z_2$ and $Z_1$ at $u=1$, we identify $C(0)=S_0$, $C'(0)=S_1$, and so on. In other words, nontrivial boundary conditions at the horizon 
for functions $F_i$ of the solution $Z_1$ can be understood as coefficients of the small-$\wfr$ expansion of a multiplicative factor $C(\wfr)$. Since in holography we work with bulk solutions normalized to one at the boundary, i.e. $f_k(u)=Z(u)/Z(\epsilon)$, such a multiplicative factor cancels. This justifies  always using the expansion \eqref{expa1} 
with the boundary conditions  \eqref{expa2}.
\section{The coefficients $A_i$ and $B_i$ of the differential equation (\ref{eq:eom_GB_ginv})}
\label{appendixC}
{\bf Scalar channel}
\begin{align}
A_1 = \, & -\frac{1}{u}   - u \left[ \frac{1}{\left(\ggb ^2-1\right) \left(1-u^2\right)^2 + 1 - u^2} + \frac{1}{\left(1-u^2\right) \sqrt{\ggb ^2-\left(\ggb^2-1\right) u^2}}\right], \\
B_1 = \, &\frac{(\ggb -1) (\ggb +1)^2 \left(3 \left(\ggb ^2-1\right) u^2-\ggb ^2\right)\left(-\ggb ^2+\left(\ggb ^2-1\right) u^2+U\right)}{4 u \left(\ggb^2-\left(\ggb ^2-1\right) u^2\right)^{3/2} \left(-\ggb ^2+\left(\ggb
   ^2-1\right) u^2+2 U-1\right)} \qfr^2 \nn
\, & + \frac{\left(\ggb ^2-1\right)^2 \left(-\ggb ^2+\left(\ggb ^2-1\right)u^2+U\right)}{4 u (U-1) \sqrt{\ggb ^2-\left(\ggb ^2-1\right) u^2} \left(-\ggb^2+\left(\ggb ^2-1\right) u^2+2 U-1\right)}\wfr^2,
\end{align}
\\
{\bf Shear channel}
\begin{align}
A_2 = \,&-\frac{2 \ggb ^4 (\ggb +1)  \left[\frac{1}{2} \left(1-\ggb ^2\right) \left(u^2-1\right) (U-2)+U-1\right]}{u (U-1) U^3 \left[\ggb ^2 (\ggb +1) (U-1)  \qfr^2  -\left(\ggb ^2-1\right) U^2 \wfr ^2\right]}  \qfr^2 \\
\,&-\frac{\left(1-\ggb ^2\right) \left(\ggb ^4+\left(1-\ggb^2\right)^2 u^4- 2 \left(1-\ggb ^2\right) u^2 \left(U-\ggb^2\right)-\ggb ^2 U\right)}{u (U-1) U \left[\ggb ^2 (\ggb +1)(U-1)  \qfr^2  -\left(\ggb ^2-1\right) U^2 \wfr ^2\right]}   \wfr^2, \\
B_2 = \,&  \frac{\ggb ^2 (\ggb +1) (U+1)}{4 u \left(u^2-1\right) U^2} \qfr^2 + \frac{\left(U^2+2 U+1\right) }{4 u \left(u^2-1\right)^2}  \wfr^2 ,
\end{align}
\\
{\bf Sound channel}
\begin{align}
A_3 =\,& \frac{3}{2 u} + \frac{3 (\ggb -1)  \left[\left(\ggb ^2-1\right) u^2-\ggb^2\right] \left[\left(\ggb ^2-1\right) u^2 (5 U-7)-5 \ggb ^2(U-1)\right]}{2 u (U-1) U^2 D_1} \wfr^2 \nn
\,&+ \frac{ \left(\ggb ^2-1\right)^2 u^4 \left(-3 \ggb ^2+5 U-7\right)+\ggb ^2 \left(\ggb ^2-1\right) u^2 \left(18 \ggb ^2-13 U+10\right) }{2 u (U-1) U^2 D_1} \qfr^2 \nn
\,&- \frac{ 15 \ggb ^4 \left(\ggb ^2-2 U+1\right) }{2 u (U-1) U^2 D_1} \qfr^2, \\
B_3 =&~ \frac{\left(\ggb ^2-1\right)^2}{D_{0}} \biggr\{ \,12 (\ggb -1)^2 \ggb ^2 (\ggb +1) \qfr^2 u^5-4 (\ggb -1) \ggb ^2 \qfr^2 u^3 \left(3 \ggb ^2-7 U+4\right) \nn
&+ \left(\ggb ^2-1\right)^3 \qfr^2 u^6 \left(3 (\ggb -1) \wfr ^2+\qfr^2\right) \nn
& -u^2 \ggb ^2 \left(\ggb ^2-1\right)  \left[\qfr^4 \left(\ggb ^2+2 U\right)+(\ggb -1) \qfr^2 \wfr ^2 \left(9 \ggb ^2-4 U\right)-6 (\ggb-1)^2 U \wfr ^4\right] \nn
&  + \left(\ggb ^2-1\right)^2 u^4 \left[\qfr^4 \left(3 \ggb ^2 (U-2)+U\right)+2(\ggb -1) \qfr^2 U \wfr ^2-3 (\ggb -1)^2 U \wfr ^4\right]\nn
& - 3 \ggb ^4 \left[\qfr^4 \left(\ggb ^2 (U-2)+U\right)+2 (\ggb -1) \qfr^2 \wfr^2 \left(U-\ggb ^2\right)+(\ggb -1)^2 U \wfr ^4\right]  \,\, \biggr\},
\end{align}
where we have defined
\begin{align}
D_1 &\equiv \left(\ggb ^2-1\right) u^2 \left(3(\ggb -1) \wfr^2+\qfr^2\right)+3 \ggb ^2 \left(\qfr^2 (U-1)-(\ggb -1)\wfr^2\right), \nn
D_0 &\equiv 4 (\ggb -1) u (U-1)^2 U^3 D_1 .
\end{align}
In the above expressions, we used $U^2 = u^2 + \ggb^2 - u^2 \ggb^2$, as well as the dimensionless 
frequency and momentum (\ref{wq-gothic}), where the Hawking temperature is given by Eq.~\eqref{EntropyAndEnergy}.
Sometimes it is preferable to use the original radial coordinate $r$. 
For convenience, we write here the equation for the scalar fluctuation $Z_1$ in this variable:
\begin{align}\label{EqScalar}
P_2 Z_1'' + P_1 Z_1' + P_0 Z_1= 0, 
\end{align}
where the coefficients are given by
\begin{align}
&P_2 = r f \left(\lgb  f' - r \right) , \\
&P_1 = r f' \left(\lgb  f' - r \right) - 3 r f + \lgb f \left( r f'' + 2 f' \right), \\
&P_0 = \frac{2}{f \left(1+\sqrt{1-4\lgb}\right)} \biggr[ r \omega ^2 \left(\lgb  f'-r\right) -\left(1+\sqrt{1-4 \lgb}\right) f^2 
\left(\lgb    f''-1\right)  \nn
&~~~~~ +\frac{1}{2} \left(1+\sqrt{1-4 \lgb}\right) f \left(f'' \left(r^2- \lgb q^2\right)-2 \lgb f'^2+4 r f'+q^2 -12 r^2\right)  \biggr]\,,
\end{align}
where $f(r)$ is given by Eq.~(\ref{f-black}). To solve Eq.~\eqref{EqScalar} with the incoming wave boundary condition at the horizon, it is convenient to write the solution as
\begin{align}\label{ZsolScalarApp}
Z_1 = \tilde f(r)^{-i\wfr/2} \left(1 + g(r) \right),
\end{align}
where
\begin{align}
\tilde f(r) = \frac{1}{2\lgb} \left[1 - \sqrt{1-4\lgb \left(1 - \left(r_+/r\right)^4 \right) } \right].
\end{align}
This coordinate is more convenient for taking the limit of zero temperature.

\section{Equations of motion of  Einstein-Maxwell-Gauss-Bonnet gravity}
\label{sec:EMGB}
The equations of motion of Einstein-Maxwell-Gauss-Bonnet gravity following from the action \eqref{emgb-action} form a system of two coupled PDEs:
\begin{align}\label{EINeom}
&R_{\mu\nu} - \frac{1}{2} g_{\mu\nu} R + g_{\mu\nu} \Lambda = \CT^{GB}_{\mu\nu} +  2 \kappa^2_5 \CT^{A}_{\mu\nu}, \\
&\nabla_\nu F^{\mu\nu}  =  4 \alpha_4 \nabla_\nu \left(R F^{\mu\nu} \right) + 2 \alpha_5 \nabla_\nu \left(R^{\mu\rho} F_{\rho}^{~\nu} - R^{\nu\rho}F_\rho^{~\mu} \right) + 4 \alpha_6 \nabla_\nu \left(R^{\alpha\beta\mu\nu} F_{\alpha\beta} \right) \nn
&+ 8 \alpha_7 \nabla_\nu \left( F_{\alpha\beta} F^{\alpha\beta} F^{\mu\nu} \right)  -  4 \alpha_8 \nabla_\nu \Box F^{\mu\nu} - 2 \alpha_9 \nabla_\nu \nabla_\rho \left( \nabla^\mu F^{\rho\nu} - \nabla^\nu F^{\rho\mu} \right)   \nn
&+ 2 \alpha_{10} \nabla_\nu \left(\nabla^\nu \nabla_\rho F^{\rho\mu} - \nabla^\mu \nabla_\rho F^{\rho\nu}  \right)  +  8 \alpha_{11} \nabla_\nu \left( F^{\nu\rho} F_{\rho\sigma} F^{\sigma \mu} \right).
\label{MAXeom}
\end{align}
Here, the gravitational energy-momentum tensor term is given by
\begin{align}
\CT^{GB}_{\mu\nu} =& \; \frac{\lgb L^2}{4} g_{\mu\nu} \left( R^2 - 4 R_{\mu\nu} R^{\mu\nu} + R_{\mu\nu\rho\sigma} R^{\mu\nu\rho\sigma}  \right) \nn  
&- \lgb L^2 \left( R R_{\mu\nu} - 2 R_{\mu\alpha} R_{\nu}^{~\alpha} - 2 R_{\mu\alpha\nu\beta} R^{\alpha\beta} + R_{\mu\alpha\beta\gamma} R_\nu^{\alpha\beta\gamma}  \right),
\end{align}
and the Maxwell field contribution has the form
\begin{align}
\CT^{A}_{\mu\nu} &= - \frac{1}{8}\left( g_{\mu\nu} F^2 - 4 F_{\mu\lambda} F_{\nu}^{~\lambda} \right) \nn
&+\frac{\alpha_4}{2} \left[ g_{\mu\nu} R F^2 - 4 R F_{\mu\alpha} F_\nu^{~\alpha} - 2 R_{\mu\nu} F^2 + 2 \nabla_\mu \nabla_\nu F^2 - 2 g_{\mu\nu} \Box F^2 \right] \nn
&+ \frac{\alpha_5}{2} \left[ g_{\mu\nu} R^{\alpha\beta} F_{\alpha\lambda} F_\beta^{~\lambda} -4 R_{\mu\alpha} F_{\nu\beta} F^{\alpha\beta} - 2 R_{\alpha\beta} F_\mu^{~\alpha} F_\nu^{~\beta} - \Box \left(F_{\mu\alpha} F_\nu^{~\alpha} \right) - g_{\mu\nu} \nabla_\alpha \nabla_\beta \left( F^\alpha_{~\lambda} F^{\beta\lambda}\right) \right. \nn
&~~~~~~~  \left.~+ \nabla_\alpha \nabla_\mu \left(F_{\nu\beta} F^{\alpha\beta} \right) + \nabla_\alpha \nabla_\nu \left( F_{\mu\beta} F^{\alpha\beta} \right) \right]  \nn
&+ \frac{\alpha_6}{2} \left[ g_{\mu\nu} R^{\alpha\beta\gamma\delta} F_{\alpha\beta} F_{\gamma\delta} -6 R_{\mu\alpha\beta\gamma} F_\nu^{~\alpha} F^{\beta\gamma} - 4 \nabla^\beta \nabla^\alpha \left(F_{\mu\alpha} F_{\nu\beta} \right) \right] \nn
&+ \frac{\alpha_7}{2} \left[ g_{\mu\nu} \left(F^2\right)^2 - 8 F^2 F_{\mu\lambda} F_\nu^{~\lambda} \right] \nn
&+ \frac{\alpha_8}{2} \left[ g_{\mu\nu} \nabla_\alpha F_{\beta\gamma} \nabla^\alpha F^{\beta\gamma} - 2 \nabla_\mu F_{\alpha\beta} \nabla_\nu F^{\alpha\beta} - 4 \nabla_\alpha F_{\mu\beta} \nabla^\alpha F_\nu^{~\beta}  + 4 \nabla_\alpha \left( \nabla_\mu F^{\alpha\beta} F_{\nu\beta}  \right)   \right. \nn
&~~~~~~~  \left.~+ 4 \nabla_\alpha \left( \nabla^\alpha F_{\mu}^{~\beta} F_{\nu\beta}  \right)  - 4 \nabla_\alpha \left( \nabla_\mu F_\nu^{~\beta} F^\alpha_{~\beta}  \right)    \right]  \nn
&+ \frac{\alpha_9}{2} \left[ g_{\mu\nu} \nabla_\alpha F_{\beta\gamma} \nabla^\beta F^{\alpha\gamma} - 2 \nabla_\alpha F_{\mu\beta} \nabla^\beta F_\nu^{~\alpha} - 4 \nabla_\mu F_{\alpha\beta} \nabla^\alpha F_\nu^{~\beta}  + 2 \nabla_\alpha \left( \nabla^\alpha F_\mu^{~\beta} F_{\nu\beta}  \right)   \right. \nn
&~~~~~~~  \left.~+ 2 \nabla_\alpha \left( \nabla_\mu F^{\alpha\beta} F_{\nu\beta}  \right)  - 2 \nabla_\alpha \left(F^\alpha_{~\beta}  \nabla_\nu F_\mu^{~\beta}   \right)    \right]  \nn
&+ \frac{\alpha_{10}}{2} \left[ g_{\mu\nu} \nabla_\alpha F^{\alpha\gamma} \nabla^\beta F_{\beta\gamma} -2 g_{\mu\nu} \nabla_\alpha \left(F^{\alpha\gamma}\nabla^\beta F_{\beta\gamma}    \right)  - 4 \nabla_\mu F_{\nu\beta} \nabla_\alpha F^{\alpha\beta} - 2 \nabla^\alpha F_{\mu\alpha} \nabla^\beta F_{\nu\beta}    \right. \nn
&~~~~~~~  \left.~+ 4 \nabla_\mu \left(  F_{\nu\beta} \nabla_\alpha F^{\alpha\beta} \right)  +4 \nabla^\alpha \left( F_{\mu\alpha} \nabla^\beta F_{\nu\beta}  \right)    \right]  \nn
&+ \frac{\alpha_{11}}{2} \left[ g_{\mu\nu} F^{\alpha\beta} F_{\beta\gamma} F^{\gamma\delta} F_{\delta\alpha} -  8  F_{\mu\alpha} F_{\nu\beta} F^{\alpha\gamma} F^\beta_{~\gamma}  \right]   .
\end{align}

\bibliographystyle{JHEP}
\bibliography{refs}

\end{document}